\documentclass[printer]{aa}  

\usepackage{graphicx}
\usepackage{txfonts}
\usepackage{bm}
\usepackage[]{hyperref}%draft
\usepackage[dvipsnames]{xcolor}
\usepackage{color}
\definecolor{nico_color}{RGB}{47, 143, 158}
\usepackage{array}
\usepackage{bbm}
\usepackage{tikz}
\usepackage{subcaption}
\renewcommand*{\vec}[1]{\bm{#1}}

\newcolumntype{C}[1]{>{\centering\let\newline\\\arraybackslash\hspace{0pt}}m{#1}}

\newcommand{\ii}{\mathrm{i}}
\DeclareMathOperator{\e}{e}

\DeclareMathOperator{\cov}{\mathrm{cov}}

\DeclareMathOperator{\sinc}{\mathrm{sinc}}

\renewcommand*{\vec}[1]{\bm{#1}}
\newcommand{\dd}{{\rm d}}

\def\E{\mathbb{E}}
\def\d{\mathrm{d}}
\def\spleaf{\textsc{s+leaf}}

\def\celerite{\textit{celerite}}

\begin{document} 

    %\title{An analytical model of stellar activity}
    %\title{An analytical model of photometric and radial velocity variations induced by stellar activity}
    %\title{Analytical kernels for photometric and radial velocity variations induced by stellar activity}
	%\title{A stochastic model of stellar activity}
	%\title{Stellar activity as a stochastic process}
	%\title{Stellar observables as stochastic processes}
	%\title{Stellar variability as a stochastic process}
	\title{A statistical model of stellar variability}
	%\subtitle{I. Physics-based, fast models of stellar activity signals with FENRIR }
	\subtitle{FENRIR: a physics-based model of stellar activity, and its fast Gaussian process approximation}
	%\subtitle{I. Transferring physical hypotheses on }	
	\titlerunning{Analytical kernels for stellar activity}

%An 
 %An analytical model of signals induced by g
	
	\author{Nathan C. Hara 
		\inst{\ref{i:geneve}}, 
		Jean-Baptiste Delisle 
		\inst{\ref{i:geneve}}}
		
	\institute{
		Observatoire Astronomique de l’Université de Genève, 51 Chemin de Pegasi, 1290 Versoix, Switzerland\label{i:geneve}  \email{nathan.hara@unige.ch}
	}

	%       \date{Received September 15, 1996; accepted March 16, 1997}
	
	% \abstract{}{}{}{}{} 
	% 5 {} token are mandatory
	
	\abstract
	% context heading (optional)
	{The detection of terrestrial planets by radial velocity and photometry is hindered by the presence of stellar signals. Those are often modeled as stationary Gaussian processes, whose kernels are based on qualitative considerations, which do not fully leverage the existing physical understanding of stars.
	%It is not clear how this model relates to the physical phenomena occurring on the star. %Terrestrial exoplanets can be detected with the transit and radial velocity techniques. The data are affected by complex stellar signals, which are not yet modelled well enough to allow the detection of Earth-like planets in the habitable zone.
	%Stellar signals are often modelled as stationary Gaussian processes, whose kernels are chosen based on qualitative considerations.
% %As such, they are fully characterised by autocovariance functions, or kernels, whose form stems from qualitative considerations on stellar rotation, spots and faculae appearance and convection in the photosphere. Some kernel were developed to speed up the computations and make these stellar activity models scalable.
	}       % \keywords{Exoplanets}
	{
	%We define a new formalism building the model of photometric and spectroscopic 
	%Our aim is improve the accuracy and interpretability of the models of stellar signals.
 Our aim is to build a formalism which allows to transfer the knowledge of stellar activity into practical data analysis methods. In particular, we aim at obtaining kernels with physical parameters. This has two purposes: better modelling signals of stellar origin to find smaller exoplanets, and extracting information about the star from the statistical properties of the data. 
 % To that end, we build a statistical framework allowing to transfer the physical knowledge of stars into practical data analysis methods.   
	}
	{ 	
	%To bridge the gap between the physical understanding of stellar signals and the methods used to analyse the data, 
	%We build a physics-driven model handling simultaneously stellar induced signals in photometry, radial velocity and other types of spectroscopic indicators, called finite energy random impulse response (FENRIR). 
	%Our framework consists in modelling the effect of a stellar feature (spot, facula, granulation cell) on several time series of observations of a given star: RV, photometry, spectroscopic indicators, we build a model of
	We consider several observational channels such as photometry, radial velocity, activity indicators, and build a model called FENRIR to represent their stochastic variations due to  stellar surface inhomogeneities. %which appear following a Poisson process. 
	%Given the effect of a single feature as a function of its parameters (maximum size, latitude, brightness, lifetime...) and the probability distribution of these parameters, 
	We compute analytically the covariance of this multi-channel stochastic process, and implement it in the S+LEAF framework to reduce the cost of likelihood evaluations from $O(N^3)$ to $O(N)$. We also compute analytically higher order cumulants of our FENRIR model, which quantify its non-Gaussianity.
	%In the radial velocity effects, we include both the photometric breaking of approaching and receding limb, convective blueshift inhibition and granulation. Differential rotation can be included at the cost of losing the  $O(N)$ computational cost.
	}
	{%We show that the properties of our processes are governed by the effect of a single feature.  closed-loop relation between the radial velocity, photometry and spectral indicators, and
%The Gaussian process model we obtained for radial velocities and photometry can be used as other Gaussian process models, but have physically motivated parameters.
 We obtain a fast Gaussian process framework with physical parameters, which we apply to the HARPS-N and SORCE observations of the Sun, and constrain a solar inclination compatible with the viewing geometry.
 We then discuss the application of our formalism to granulation. We exhibit non-Gaussianity in solar HARPS radial velocities, and argue that information is lost when stellar activity signals are assumed to be Gaussian. We finally discuss the origin of phase shifts between RVs and indicators, and how to build relevant activity indicators.
 %We argue that stellar activity is non stationary because of magnetic cycles, and that some information is lost when stellar activity signals are assumed to be Gaussian. We suggest an explanation for the non-Gaussianity of solar RVs based on the inhibition of convective blueshift.   Our formalism offers insight on how to build relevant statistical indicators, and we discuss its application to granulation. 
 We provide an open-source implementation of the FENRIR Gaussian process model with a Python interface. 
 %We discuss the application of our framework to observation strategies and Doppler imaging.
	%Our framework yields the cross-correlation of photometry, radial velocity, and other spectral indicators such as $logR'_{HK}$, H$\alpha$, FWHM, etc. This generalises in particular the $FF'$ method to model jointly photometric and radial velocity variations to arbitrary inclinations and limb-darkening laws. %The model requires only an analytical expression of one stellar spot and facula as a function of its parameters. The calculation of the signal shape with limb-darkening covariance and its Fourier expansion are done automatically. 
	%The variations are non stationary due to the presence of magnetic, and is found non Gaussian based on our calculation of high order cumulants.  We show that the analytical kernels can help interpret the parameters of existing kernels. Furthermore they can be used, in principle, to infer the stellar inclination through a statistical Rossiter-McLaughlin effect. We show that the phase shift between the radial velocity and indicators can be linked to the ratio of photometric to convective blueshift inhibition effect. 
	}
	{}
	
	\maketitle
	\section{Introduction}
	\label{sec:introduction}
	%\subsection{Stellar activity as a limitation to Earth-like planet detection}
	
   % \paragraph{Science case: why it is important to detect small planets} %As of September 2022, over 5000 exoplanets are known. Their sizes ranges from smaller than Mars to several Jupiter masses. %The study of their demographics has greatly enhanced our knowledge of the planet formation mechanisms. 

    %The photometric mission Kepler~\citep{borucki2011} has allowed to do the statistics of small planets down to ..... With the latest generation spectrographs, one can detect and measure the mass of exoplanets down to ... . 
    %In both cases, the farther exoplanets are from their host star, and the harder they are to detect. As the orbital period increases, the probability of transit decreases, and there are less and less transits or RV oscillations to stack upon one another, and the signature of ingress and egress becomes less and less sharp in the photometry, the RV variation is also at lower and lower frequency, making it more difficult to distinguish from low frequency instrumental or astrophysical noise. The photometric mission Kepler~\citep{borucki2011} has allowed to do the statistics of small planets down to ..... With the latest generation spectrographs, one can detect and measure the mass of exoplanets down to ... . 
    
    %For both techniques, the farther exoplanets are from their host star, and the harder they are to detect, in particular because the signature of planets becomes harder to disentangle from low frequency instrumental or astrophysical noise.
    
    Besides a few exceptions, the smallest known exoplanets have been detected either with the transit or the radial velocity (RV) observational methods. Unfortunately, both techniques  have not yet given detections of Earth twins. %: Earth mass planets orbiting a Sun-like star at a separation of 1 AU.  
    In the coming decade it will be crucial to push the detection limits with RV and photometry. % Such planets have been detected in M and K dwarves, but true Earth twins around G stars are currently inaccessible. Pushing the detection limits beyond them serves several purpose.  
    The PLATO mission, to be launched in 2026, will search for Earth twins with photometry. Measuring their mass with a precision of 10\% through RV follow up is also part of the core science objective of the mission~\citep{rauer2016}. A good precision on the mass is also required to interpret robustly the observation of their atmosphere \citep[around 20\%][]{batalha2019}. The existence of a population of Earth like planets at a few AUs can be an outcome of the pebble accretion formation model, and accessing this parameter space would help further test planetary formation scenarios \citep[][]{lambrechts2019}. Finally, the detection of Earth twins with radial velocity within 20 pc would pave the way for the search for life outside the solar systems, through their atmospheric characterization with Habitable Worlds Observatory \citep{crass2021}, and LIFE \citep{quanz2021}.

\begin{figure*}
  \centering
  \includegraphics[width=0.93\linewidth]{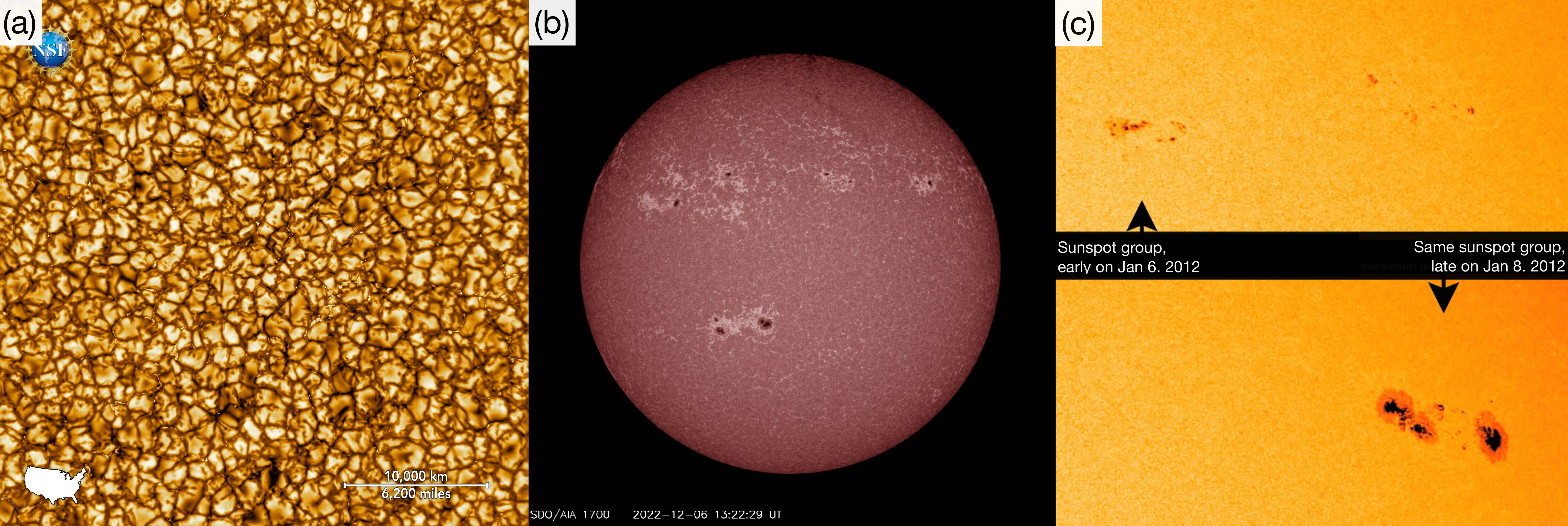}
  \caption{(a) Inouye Solar Wave Front Correction (WFC) image, captured Jan 28, 2020, at 789nm. The granulated structures around the spot are due to convection: hot plasma moves upwards at the center of granules, cools down and goes downwards between granules (darker inter-granular regions). The contribution of brighter intra-granular region exceeds that of intergranular ones, creating a so-called convective bluedhift on observed spectra.  Credit: NSO/AURA/NSF. (b)  SDO observation of the Sun at wavelength 1700 $\odot{A}$. The bright regions are called faculae, notice that faculae areas surround the spots (darker regions). Both faculae and spots inhibit the upward convective motion of the gas. Faculae cover more area than spots, but have a smaller temperature contrast with the continuum. The latitude, number and lifetime of the spots varies along stellar magnetic cycles, on the timescale of a few years (11 years for the Sun).  Courtesy of NASA/SDO and the AIA, EVE, and HMI science teams. (c) Group of sunspots observed at two different dates, top: Jan 6 2012, bottom: Jan 8 2012. Courtesy of NASA/SDO.
 }
  \label{fig:granmag}
  \end{figure*}

 %with RV \citep{crass2021}

Both transits and RV rely on the observation of the effect of the planet on a star, and  time-dependent inhomogeneities on the stellar surface cause complex signals which constitute a major limitation to the detection of Earth analogs. More generally, these signals hinder the detection of exoplanets and corrupt the estimate of their orbital elements \citep{hara2019, damasso2019, luhn2022}. It is crucial to understand very precisely the effect of the star on the data to better disentangle the signatures of the star, instrument and planets. 

Several types of stellar signals are known to affect spectroscopic and photometric data, here listed from high to low frequency variations.  %Asteroseismic signals, due to acoustic waves propagating in the star and creating oscillations of a few minutes. The physics behind asteroseismology is a well studied field with very detailed models \citep[][]{christensendalsgaard2014}. 
Acoustic waves propagating in the star and creating oscillations of a few minutes.
In RV surveys, asteroseismic signals are averaged out by tuning the integration time to a few oscillations \citep[$\sim 15$ min][]{dumusque2011i, chaplin2019}. It has also been suggested to model them with Gaussian processes with quasi-periodic kernels \citep{luhn2022}. Second, convection at the surface of the star creates a so-called granulation pattern: hot plasma rises to the surface, cools down and goes downwards creating a corrugated aspect of the stellar surface made of contiguous granules. Hot, upward moving gas is globally brighter than cooled, downward moving gas creating a so-called convective blueshift effect \citep[][see Fig. \ref{fig:granmag}.a]{dravins1981}. There are at least two granulation time-scales, due to granules themselves and so-called super-granulation, corresponding to a global motions of structures of $\approx3 \times 10^7$m.  Both in photometry, RV and other quantities derived from the spectrum, granulation appears as a correlated noise whose power spectrum density decays asymptotically as a power law, characterized by a time-scale and an amplitude \citep{cegla2013, kallinger2014, cegla2019_review, sulis2020, dravins2021}. %affecting the RV, but also the spectral shape \citep{cegla2019_review}. 

 Furthermore, the star can exhibit regions of enhanced magnetic flux, which manifest as spots or faculae, respectively darker and brighter than the continuum of the stellar surface (see Fig. \ref{fig:granmag}.b and c). The lifetime of these structures is correlated with their size, and can range from a few days to months depending on the stellar type \citep{meyer1974,martinezpillet1993}. Magnetic regions change the total flux as measured by photometry. Furthermore, they break the symmetry between the approaching and receding limb of the star and inhibit locally the convective blueshift. As a result, the presence of magnetic region alters the shape of the spectrum and in particular the measured radial velocity of the star~\citep[][]{saar1997, desort2007, meunier2010a, boisse2012, dumusque2014,borgniet2015, meunier2019I, meunier2019II, meunier2019III}. The rate of apparition and the properties of spots and faculae vary with the magnetic cycles of stars on the timescales of several years ($\sim$ 11 years on the Sun).  
Stellar meridional winds \citep{becker2011, meunier2020} and relativistic effects \citep{cegla2012} also have a RV signature, but of a much smaller amplitude. 

In the present work, we aim at modelling in detail the effects of stellar activity: spots and faculae and their interplay with magnetic cycles. We also present a model for granulation (see Section \ref{sec:granulation}), but do not discuss acoustic oscillations.

%On the Sun: the RV measurements taken by HARPS-N exhibit two granulation timescales, 6 min and 13h with respective RV amplitudes 32 and 68 cm/s, and the contribution of magnetic activity has an estimated amplitude of \citep[74 cm/s][]{almoulla2023}. The effect of granulation on the VIRGO photometric observations of the Sun binned on 1h have a standard deviation of $\approx$ 20-30 ppm \citep{sulis2020}. Compared to the expected signal of the Earth, 9 cm/s and 80 ppm in RV and photometry, we see that stellar activity signal pose serious challenges to the detection of Earth analogs  in RV, or planets with $\approx 1/2$ the size of the Earth. As mentioned earlier, stellar effects generally hinder planet detection and corrupt the estimates of orbital elements. 

To disentangle stellar and planetary signals in RV and photometry, we can leverage the fact that stellar signals have a certain temporal structure, and it is now standard practice to model these signals with a stationary Gaussian process. The process is characterized by the covariance of the stellar signal sampled at two epochs separated by a time interval $\Delta t$, or kernel -- or equivalently its Fourier transform, the power spectral density -- which expresses the self-similarity of the stellar signal. Signals due to magnetic activity are described kernels which are a product of a decaying function and a periodic one at the mean rotation period of the star, conveying the idea that the stellar surface is similar to itself after one rotation \citep{aigrain2012, haywood2014, foremanmackey2017,perger2021}. These models are not directly mapped to physical parameters, and \cite{luger2021_mappingI} provides kernels directly stemming from a statistical model of the stellar surface for photometry.  

%Granulation is often described by a Gaussian noise whose power spectral density decays asymptotically as the inverse frequency to the power 4~\citep[e.g.][]{dumusque2011i, kallinger2014, cegla2013, sulis2020}. Unfortunately, Gaussian processes act as frequency filters, and might filter out planetary signals \citep[][]{delisle2019a}. 

To analyse RV, we can use an additional fact: while planets cause a pure Doppler shift, stellar signals also affect the spectral shape \citep{haraford2023}. In particular, as a stellar spot passes in the visible hemisphere, we expect variations of the asymmetry of spectral lines or their width \citep{queloz2001, queloz2009}. These are two examples of so-called spectral indicators: quantities computing from the spectrum which characterize its shape change. Spectral indicators can be used as linear predictors to fit to the RV time series \citep{haywood2022, cretignier2022, zhao_expres_2022}. However, on the one hand indicators are noisy themselves and fitting them linearly does not propagate their uncertainty. Furthermore, it is not clear whether the RV should be expected to be in the vector space spanned by a few indicators. In particular we expect non-linear dependencies such as phase shifts between RVs and indicators \citep{bonfils2007, forveille2009, santerne2015, lanza2018}. \cite{aigrain2012} models the RV variation as a linear combination of the square of the photometric signal and the product of itself and its first time derivative. 
 
 %, it accounts for measurement uncertainties and non linear dependencies between the observations considered.
 
A way to circumvent these issues is to model simultaneously the RV, indicators, and photometry if available. This has been done in the Gaussian process framework \citep{rajpaul2015,gilbertson2020, jones2022,barragan2022, delisle2022}.  Typically each time series is modelled as a linear combination of a Gaussian process and its first, possibly second derivative. The Gaussian process is then less likely to filter out planets because it is better constrained. %This approach can be easily included in the Bayesian framework, because  the probability of the data available knowing the model parameters is explicitly defined.
However, neither the hypothesis that the effect of stellar activity on the different channels is a linear combination of the process and its derivatives, nor the kernels used to describe the process are rooted in a physical model. \cite{luger2021_mappingI, luger2021_mappingII, luger2021_mappingIII} builds a physics-based Gaussian process model of photometry and spectra, but does not take into account the inhibition of convective blueshift, important in radial velocity measurements. Finally, let us note that the hypothesis that the signal is Gaussian, stationary is seldom discussed.  
%Since the observer is uncertain as to when granules, spots and faculae appear, we describe the stellar effects on the channels jointly as a stochastic process. 

We consider several time series: photometry, RV, spectroscopic indicators, or the time series of the whole spectra at different wavelength, which we call channels. These need not be sampled at the same epochs. The representation of different channels as a joint stochastic process is adapted, because of the stochastic nature of stellar surface processes. Our aim is here to build a formalism to allow to transfer physical assumptions on the stellar processes into practical, fast data analysis methods of the channels available. Our model does not start from an assumption of Gaussianity, but we compute analytically its mean and covariance, to approximate it with a Gaussian process. We also compute higher order cumulants of the model, and show that there is information to be harvested in the non Gaussianity of stellar signals.

Besides exoplanet detection and characterization, a more obvious motivation to understand stellar signals is to gain information on the star. In the context of Doppler imaging, one can infer the position and size of magnetic regions from the spectral shape change they induce. The inference is usually done on a weighted average of the spectral lines \citep{deutsch1958, khokhlova1976, goncharskii1977, goncharskii1982, vogt1983, vogt1987}. In a recent work, \citep{luger2021_mappingIII} showed the surface brightness decomposed in spherical harmonics can be inverted from spectral profiles in a Gaussian process framework. To constrain the stellar surface at a given time with Doppler imaging, the effect of magnetic region must be important enough, and the star must rotate fast enough. Similarly to \cite{luger2021_mappingIII}, our framework allows, in a sense that will be made precise to perform ``statistical Doppler imaging'', that is to retrieve statistical properties of the magnetic regions even if their individual signals are too faint to retrieve their instantaneous position. 

%to retrieve information not from the instantaneous, but the statistical distribution of the magnetic region. %We explore in particular the ability of our framework to estimate the solar inclination and average latitude of magnetic regions from HARPS-N radial velocity measurements taken on the Sun. 

Our article is organized as follows. In Section \ref{sec:methods}, we present our general statistical formalism. In Section \ref{sec:physmodel}, we discuss the physical assumptions that are adopted to model the effects of magnetic activity. We apply our formalism to the analysis of HARPS-N and SORCE observations of the Sun in Section \ref{sec:sun}, and discuss its ability to retrieve stellar inclinations. 
%We discuss the computation of physically motivated kernels in Section \ref{sec:computations} and how it can be implemented numerically in an efficient way within the S+LEAF framework of \cite{delisle2022}. \citep[e.g.][]{nemec2022}
In Section \ref{sec:discussion}, we discuss various extensions of our work. We  discuss the link between the granulation signal and the properties of individual granules in \ref{sec:granulation}. We show that our model can be leveraged to interpret non-Gaussianity in the observations of Solar RV in Section \ref{sec:testing_gaussianity}. We suggest a link between the phase shifts between RV and photometry and the ratio of spots to plages in Section \ref{sec:phaseshifts}. We discuss closed-loop relationships between indicators in Section \ref{sec:closed_loop}, and the relationship of our work with Doppler imaging in Section \ref{sec:doppler_im}. We conclude in Section  \ref{sec:conclusion}. An open-source reference implementation of our algorithms is publicly available as python package \footnote{\url{https://gitlab.unige.ch/jean-baptiste.delisle/spleaf}}

	\section{Statistical framework}
	\label{sec:methods}

 %\subsection{Specifications}

%Our goal is to find a statistical model that allows to use as much detail as possible from 

	\subsection{Finite energy random impulse response (FENRIR)}
	\label{sec:FENRIR}
	
%Our goal is to have a joint model of stellar activity effects on the different channels, which has the potential to incorporate as much information as possible from the empirical and theoretical understanding of stars. 
 
 	Let us consider that the stellar activity affects several observables: radial velocity, photometry, activity indicators. The time series corresponding to one observable is called a channel. Because the stellar surface changes in an unpredictable way, the channels can be considered as a random processes, which we here aim to describe. 
 Our goal is to transfer physical knowledge of  magnetic activity and granulation into practical data analysis methods. We consider that the effect of stellar activity can be modelled as features (magnetic regions or granulation cells.   To be as realistic as possible our model must satisfy a few specifications. 
 (i) the effect of the stellar feature on the channels shall depend on its physical parameters (size, position...) as well as stellar parameters. (ii) The feature parameters shall be allowed to be drawn from a parametrized distribution . (iii) The distribution of stellar features and the rate at which they appear shall depend on time, as it may vary in particular with the magnetic cycle. (iv) the properties of stellar features shall be allowed to depend on the features already present. 
 
\begin{figure}
    \centering
	\includegraphics[width=9cm]{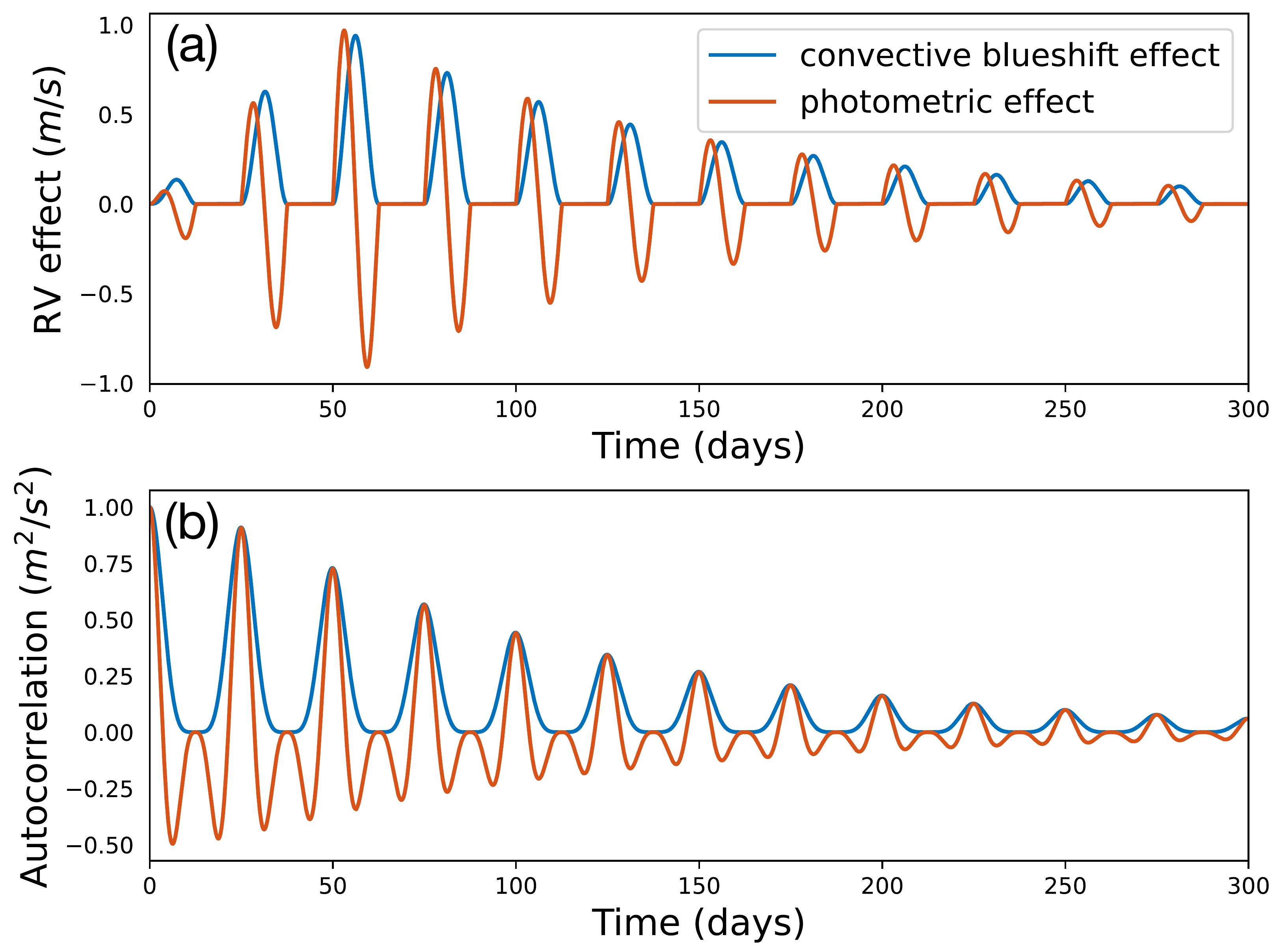}
	\caption{ (a) Effect in radial velocity $g(t)$ as a function of time for a star with an equatorial stellar spot whose size increases then decreases. Blue and red curve correspond to the RV effect due to the inhibition of the convective upwards convective motion due to the magnetic field in the spot, and red to the photometric effect breaking the imbalance between the approaching and receding limb.  (b) Autocorrelation of the stochastic process defined in Eq.~\eqref{eq:ytf} for functions $g$ corresponding to (a). }
	\label{fig:stellar_activity_g}
	\end{figure}

	For the sake of clarity, we assume there are two channels, for instance RV and photometry or RV and an activiy indicator such as the $\log R'_{HK}$ \citep{noyes1984}.  The effect of stellar activity on these two channels is denoted by $y(t)$ and $z(t)$. We assume that a stellar feature has parameters $\gamma$, and affects channels $y$ and $z$ through functions $g(t, \gamma)$ and $h(t, \gamma)$, respectively. The vector of parameters $\gamma$ includes, but is not necessarily limited to, a longitude, area, and lifetime. In Fig.~\ref{fig:stellar_activity_g} (a), we show an example of $g(t, \gamma)$ as a function of $t$ where $g$ models the RV variation due to an equatorial spot as a function of time.
    
 	Stellar features are transient, supposing they appear at time $t_0$, their effect on $y$ and $z$ as a function of time is modelled by functions $g(t-t_0, \gamma(t_0))$ and $h(t-t_0, \gamma(t_0))$. If the features appear at times $t_k$, the effect on the channel $y$ and $z$ are
	\begin{align}
    y(t) & = \sum\limits_{k=-\infty}^{+\infty}  g(t-t_k, \gamma(t_k))  \label{eq:ytf} \\
     z(t) & = \sum\limits_{k=-\infty}^{+\infty}  h(t-t_k, \gamma(t_k))  \label{eq:ztf}
     \\
     \gamma(t_k) &\sim p(\gamma, \mid t_k, \eta) \label{eq:dist}
\end{align}
where Eq. \eqref{eq:dist} means that if a feature appears at time $t_k$, its parameters are drawn from a probability distribution $p(\gamma, \mid t_k, \eta)$ where $\eta$ is a vector of parameters describing the distribution which will be later interpreted as the vector of hyperparameters of the Gaussian process modelling $y(t)$ and $z(t)$.

%To simplify notations, we simply denote $p(\gamma , t_k) = p(\gamma , t_k, \eta)$, such that the dependence of $p$ on $\eta$ is implicit. 

We assume that the features appear with a non stationary rate $\lambda(t)$, meaning that assuming a feature appeared at time $t_0$, the probability that the next feature appears at $t_0 +\Delta t$ is
\begin{align}
p(t\mid t_0) = \lambda(t_0+\Delta t) \e^{-\int_{t_0}^{t_0+\Delta t}\lambda(t) \dd t}. \label{eq:lambda}
\end{align}
Equivalently, the distribution of times of appearance can be defined as follows. The number of features $N$ in a time interval $[t_1, t_2]$ follows a Poisson distribution of parameter $\Lambda = \int_{t_1}^{t_2}\lambda(t) \dd t$, and the times of appearance of a spot $t_i, i=1..N$ are drawn independently from the distribution $\lambda(t)/\Lambda $. 	A non constant rate $\lambda(t)$ might be useful to model magnetic cycles, because spots are known to appear more frequently at the peak of the cycle. So far, specifications (i), (ii) and (iii) are satisfied. We will see that the correlation between features (requirement (iv)) can be modelled in the choice of the impulse response $h$ and $g$. 

Our choice of the name ``finite energy random impulse response'' stems from the fact that, if $\gamma$ were constant, we could write Equations \eqref{eq:ytf} and \eqref{eq:ztf} as $\sum_k \delta(t-t_k) \ast g$ and \eqref{eq:ztf} as $\sum_k \delta(t-t_k) \ast h$, where $\ast$ is the convolution product and $\delta(t)$ the Dirac function. In this formulation, the sum of Dirac is the input signal,  $y(t)$ and $z(t)$ are the outputs, and $g$ and $h$ are the impulse responses. We allow them to be random but impose that they have finite energy so that the covariance of the output is always finite, hence our choice of the method name. %Other name options would have been ``random transfer function'' or ``continuous random moving average''. 

%Restricting to finite energy responses guarantees that the covariance can always be computed.
%Indeed, if time is discrete $y(n)$ and $z(n)$ would be moving average processes, and $g(n)$ and $h(n)$ would be their parameters, or finite impulse response. 

%We now build the model in two steps. First, assuming that the features appear following a Poisson process with rate $\lambda(t)$, we establish the analytical form of the autocovariance of $y(t)$, $z(t)$ as well as their covariance, and higher order cumulants of the processes. Second, we specify expressions for $h$ and $g$ corresponding to stellar activity assuming that the stellar features are small, and compute the corresponding kernels when the rate of the Poisson process is constant. Finally, we show how the covariance can be closely approximated with a Fourier expansion, leading to a representation of the covariance matrix as a CELERITE kernel \citep[][]{foremanmackey2017}, thus the likelihood evaluation cost is linear in the number of datapoints. 

\subsection{Using FENRIR models}
\label{sec:usingFENRIR}

When searching for exoplanets, stellar contributions must be estimated as precisely as possible to be removed. Conversely, stellar signatures in the data can be used to gain information on the star. FENRIR models can be used for both purposes.

 In principle, we could try to determine exactly how many features $N$ affect the dataset, find their their times of appearance $(t_k)_{k=1,..,N}$ and estimate their parameters $\gamma(t_k)$, and have the most of what our model can give both for exoplanet detection and inference of the stellar properties. This would give maximum information on the star but as shown in \cite{luger2021_mappingI,luger2021_mappingIII}, several spot structures can correspond to the same data. Handling this degeneracy and the large number of parameters is impractical computationally speaking. 
 
 On the other hand, we can simplify the problem, and try to find a spectral variability indicator whose effect on the data   $h(t,\gamma)$ is such that the effect of the feature on the signal of interest (photometry or radial velocity), $g(t, \gamma)$, is proportional to $h(t,\gamma)$. If the signal to noise ratio on such an indicator is good enough, we could potentially estimate the effect of stellar variability on RV or photometry simply with a linear scaling with this indicator. The unsigned magnetic field is known to be a good variability indicator \citep[][]{haywood2022}, and we will see that it is because it is approximately proportional to the RV effect due to the inhibition of convective blueshift. Additionally, let us note that if there exists a phase shift between $g(t, \gamma)$ and $h(t,\gamma)$ for all feature parameters $\gamma$, then there is a phase shift between $y(t)$ and $z(t)$. Such phase shifts, and more generally closed-loop relations, are known to exist between activity indicators and RV \citep{bonfils2007,forveille2009,  santerne2015,lanza2018, colliercameron2019}, and as we shall see they can be interpreted physically through the FENRIR framework.  

It is not clear that representing stellar signals as a linear combinations of activity indicators is realistic enough. Furthermore, if the activity indicator is noisy, using them as linear predictors introduces noise and this uncertainty must be accounted for. A more principled way to account for stellar signals is  % the posterior distribution $p( \eta \mid (y_j(t_i))_{i,j}  )$. %Supposing now the parameters of interest are , our likelihood becomes 
%A middle ground between the two approaches described above is
to  model simultaneously the channels $y(t)$ and $z(t)$, and more generally $M$ channels $y_j(t)$, $j=1...M$ with a likelihood function. Given observation times $t_i,i=1...N$, we want to characterize the joint statistical distribution of the  vector with $MN$ components, $Y = (y_j(t_i))_{i=1...N, j=1...M}$,  as a function of the statistical properties of the features, $\eta$ in Eq. \eqref{eq:dist}. Ideally, we would want a likelihood function $p(Y \mid \eta)$, or an approximation of this function. Because the signal due to the planets and to the star are additive, we can further generalize the expression of the likelihood to $p(Y \mid \eta, x )$ where  $x$ (number of planets, periods, masses, radii, eccentricities...). 
To gain information on $\eta$ and $x$, we can compute the posterior distribution $p( \eta, x \mid Y  )$ with an appropriate numerical methods. Depending on whether our objective is to gain information on the planets or the star, the posterior will be marginalized (integrated) with respect to $\eta$ or $x$, respectively, obtaining $p(  x \mid Y  )$ and $p(  \eta \mid Y  )$. 

%Gaining information on the star could then be done for instance by . Supposing now the parameters of interest are the planetary parameters goal is to marginalize  $p( \eta, )$ 

If the stellar features to be studied are magnetic regions, trying to find the number of regions and their parameters ($\gamma_k$ in \eqref{eq:ytf}) is the objective is Doppler imaging. However, for quiet stars, or stars not rotating fast enough, Doppler imaging cannot resolve the stellar surface. Here, we do not aim directly at finding the individual feature properties, but at characterizing the statistical distribution of their parameters (lifetime, size, position, etc), and perform in some sense a ``statistical Doppler imaging''.

% with well propagated uncertainty.
%This is easily seen if we can map the different channels to the surface of the star in spherical harmonics. This aspect is further discussed in SECTION DOPPLER.

%\textsc{fenrir} Fenrir FENRIR

In \cite{rajpaul2015, jones2022, gilbertson2020, barragan2022, delisle2022}, the different channels are described by a multivariate Gaussian process. In that case, the likelihood is a Gaussian multivariate distribution for any collection of observation times.
Denoting by $\mathbb{E}$ the mathematical expectancy and by $y_i(t)$, $i=1..n$ the different channels, Gaussian processes are fully characterized by the mean functions, $\mathbb{E}\{y_i(t)\}$, and the covariance function. Considering two times $t$ and $t'$, the covariance is $\mathbb{E}\{y_i(t) y_j(t')\}-\mathbb{E}\{y_i(t)\} \mathbb{E}\{y_j(t')\}$, where $(i,j)$ take all possible combination of pairs of channels. %In the FENRIR models, mean functions and covariances would be characterized by the parameters of the distribution of feature $\eta$ characterizing  ($\gamma(t) \sim p(\gamma, t, \eta)$).
%The parameters $\eta$ are the hyperparameters of our Gaussian process, and could give some information about the star (inclination, typical latitude of spots, etc).
With the hypotheses listed in section \ref{sec:FENRIR}, in Appendix \ref{app:nonstationary}, we show that the mean of the $y(t)$ process is
\begin{align}
	\mathbb{E}_\eta\{ y(t_a) \} = \iint  g(t_a  - t, \gamma) \lambda(t)  p(\gamma \mid t, \eta)   \dd t \dd \gamma \,  \label{eq:mean}.
%	\mathbb{E}_\eta\{ z(t_a) \} = \iint  h(t_a  - t, \gamma) \lambda(t)  p(\gamma \mid t, \eta)  \dd t \dd \gamma \	
\end{align}
	The covariance of $y(t)$ at times $t_a$ and $t_b$ is
\begin{align}
 \mathrm{Cov}_\eta(y(t_a), y(t_b)) = \iint  g(t_a  - t, \gamma) g(t_b  - t, \gamma) \lambda(t)  p(\gamma\mid t, \eta)  \dd t  \dd \gamma  \label{eq:covar}.
\end{align}
The mean and covariance of $z(t)$ are obtained by replacing $g(t)$ by $h(t)$ in Eqs. \eqref{eq:mean} and \eqref{eq:covar}, respectively. The covariance of $y(t_a)$ and $z(t_b)$ is
\begin{align}
	\mathrm{Cov}_\eta(y(t_a), z(t_b)) = \iint  g(t_a  - t, \gamma) h(t_b  - t, \gamma) \lambda(t)  p(\gamma \mid t, \eta) \dd t \dd \gamma .
	\label{eq:crosscova_body}
\end{align}
Note that both the mean and variance are proportional to $\lambda$, reproducing the fact that as stellar activity increases along the magnetic cycle, there is both a systemic effect in \eqref{eq:mean} and an increase in variance in \eqref{eq:covar}.
Assuming that $\lambda(t)$ is constant and $p(\gamma\mid t, \eta) $ does not depend on $t$, the process is stationary and the covariance can be written $ \mathrm{Cov}(y(t_a), y(t_b)) = k(|t_a-t_b|)$. With this assumption, the kernel $k(\tau)$ corresponding to the functions $g$ shown in Fig.~\ref{fig:stellar_activity_g} (a) are shown in Fig.~\ref{fig:stellar_activity_g} (b). 

%Equation \eqref{eq:crosscova_body} show that the covariance is in fact the cross correlation of a single feature averaged over its possible parameters. It gives a rigorous framework to the $FF'$ approach taken in \cite{aigrain2012}. 
%By choosing a physical model for the effect of stellar features (spots, faculae or group of them, granules), whose appearance is controlled by parameters $\eta$ such as the stellar inclination, limb darkening law, typical latitude of spots,

 We can build a Gaussian process model of stellar variability signals, but with hyperparameters $\eta$ with a physical interpretation  from Eq. \eqref{eq:mean}  and Eq. \eqref{eq:covar}. However, this is only an approximation of the FENRIR model.  \cite{jenkinswatts1969} provide an explanation as to why, in our case, a Gaussian approximation may lead to lost information. They consider a moving average process, which corresponds to our FENRIR process with discrete time and fixed $g$. In that case, estimating $g(n)$ from a realization of $y(n)$ is a problem of identification of a moving average process, and the covariance of $y(n)$ only gives the cross correlation of $g$. However, several functions might have the same autocorrelation while being different. Even if $g$ in Eq. \eqref{eq:ytf} is constant, modelling $y(t)$ as a Gaussian process characterized by its covariance does not allow to determine $g$ unambiguously. 
To break the degeneracy, not only the covariance but higher order cumulants of $y(t)$ have to be computed. This notion is defined and discussed in Section \ref{sec:testing_gaussianity}. 
In Appendix \ref{app:cumulants}, we show that for our FENRIR model, the cumulant of order $n+m$ of process $y(t)$ and $z(t)$ sampled respectively at $n \geqslant 0$ times $(t_i)_{i=1..n}$ and $m\geqslant 0$ times $(t_i')_{i=1..m}$ ,
%assuming that the features appear with a constant Poisson rate $\lambda$, and $p(\gamma)$ does not depend on $t$, we obtain the formula
\footnotesize
\begin{align}
\begin{split}
 &   \kappa_{ \eta}(y(t_1),y(t_2),...,y(t_n),z(t_1'),z(t_2'),...,z(t_m') )  =\\ & \iint g(t-t_1, \gamma)  ... g(t-t_n, \gamma) h(t-t_1', \gamma)  ... h(t-t_m', \gamma)\lambda(t) p(\gamma \mid  t, \eta) \dd t \dd \gamma. 
    \label{eq:kappan}
\end{split}
\end{align}
\normalsize
%\begin{align}
%    \kappa_n(\tau_1, ..., \tau_{n-1}) = \lambda \iint g(t, \gamma) g(t+\tau_1, \gamma)  ... g(t+\tau_{n-1}, \gamma) p(\gamma) \dd t \dd \gamma,
%    \label{eq:kappan}
%\end{align}
of which Eqs.~\eqref{eq:mean}, \eqref{eq:covar} and \eqref{eq:crosscova_body}  are particular instances, when $n=1$ and $n=2$ and $(m,n) = (1,1)$. If $m=0$, Eq. \eqref{eq:kappan} considered as a function of $t_1,...,t_n$ is the so-called $n$-point correlation function of $y(t)$.  %We can also generalize the covariance of different channels, Eq. \eqref{eq:crosscova_body}, to higher order cumulants (see Appendix \ref{app:cumulants}).
%\footnotesize
%\begin{align}
%    \kappa_n(q_1(t_1),q_2(t_2),...,q_n(t_n))  =  \iint f_1(t-t_1, \gamma) f_2(t-t_2, \gamma)   ... f_n(t-t_n, \gamma) \lambda(t) p(\gamma \mid t) \dd t \dd \gamma. 
%    \label{eq:kappan_app}
%\end{align}
\normalsize
Understanding the non Gaussianity of stellar signals might lead to improve for instance the Gaussian network regression methods used to model spectroscopic signals \citep{camacho2022}.
%where $q_i(t) = y(t)$ or  $q_i(t) = z(t) $. If $q_i(t) = y(t)$, then $f_i(t) = g(t=$ and if $q_i(t) = z(t)$, then $f_i(t) = h(t)$.

%In Section \ref{sec:testing_gaussianity}, we suggest a way to study non Gaussianity with the Fourier transform of cumulants, called the polyspectra. 

To model stellar variability signals in RV and photometry, we need a sum of at least two FENRIR processes, one for granulation and the other for the effect of spots and faculae. If the processes are independent, the cumulants of their sum is the sum of their cumulants. This is in particular true for the covariances. %$$, and in particular the covariance of a sum of independent process is the sum of their cumulants. 

%The covariance in Eq. \eqref{eq:covar} and higher order cumulants in Eq. \eqref{eq:kappan} depend on the hyperparameter $\eta$. Inference based on those thus gives the properties of the statistical distribution of stellar features, as opposed to the individual parameters of a  
 
%To further discuss the stochastic process approach, let us consider the Doppler imaging problem.Let us suppose that we have several functions $g_1(t, \gamma),...,g_n(t, \gamma) $ which represent the value of intensity of a least square deconvolution profile 

We now have several avenues to explore:
(1) What are the expressions of the mean and covariance of our channels, to have a physically motivated Gaussian process model?
(2)  what can we learn from this model about the star? (3) are stellar signals Gaussian? (4) what activity indicators are affected by a stellar feature proportionally to the RV or photometric effect? We show that our formalism allows to contrain the inclination of the star relative to the plane of the sky in Section \ref{sec:sun}.  
Questions 3-4 are deferred to Section \ref{sec:discussion}. Below, we focus on the Gaussian process approximation of FENRIR models. 

%In any case, we need to express an explicit model of the effect of one feature on the different channels, $g$ and $h$ in Eq. \eqref{eq:ytf} and

\subsection{The Gaussian process representation}

Supposing we have time series of $y(t)$ and $z(t)$ sampled at times $(t_i)_{i=1..q}$. If we want to analyze these time series jointly, we must
compute the likelihood of the vector $Y = (y(t_i),z(t_i))_{i=1..q}$ with $2q$ components, obtained by stacking vertically the two vertical vectors $y(t_i)_{i=1..q}$ and $z(t_i)_{i=1..q}$. 
\begin{align}
    p(Y \mid \eta) = \frac{\e^{-\frac{1}{2}Y^T V^{-1}(\eta) Y  }}{\sqrt{2\pi}^{2q} |V(\eta)|} 
    \label{eq:gauss_likelihood}
\end{align}
The covariance matrix $V$ is a $2q \times 2 q$ matrix.

More generally, if we analyze jointly channels $y(t)$ and $z_j(t)$ for $j=1..p$ with impulse responses $g$ and $h_j$, $j=1..p$ respectively, our dataset $Y$ is the stacked $m+1$ vectors, and we want to compute its  $pq \times pq$ covariance matrix which can be seen as made of $p \times p$ blocks of $q \times q$ matrices. Each block parametrized by $i,j$ is such that its element on row $n$, column $m$ is, from Eq. \eqref{eq:crosscova_body},
\begin{align}
    V^{ij}_{nm}(\eta) = \iint  g(t_n  - t, \gamma) h_j(t_m  - t, \gamma) \lambda(t)  p(\gamma \mid t, \eta) \dd t \dd \gamma .
	\label{eq:crosscova_body}
\end{align}
In practice, using the new covariances will be exactly the same as using existing kernels, for instance the so-called quasi periodic kernel \citep{aigrain2012, haywood2014}, 
\begin{equation}
    k^\mathrm{QP}(|t_i-t_j|, \eta) =\eta_1^2 \exp \left[- \frac{(t_k-t_l)^2}{2 \eta_2^2}  - \frac{2 \sin^2  \frac{\pi (t_k-t_l)}{\eta_3} }{\eta_4^2}\right] .
   \label{eq:kernel_qp}
\end{equation}
except that our hyperparameters $\eta$ will be the parameters of our distribution of stellar feature: the inclination of the star, the typical longitudes where magnetic regions appear, their lifetime, the lifetime of granulation cells etc (the vector of parameters $\eta$ in Eq. \eqref{eq:crosscova_body}).

To explore the parameter space of $\eta$, either to compute Bayesian evidences for exoplanet detection or planet parameter estimation, we need to evaluate the likelihood many times. Evaluating the likelihood requires to invert the covariance matrix $V$ in Eq. \eqref{eq:gauss_likelihood}, which scales in general as the cube of the number of rows. To analyze efficiently the tens of thousand of points of solar RVs, we need to reduce this cost. 
The covariance matrix is represented in the S+LEAF framework \citep{delisle2019b, delisle2022}, so that the calculation of its inverse and determinant scales linearly with its number of rows: the number of datapoints times the number of time series considered, and not with the cube, the general cost of matrix inversion. 

To make our formalism applicable to a wide range of assumptions, we want the calculation process to have as many automatic step as possible, where the user-defined part pertains only to the physical assumptions. In Appendix \ref{app:gaussian_process}, we show that the likelihood can be evaluated in linear time if the following conditions are met. These conditions can be dropped, at the cost of a higher computational time. 
\begin{enumerate}
    \item The effect of a single feature without limb-darkening is represented in the form
    \begin{equation}
            g_0(t, \gamma) = W(t) \mathbf{1}_{\mathrm{vis}}(t) \sum_{k=0}^d a_k(\gamma) \cos{k \omega t} + b_k(\gamma) \sin{k \omega t}, \label{eq:feature_g0}
    \end{equation}
    where $\omega$ is the rotation frequency of the star, $W(t)$ modulates the intensity of the signal, $\mathbf{1}_{\mathrm{vis}}(t)$ is a function equal to one when the feature is visible and 0 otherwise.  This is not a strong assumption, as this form naturally emerges from a physical model (see Section \ref{sec:physmodel}).
    
    \item The longitude at which the feature attains its maximal size is random on $[0, 2\pi]$. 
    
    \item The effect of differential rotation is neglected. 
    
    %\item A limb darkening-law $l(\gamma, \eta)$, whose form will be made clear in the next sections. The expression of $g$ is taken as $g(t, \gamma) = g_0(t,\gamma) l(\gamma)$.

    %\item An assumption on whether features appear in groups on different latitudes.  

   % \item An assumption for the form of the distribution of parameters of the features $p(\gamma, \eta)$. We recall that $\gamma$ are the parameters that are drawn randomly each time a feature appears, and $\eta$ are the hyperparameters describing the distribution (for instance, expressing that spots appear preferentially close to a certain latitude). 
    
   % \item An binary assumption on whether the variations of $W(t)$ have  frequencies lower or higher than $\omega$. The first case is discussed in the context of long-lived spots and faculae in Section \ref{sec:spots}, and the second in the case of granulation cells, whose lifetime is small compared to stellar rotation period, in section \ref{sec:granulation}.  The case where the spot lifetime is comparable to the stellar rotation period is discussed in \cite{gilbertson2020}, and we do not discuss it in the present work. 
\end{enumerate}
The different steps we take to compute the likelihood are detailed in Appendix \ref{app:gaussian_process}.

 %In the next section, we give expressions modelling the effect of spots, faculae, and granulation. 

%\subsection{Mean, covariance and cumulants}
%	\label{sec:methods_overview}
	
%Before diving into these questions, let us give the expressions of the mean, covariance, and higher order cumulants. 

% For a stationary process, cumulants are written as a function of time  differences with $t_1$. We write $t_i = t_1 + \tau_i $. and the cumulant of order $n$ as $\kappa_n(\tau_1, ..., \tau_{n-1})$. 

 \section{Physical model of magnetic activity signals}
\label{sec:physmodel}

A FENRIR model is defined by the impulse responses of its different channels, $g$ in Eq. \eqref{eq:ytf}. Second, we need to define the probability distribution of the impulse response parameters $\gamma$ (see Eq. \eqref{eq:dist}), in particular what is the size and lifetime of the feature? Finally, we need a rate of appearance of features $\lambda(t)$ (see Eq. \eqref{eq:lambda}). In the present section, we model spots and faculae in section \ref{sec:spots}, and granulation in section \ref{sec:granulation}. 
The distribution of stellar activity is complex. The object of this section is not to have a model as realistic as possible, but rather to illustrate how several physical assumptions can be translated in our formalism. %Second, one of our goals to compute the covariance of the FENRIR to build physically motivated Gaussian processes. The general cost of likelihood evaluation is $O(N^3)$ if there are $N$ data points. If the covariance can be put in semi-separable form, we obtain a $O(N)$ computational cost, but this form cannot be obtained for all covariance kernels, which in turns restricts the choices of impulse response, distributions and rate variation. Th

 %This has been specified in sections \ref{sec:spots}, \ref{sec:limbdarkening} and \ref{sec:group}.

%Our goal is here to give an expression of the impulse response $g$ and $h$ modelling the effect of a single stellar feature, chosen to be either a spot, facula, and granulation cell. The distribution of features on the Sun is complex \citep{borgniet2015}. 

%We here provide a simplified one, which will be refined in future work. 

\subsection{Spots and faculae}
	\label{sec:specg}
		\label{sec:spots}

%To obtain a usable representation of stellar activity, we need to specify the effect of a feature on the data: the functions  $g(t,\gamma)$ and $h(t,\gamma)$ in Equations \eqref{eq:ytf} and \eqref{eq:ztf}, where $\gamma$ is the vector of parameters of the feature. As mentioned in Section \ref{sec:introduction}, there are several types of physical phenomena occurring on the surface of the star known to affect the data. Based on the physical models of the effect of these phenomena on photometry and spectroscopic data \citep{saar1997, meunier2010a, meunier2010b, dumusque2011i, dumusque2012ii, dumusque2014, cegla2018, cegla2019} we discuss effect of spots and faculae. 

\begin{figure}
    \centering
    \includegraphics[width=\linewidth]{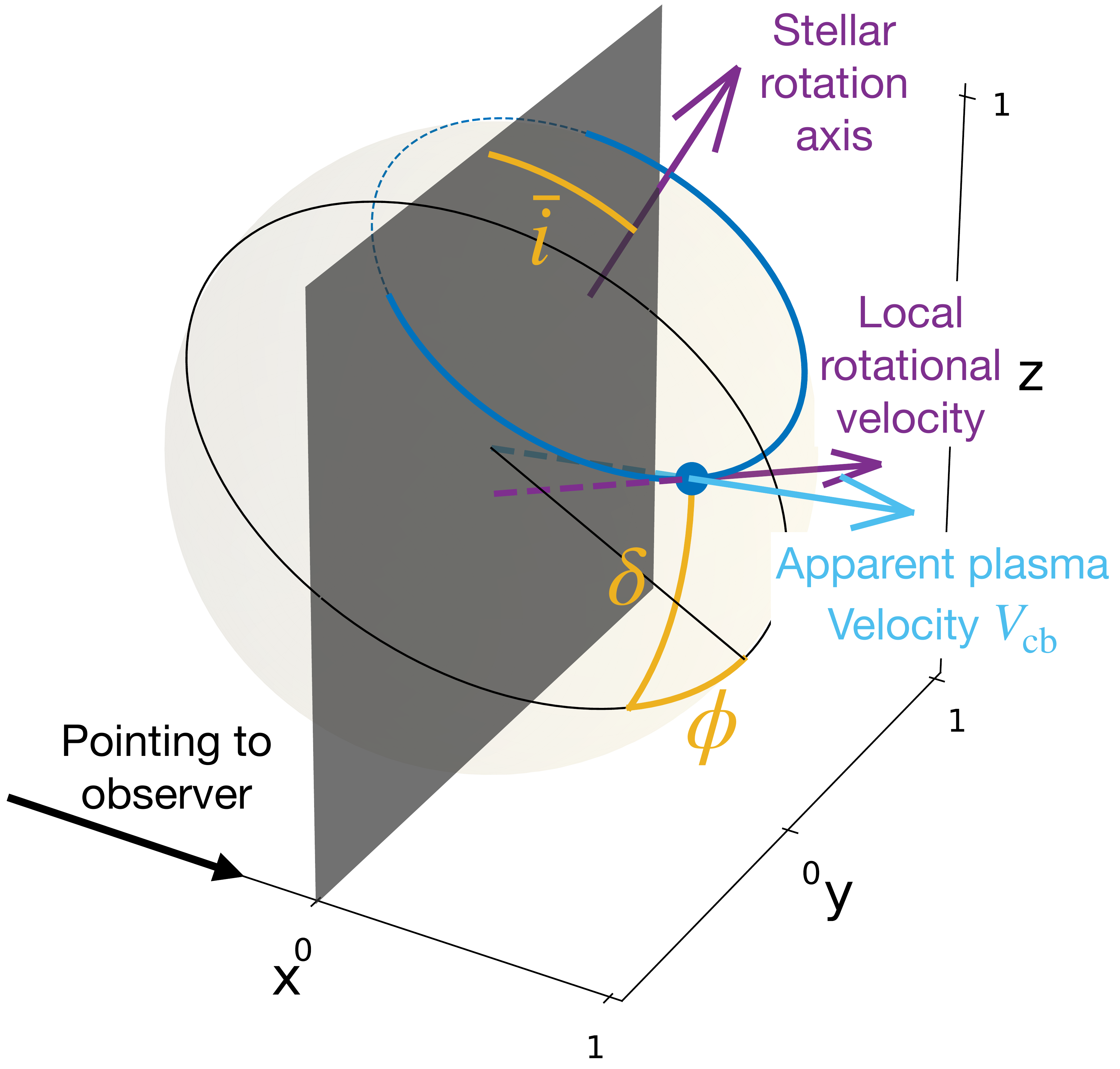}
    \caption{Geometry of the problem: a stellar feature represented by the dark blue point moves as the star rotates and is visible only a fraction of the time. It modifies the local emission, and velocity of the plasma, which appears to move outwards because of its convective motion. We parametrize the position of the feature with the angle between the sky plane and rotation axis $\bar{i}$, the longitude of the feature $\phi$ and its latitude $\delta$. As a remark, $\bar{i} = \pi/2-i$ where $i$ is the inclination classically used in the projected mass and velocities $m\sin i$ and $V\sin i$. }
    \label{fig:geometry}
\end{figure}

Spots and faculae are regions of the star with an a magnetic field stronger than the continuum. While spots are darker than their surroundings, faculae are brighter. Consequently, as they pass accross the visible hemisphere of the star, they have an effect on the global flux of the star measured with photometry. Secondly, because they  break the imbalance of the approaching and receding limb of the star, they have a Doppler signature. Finally, their magnetic field inhibits the convection motion of the plasma in the stellar photoshere. The hot plasma has an upward motion, it cools down and moves back towards the center of the star. Because the plasma moving outwards is hotter, it represents a higher fraction of the total flux resulting in a global blueshift of the stellar light. In spots and faculae, the magnetic field tends to globally slow this motion, resulting in the so-called inhibition of the convective blueshift \citep{meunier2010b}. In Fig.~\ref{fig:geometry}, we show the geometry of our problem. 

\subsection{Impulse response}

 We first assume that the radial velocity of the star is the sum of local radial velocities weighted by their flux.  To model the impulse response $g$, following \cite{aigrain2012} we assume that the spots and faculae, referred collectively as magnetic regions, are infinitesimal surfaces on the star.  We further refine the model of \cite{aigrain2012}, and suppose that the effect of a stellar magnetic region is multiplied by a certain Limb-Darkening law.%, as well as a function modelling the intensity of the effect as the magnetic region appears and vanishes. 

\begin{figure*}
	\begin{tikzpicture}
	\path (0,0) node[above right]{\includegraphics[width=9cm]{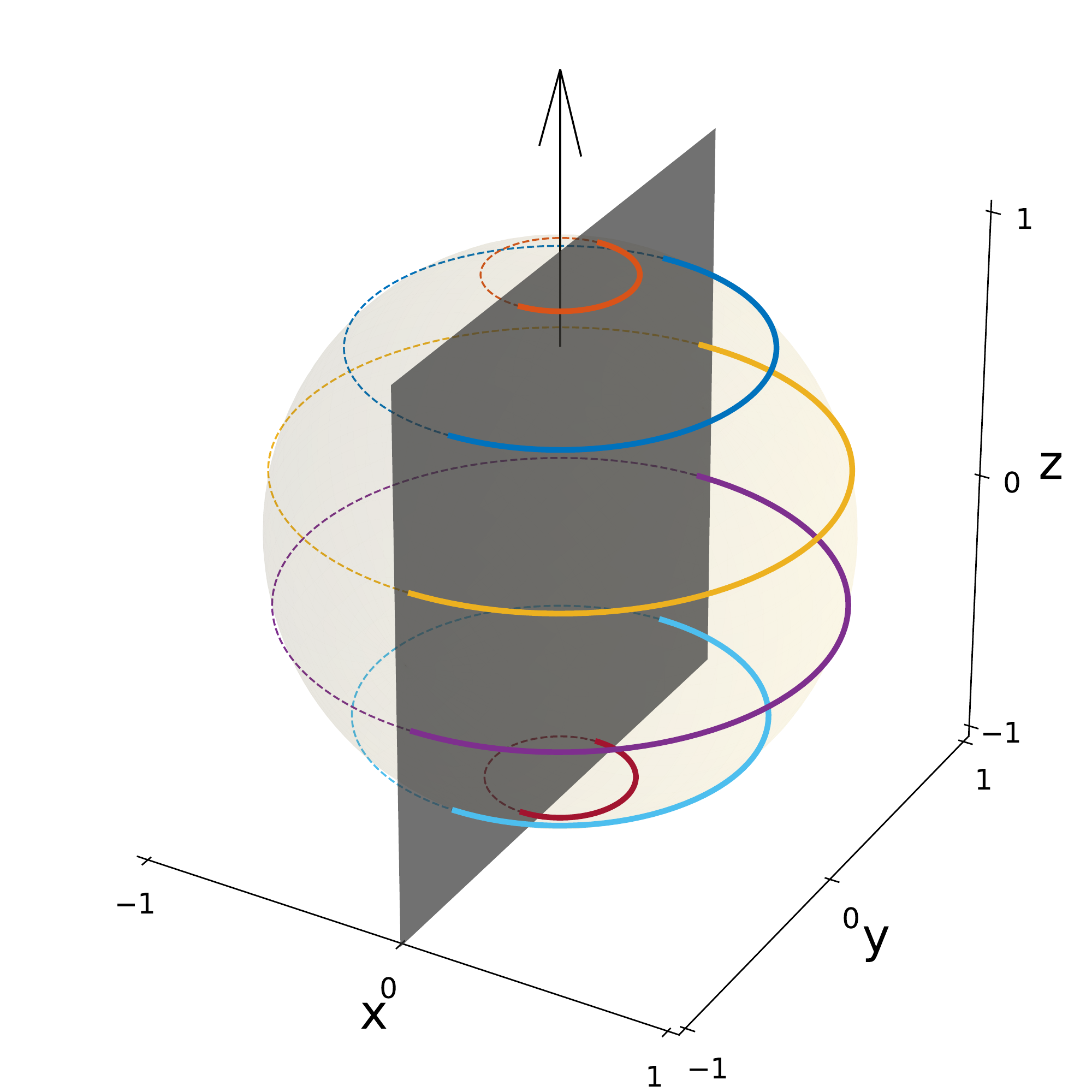}}; %
	\path (1.,5) node[above right]{\large(a)};
	\path (0,-10.8) node[above right]{\includegraphics[width=9cm]{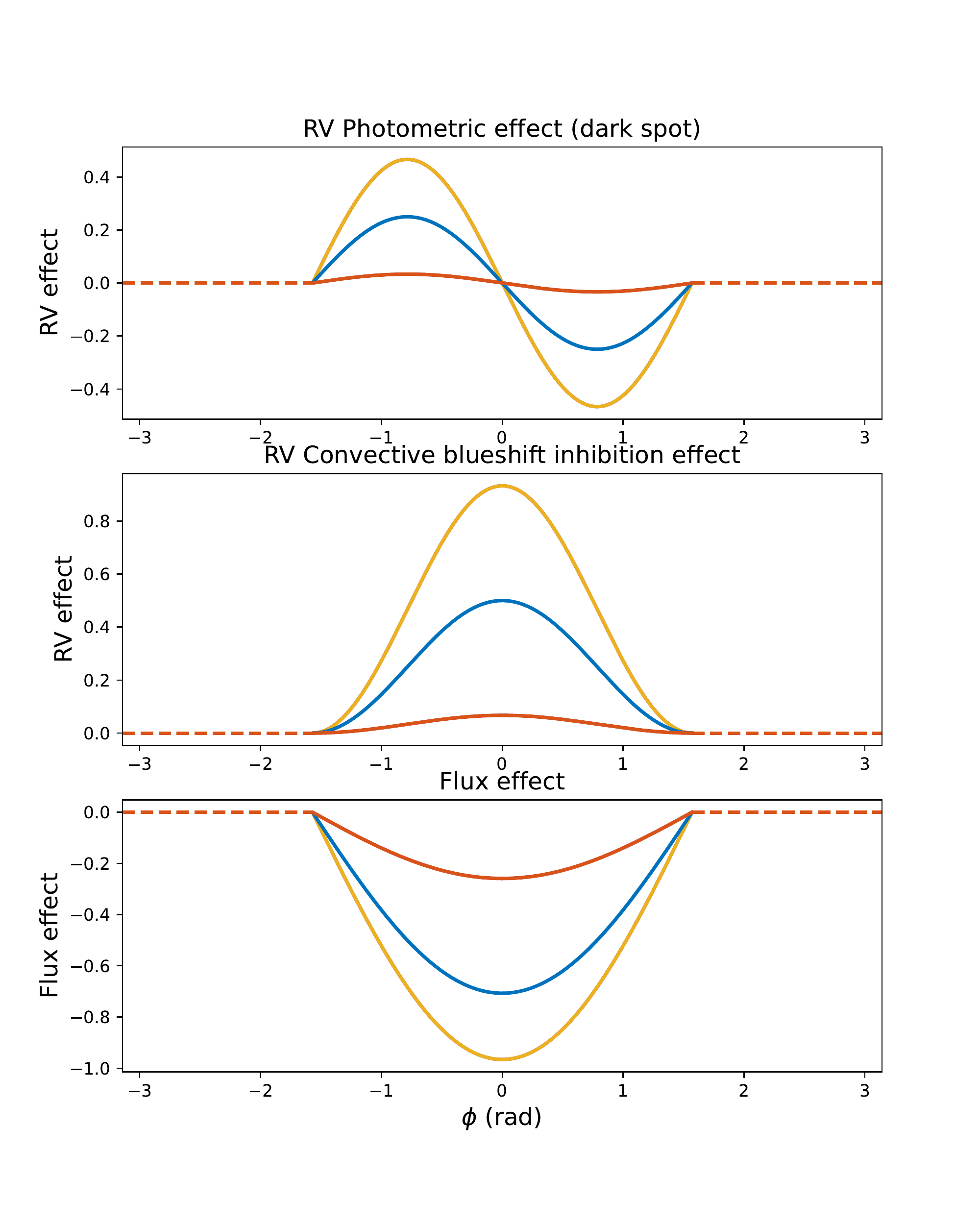}};
	\path (1.,-0.5) node[above right]{\large(b)};
 	\path (1.3,-1.5) node[above right]{\large(b1)};
  	\path (1.3,-4.5) node[above right]{\large(b2)};
  	\path (1.3,-7.5) node[above right]{\large(b3)};
	\path (9,0) node[above right]{\includegraphics[width=9cm]{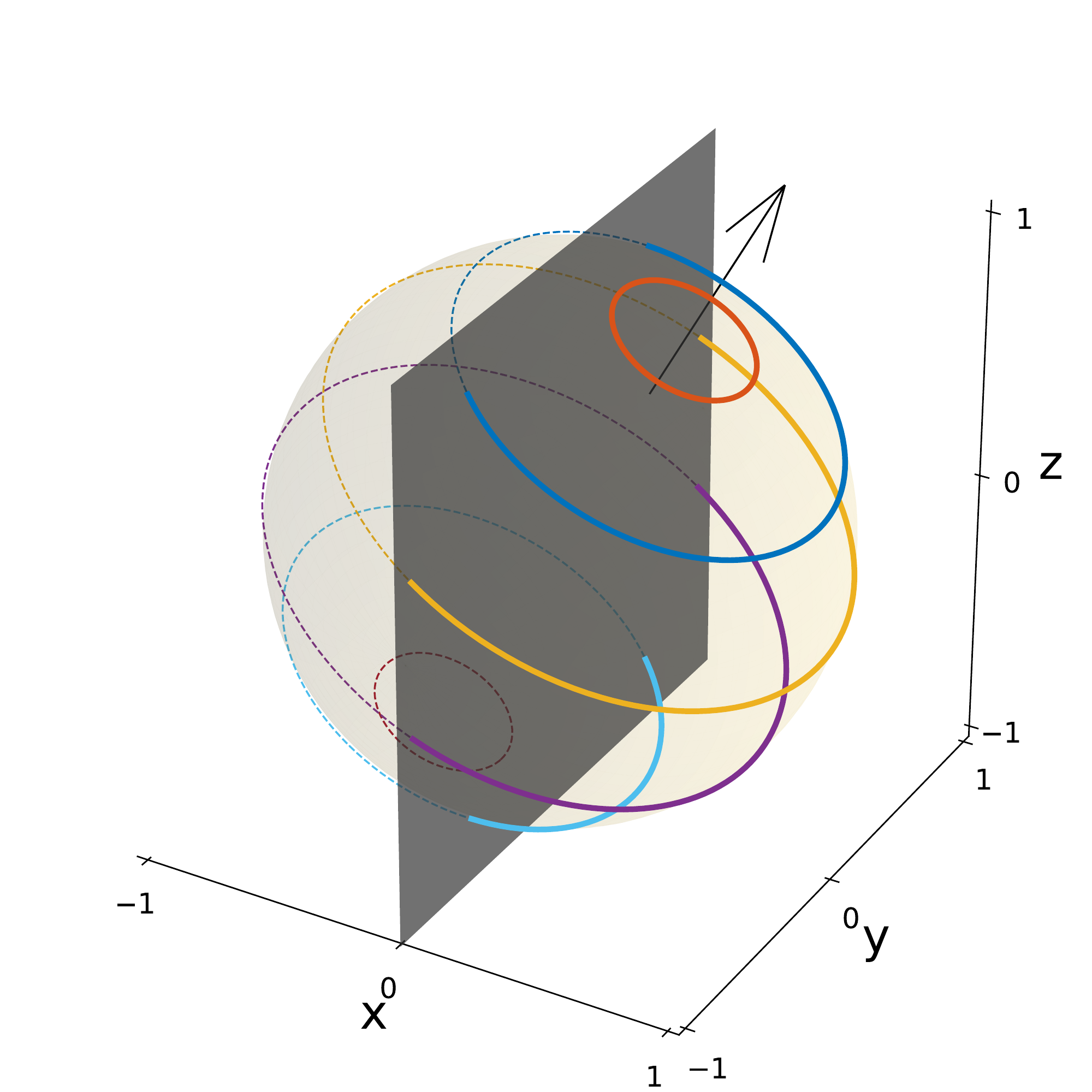}};
	\path (10,5) node[above right]{\large(c)};
	\path (9,-10.8) node[above right]{\includegraphics[width=9cm]{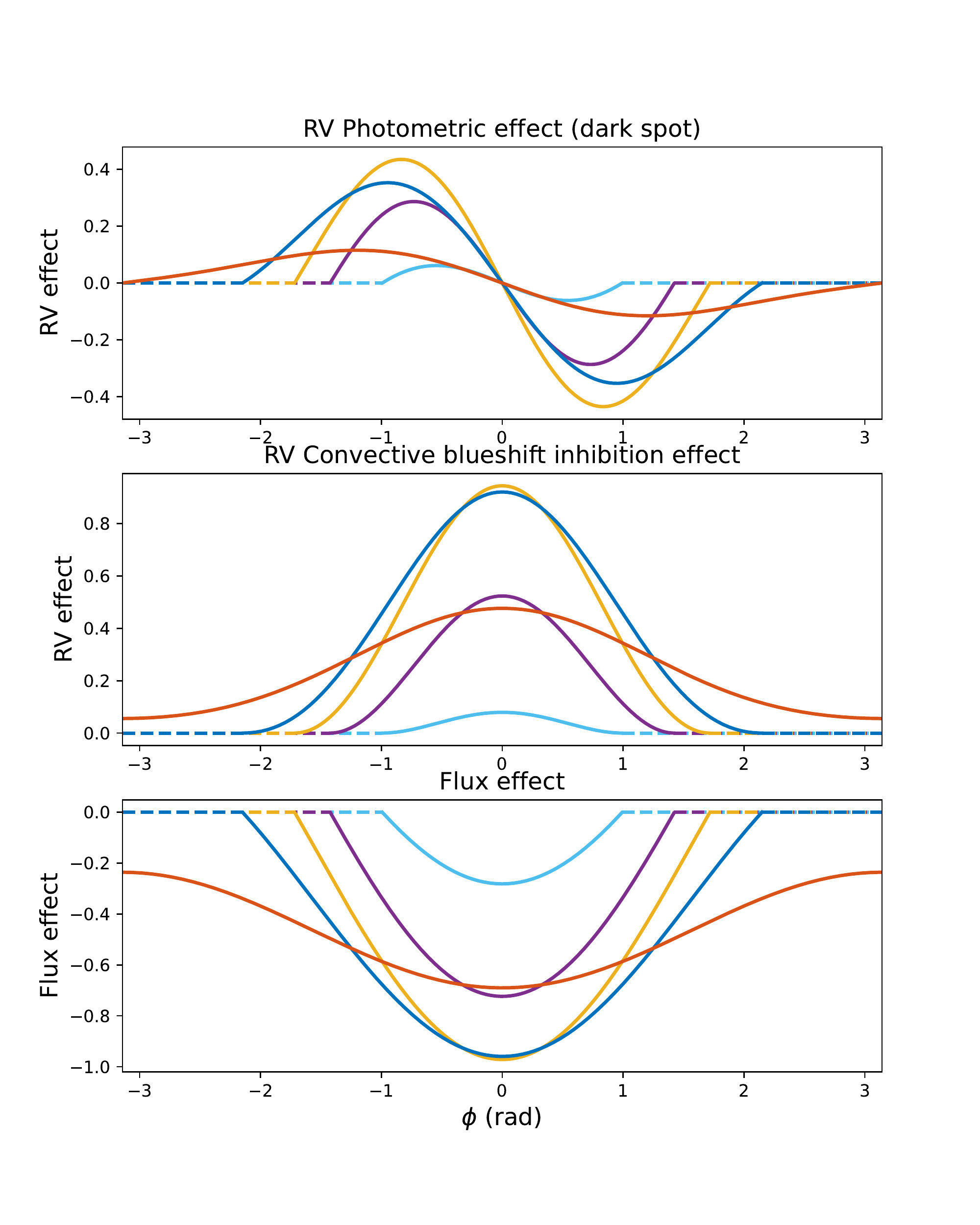}};
	\path (10,-0.5) node[above right]{\large(d)};
 	\path (10.3,-1.5) node[above right]{\large(d1)};
  	\path (10.3,-4.5) node[above right]{\large(d2)};
  	\path (10.3,-7.5) node[above right]{\large(d3)};
	\end{tikzpicture}
	\caption{(a) and (b), colored lines represent position of dark spots as the star rotates counter-clockwise with different inclinations. (c) and (d) represent the corresponding effect of dark spots with a constant limb-darkening law. Top: photometric RV effect, middle: RV convective blueshift inhibition effect, bottom: flux effect.   }
 	\label{fig:gs_example}
\end{figure*}
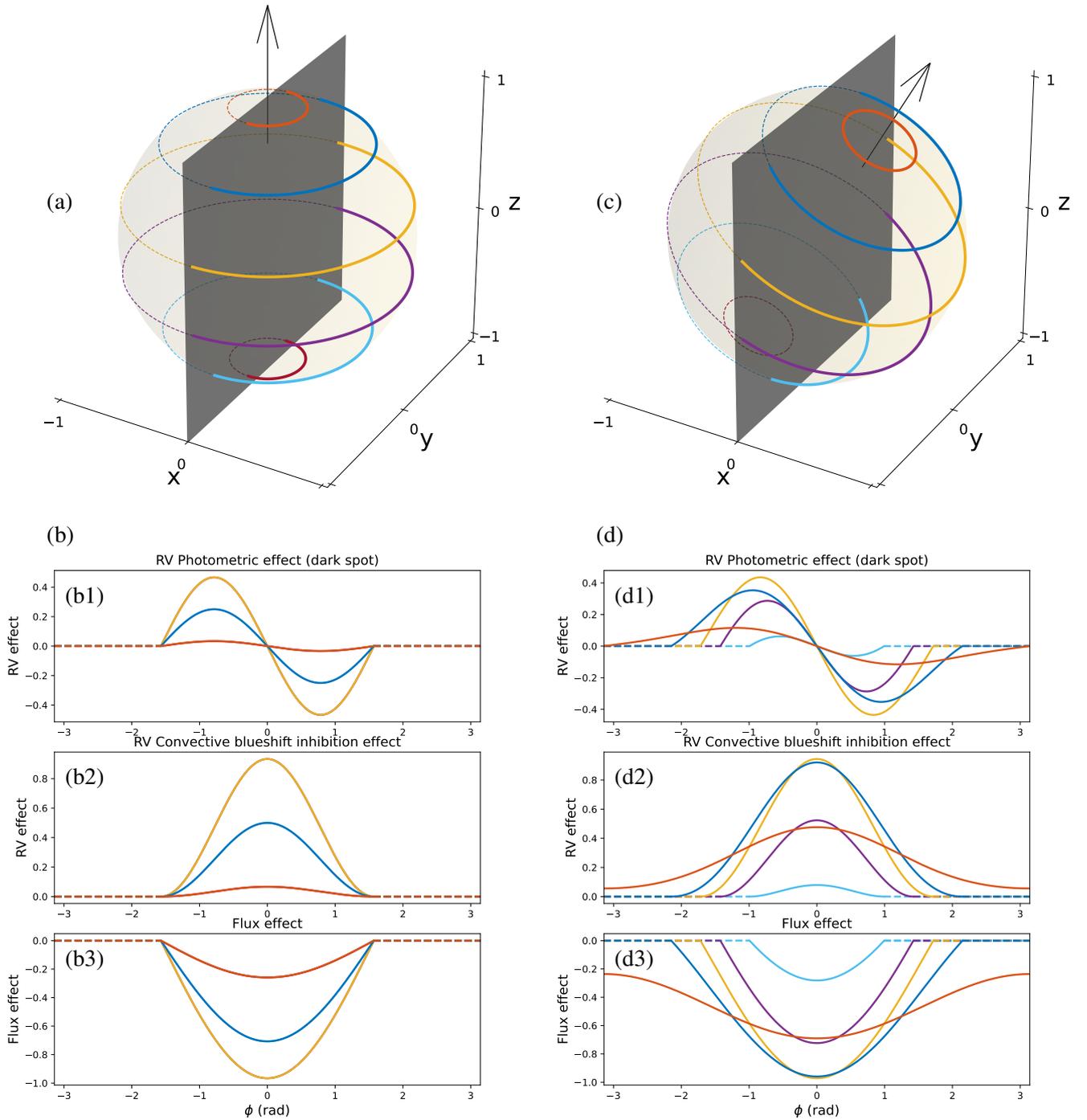

In Appendix \ref{app:spoteffect}, we show that based on these assumptions, the expression of $g$ in Eq.~\eqref{eq:ytf} for the photometric and inhibition of convective blueshift effect in radial velocity, $g_{ph}$ and $g_{cb}$, and the effect on photometry $h$ are
\small
\begin{align}
    g_{ph}(\bar{i},\delta,\phi, \omega, a, \tau) =&  W^\tau(t) l_{ph}^a(J(\bar{i},\delta,\phi)) \Delta f \omega R^\star  J(\bar{i},\delta, \phi) \cos \bar{i} \sin \phi \cos \delta \label{eq:yph_main}\\
    g_{cb}(\bar{i},\delta,\phi, \omega, a, \tau)  =& W^\tau(t) l_{cb}^a(J(\bar{i},\delta,\phi)) ( \Delta f V_{cb} + \Delta V_{cb}(f + \Delta f) ) J(\bar{i},\delta, \phi)^2 \label{eq:ycb_main} \\
    h (\bar{i},\delta,\phi, \omega, a, \tau) = & W^\tau(t) l_{f}^a(J(\bar{i},\delta,\phi))   \Delta f J(\bar{i},\delta,\phi) \label{eq:z_main}.
\end{align}
\normalsize 
where $\phi$ and $\delta$ are the latitude and longitude of the magnetic region on the stellar surface, $\bar{i}$ is the inclination of the star respective to the sky plane, $\omega$ is the angular rotational velocity of the star. Note that our definition of the inclination $\bar{i} = \pi/2-i$ where $i$ is the inclination classically used in the projected mass and velocities $m\sin i$ and $V\sin i$.
$\Delta f$ is the flux difference of the magnetic region with the continuum flux, $\Delta V_{cb}$ is the difference between the mean velocity of the flow of the continuum and of the magnetic region. The function $W(t)$ models the variation of the amplitude of the magnetic region as function of time and is parametrized by a time-scale of the appearance of spots $\tau$. The functions $l_{ph}$, $l_{cb}$ and $l_{f}$ are limb-darkening laws parametrized by coefficients $a$.  The quantity $J$ is ratio between the surface of the magnetic region projected onto the sky and its intrinsic surface, and its expression is 
\begin{align}
        J(\bar{i},\delta, \phi) = &   \sin \bar{i} \sin \delta + \cos \bar{i} \cos \delta \cos \phi . \label{eq:J_body}
\end{align}
Also, $J=\cos \psi$, where $\psi$ is the angle between the line of sight and the normal to the magnetic region. As such, it is the quantity classically used to express limb-darkening --- or limb-brightening --- laws.  
 In Figure \ref{fig:gs_example}, we show $g_{ph}$, $g_{cb}$ and $h$ for different spot latitudes and stellar inclination as a function of time with constant Limb Darkening. The longitude is taken as $\phi = \omega t$. 

 The effect of a given magnetic region in radial velocity is a weighted sum of the  photometric and convective blueshift inhibition effect.  The inhibition of convective blueshift seems to play the dominant role in Sun like stars \citep{meunier2010b,haywood2016}, but the photometric effect might dominate in others. 
 Depending on whether the magnetic region is a spot or a facula, the value of $\Delta f$ will be positive or negative, respectively, but the convective blueshift inhibition effect is always positive. As apparent in Fig. \ref{fig:granmag}.b, spots tend to appear surrounded by faculae, and we might want to model their combined effects.

\subsection{Limb-darkening}

\label{sec:limbdarkening}

%Expressions \eqref{eq:yph_main}-\eqref{eq:z_main} are the same as in \cite{aigrain2012} except that we allow for different stellar inclinations, and add a limb-darkening law 
%The limb-darkening, or limb-brightening law expresses the variation of observed 

\begin{figure}
    \centering
    \includegraphics[width=0.94\linewidth]{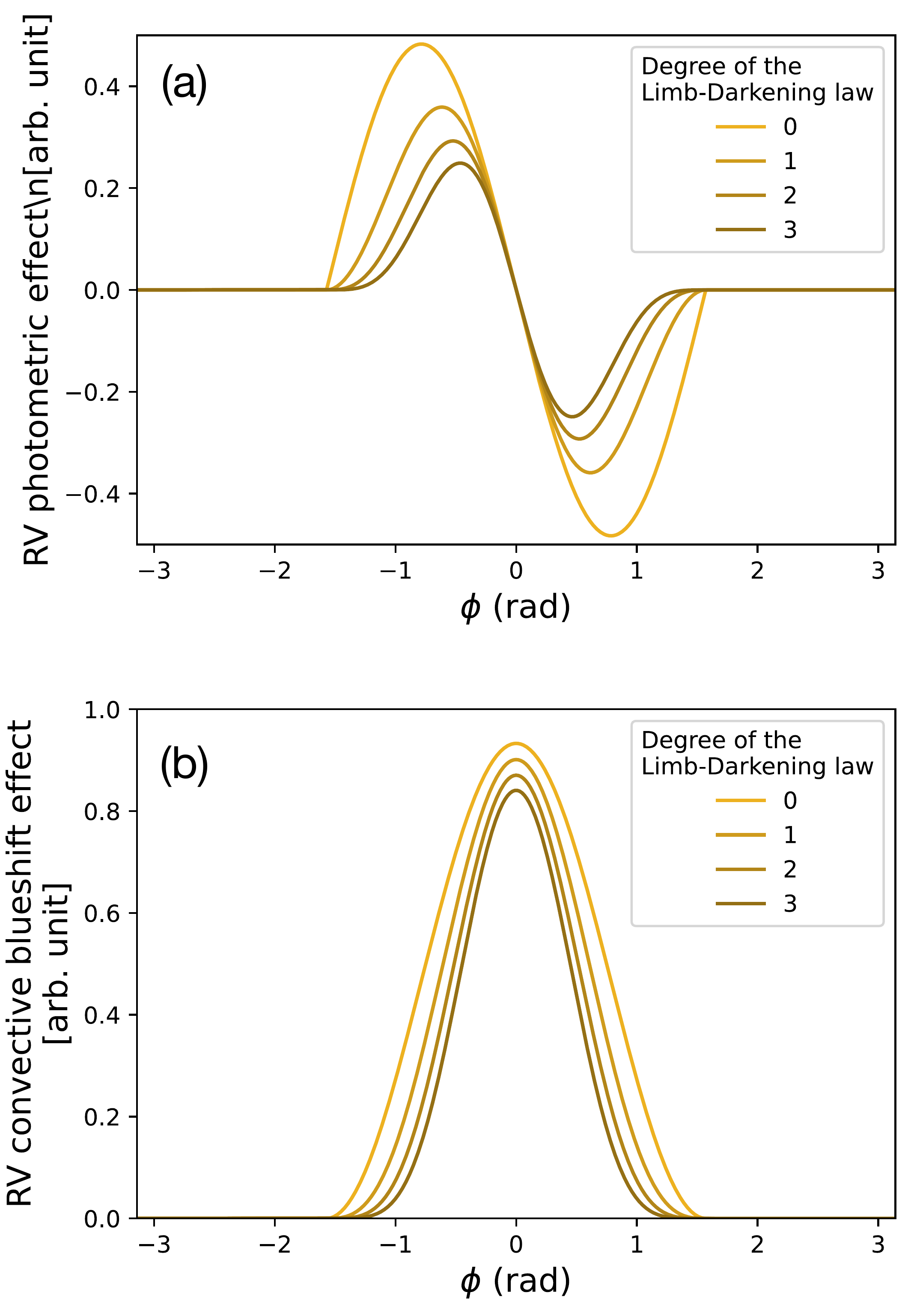}
    \caption{Different RV effects of an equatorial dark spot due to the photometric effect (Fig. a) and inhibition of the convective blueshift (Fig. b), with limb-darkening laws of different degrees. With the notations of Eq. \eqref{eq:ldlaw}, in increasingly darker yellow we take $d=0, 1, 2, 3$ and for $j<d$, $a_j=0$.  }
    \label{fig:limbdarkeffect}
\end{figure}

The central part of the Sun appears brighther than its edges, a phenomenon known as limb-darkening, also present in other stars. In our model we assume that this effect acts as a multiplicative factor that changes the amplitude of the RV and photometric signals.
Our limb-darkening law is expressed in powers of $J$, 
 \begin{align}
l(J) = \sum\limits_{k=0}^d a_k J^k .
\label{eq:ldlaw}
 \end{align}
The limb-darkening law is not necessarily the same for the convective blueshift inhibition, photometric and flux effect. For the convection pattern, the corrugated aspect of the granules blocks a fraction the flux towards the Limb \citep{beckers1978, cegla2019}. Furthermore, faculae and their counterpart in the chromosphere, plages, have a limb-brightening effect which counteracts the limb-darkening behaviour of the flux \citep{frazier1971, unruh1999, meunier2010b}.
In Fig. \ref{fig:limbdarkeffect}, we show how different choices of limb-darkening law affects the RV effect of an equatorial dark spot. The limb-darkening effect tends to smooth the effect of a spot, and shift the maximal effect towards the longitude of the observer. 

%For a degree $d=0$, we have the same effects as those of Fig. \ref{fig:gs_example} (b)

\subsection{Window function}
\label{sec:windowfunction}

%Spots and faculae grow then shrink with a certain lifetime $\tau$, which depends on the star.

In Eqs. \eqref{eq:yph_main}-\eqref{eq:z_main}, we defined a function $W^\tau(t)$ grasping the increase and decrease in intensity as spots and faculae grow and vanish. As the star rotates, the area of the feature projected onto the visible disk changes. This geometric effect is already modelled, such that the window is approximately proportional to the area of the magnetic region times the temperature difference between the feature and the continuum. The evolution of spot area on the Sun has been described in  \cite{bumba1963}, which founds a decay rate that is exponential for spots with lifetime less than a solar rotation, and linear for the others.

On the Sun, spots appear 10-11 times faster than they vanish \citep{howard1992, javaraiah2011}. For faculae, which are longer lived, this ratio is closer to 3 \citep{howard1991}. In \cite{petrovay1997}, it is argued that sunspot area has a parabolic decay, consistent with the observation of \cite{gomez2014}, that sunspots display a rapid decline in area before the rate stabilisises.    
%sunspot decay \citep{meyer1974}. 

\begin{figure}
    \centering
    \includegraphics[width=0.94\linewidth]{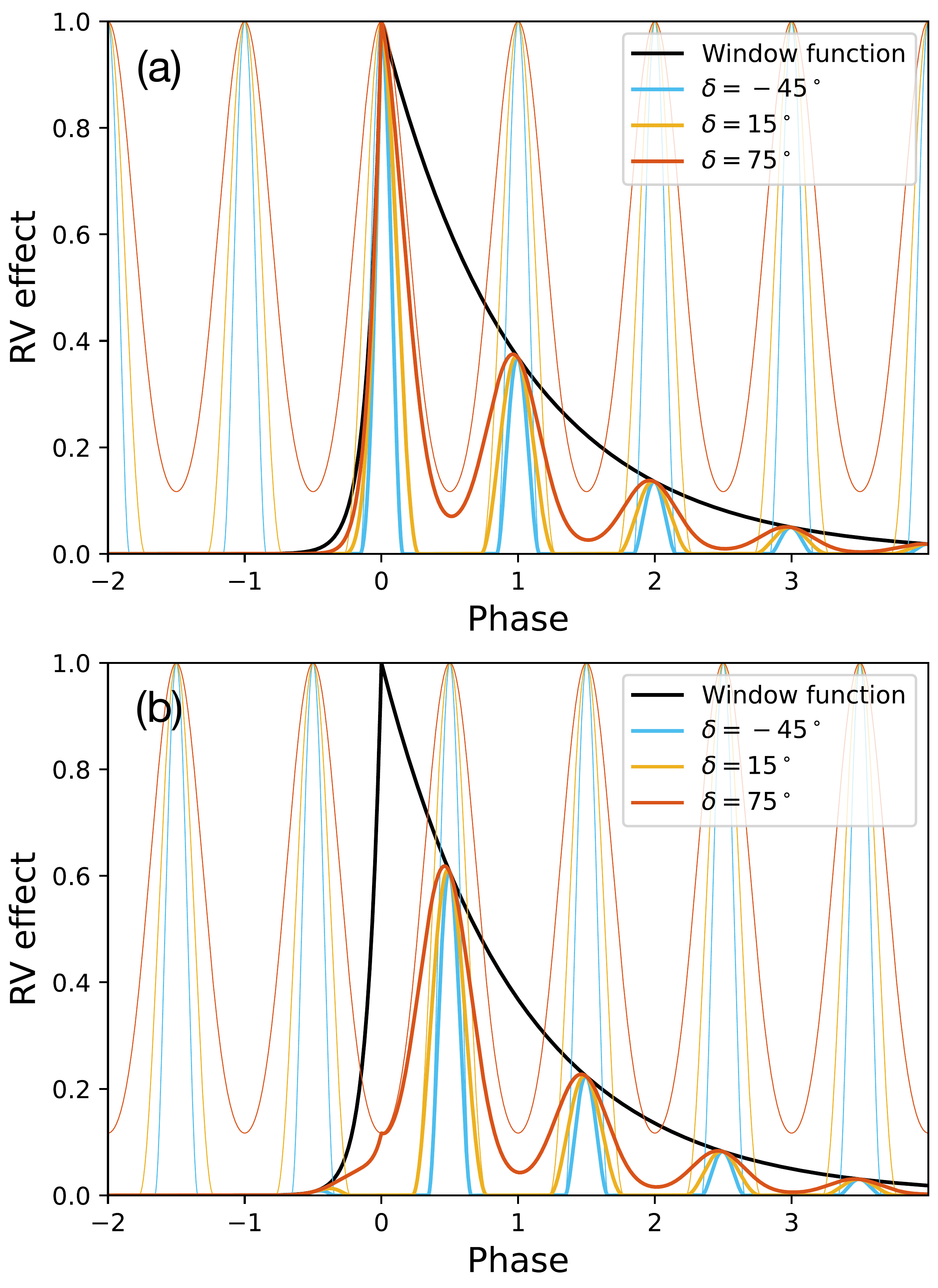}
    \caption{RV convective blueshift inhibition on the radial velocities as a function of the rotation phase with a rising and decaying amplitude with a star inclined as in Fig. \ref{fig:gs_example} (d). Black line: window function consisting of two exponentials, thin lines: effect of a spot without amplitude variation, bold lines: effect of the spot with varying amplitude. Colors correspond to the effect of spots at different latitudes (color code is the same as Fig. \ref{fig:gs_example} (d)). In (a) the maximum amplitude of the spot is attained on the visible hemisphere at the longitude of the observer, and in (b) the maximum amplitude is attained when the spot is 180$^\circ$ away from the observer. }
    \label{fig:window}
\end{figure}

In the present work, we want to represent efficiently the covariance of the different processes. To that end, we represent the covariance matrix in a semi-separable form, which restricts the possible window functions. 
We consider in particular three types of window functions: either a one sided exponential, null then with an exponential decay, asymmetric exponential, and symmetric exponential. In Fig. \ref{fig:window} we show examples of window functions with two exponential functions, such that the timescale of the growth is ten times shorter than that of the decay. Exponentials also  reproduce the observation that the decay rate decreases with time.

\begin{figure}
    \centering
    \hspace{0.1cm}
    \includegraphics[width=1.05\linewidth]{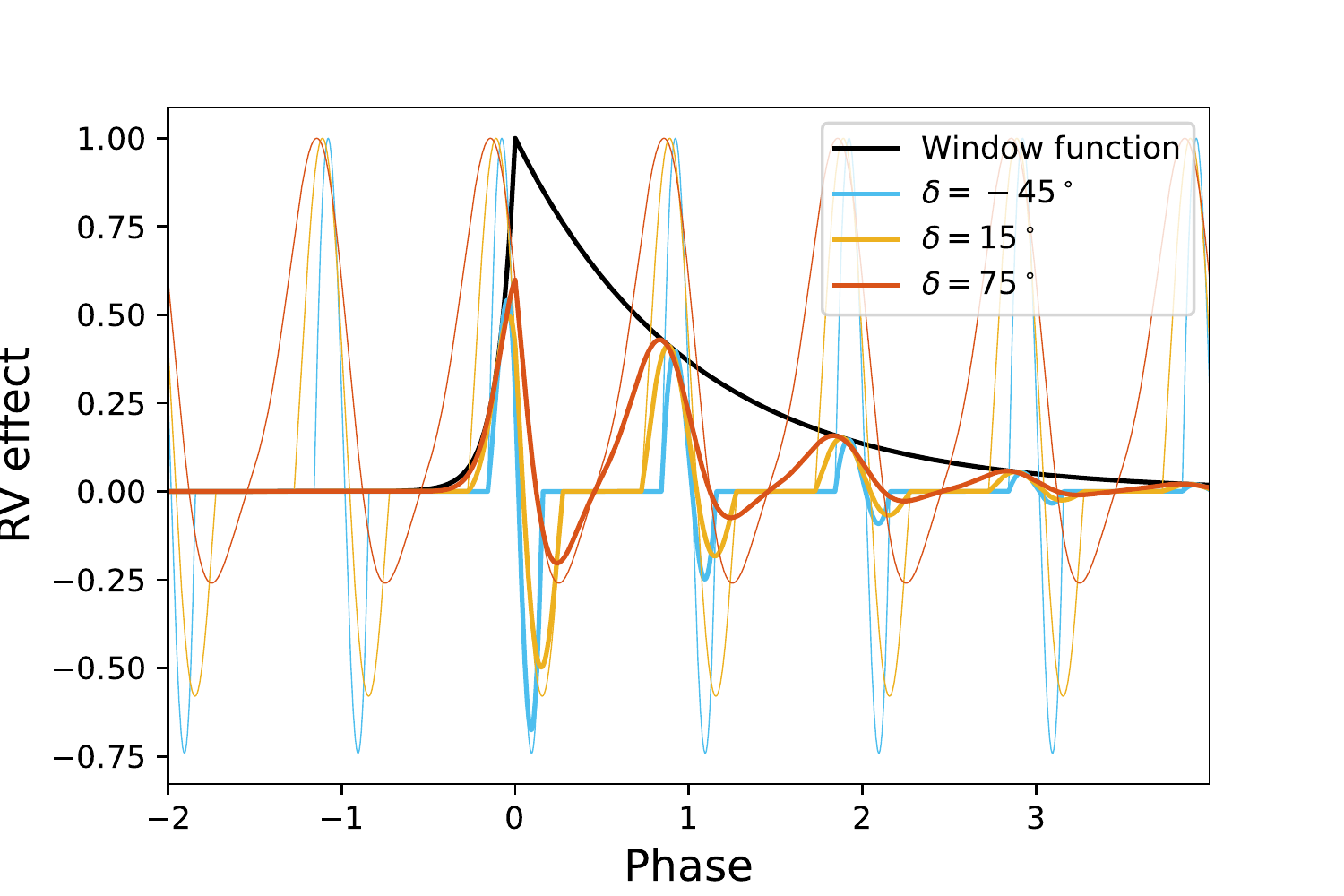}
    \caption{ RV Effect a dark spot with an equal amplitude of the convective blueshift inhibition and the RV photometric effect, with a rising and decaying amplitude with a star inclined as in Fig. \ref{fig:gs_example} (d). Black line: window function consisting of two exponentials, thin lines: effect of a spot without amplitude variation, bold lines: effect of the spot with varying amplitude. Colors correspond to the effect of spots at different latitudes (color code is the same as Fig. \ref{fig:gs_example} (d)). The maximum amplitude of the spot is attained on the visible hemisphere, at the longitude of the observer. }
    \label{fig:window_equal_contribution}
\end{figure}

\subsection{Group of spots and faculae}
\label{sec:group}

In our model, we suppose that stellar features appear potentially with a variable rate, but independently of each other. However, magnetic regions might appear in certain configuration to each other. To take this into account, we adapt our representation, and consider $g$ the effect of a group of spots. %We simply need to write $g$ as a linear combination of functions such as in eqs. \eqref{eq:yph_main}-\eqref{eq:z_main} with different phases. As a function of the impulse response of one spot $g(t;\gamma)$, our new impulse response $\tilde{g}$ is 
%\begin{equation}
 %   \tilde{g}(t; \gamma, (a_i)_{i=1..n}) = \sum\limits_{i=1}^n a_i g(t-\tau_i;\gamma_i)
%\end{equation}
 On the Sun, spots typically appear on longitudes shifted by 180$^\circ$~\cite{borgniet2015}, and there is evidence of this behaviour having effects in the Sun HARPS-N radial velocity measurements~\citep{hara2021b}. In our public code, we implemented this particular case, and assume that the impulse response is of the form 
\begin{equation}
    \label{eq:g180}
    \tilde{g}(t; \gamma, (a_i)_{i=1..n}) = g(t;\gamma_i) + \alpha g(t- P/2;\gamma_i) 
\end{equation} 
The parameter $\alpha$ can be random, controlled by a probability distribution $p(\alpha \mid \eta)$. %Another way to take into account the correlation between the appearance of spots is to assume that they appear following a process allowing for correlations, but this

\subsection{Overall impulse response}
\label{sec:overall_modell}

 The global effect of a spot or a facula on radial velocity is a linear combination of $g_{ph}$ and $g_{cb}$. Because of degeneracies between some of the parameters, we simplify the amplitude coefficients of $g_{ph}$, $g_{cb}$ and $h$. Furthermore, we assume a common limb-darkening law for all three signals, and pose $\phi =\omega t$. The effect of a spot or facula is 
 \begin{align}
    \begin{split}
   &  g_1(t; \bar{i},\delta, \omega, a, \tau, \sigma_g, \beta) = \\ & \sigma_g l_{cb}^a(J(\bar{i},\delta,\omega t)) W^\tau(t) \left(    J(\bar{i},\delta, \omega t)^2  + \beta  J(\bar{i},\delta, \omega t) \cos \bar{i} \sin \omega t  \right)\\  \end{split}
    \end{align}
The expression for the impulse response on radial velocity and flux are
 \begin{align}
 &    g(t; \bar{i},\delta, \omega, a, \tau, \sigma_g, \beta, \alpha) =  g_1(t) + \alpha g_1\left(t -\frac{\pi}{\omega}\right) \label{eq:ir_rv}  \\
  &   h(t; \bar{i},\delta, \omega, a, \tau, \sigma_h, \alpha) =  \sigma_h W^\tau(t) l_{f}^a(J(\bar{i},\delta,\omega t)) J(\bar{i},\delta,\phi)  \label{eq:ir_flux} 
 \end{align}
Our model thus depends on the stellar inclination $\bar{i}$, the latitude of the spot $\delta$, the stellar rotation frequency at latitude $\delta$, $\omega$, the limb-darkening coefficients $a$, the time scale of the lifetime of the magnetic region $\tau$, an amplitude $\sigma_g$ or $\sigma_h$ depending on whether we consider the radial velocity or flux effect,  or its and for radial velocities,a ratio between the photometric and convective blueshift effect, $\beta$. 
The angular velocity of the stellar surface depends on the latitude, a phenomenon known as differential rotation. This could be included in our model, but we do do not consider it in the present work. 

The model of Eqs. \eqref{eq:ir_rv} and \eqref{eq:ir_flux} can be used to model jointly the combined effect of spots and faculae. We can also choose to have two or more independent processes modelling isolated spots, isolated faculae, spots surrounded by faculae etc. Finding the optimal trade-off between flexibility and simplicity is left for future work.  
%Interestingly, because of the orthogonality of sine and cosines, the covariance obtained for a process with 

\subsection{Rate $\lambda(t)$}
\label{sec:lambda}
\label{sec:rate}

\begin{figure}
    \centering
    \includegraphics[width=\linewidth]{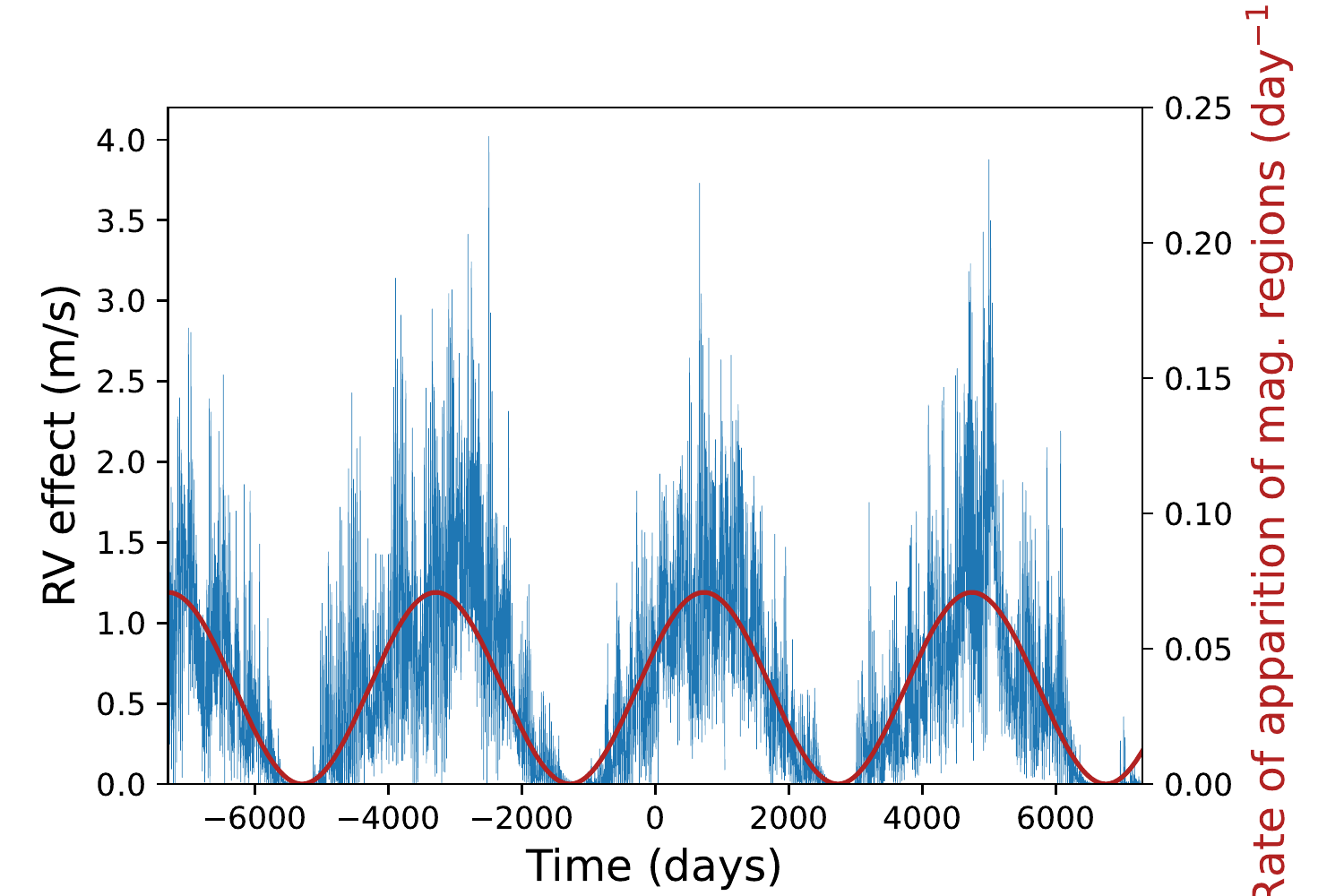}
    \caption{Simulated RV as a function of time when the RV effect is due to a combination of inhibition of convective blueshift and a RV photometric effect. The average lifetime of the magnetic regions is  15 days. The rate varies as a $A/2(1 + \cos 2\pi t/P_{magc})$ where $P_{magc}=11 years$ and $A$ is such that there are on average $50$ spots visible each day.   }
    \label{fig:mag_cycle}

    \centering
    \includegraphics[width=\linewidth]{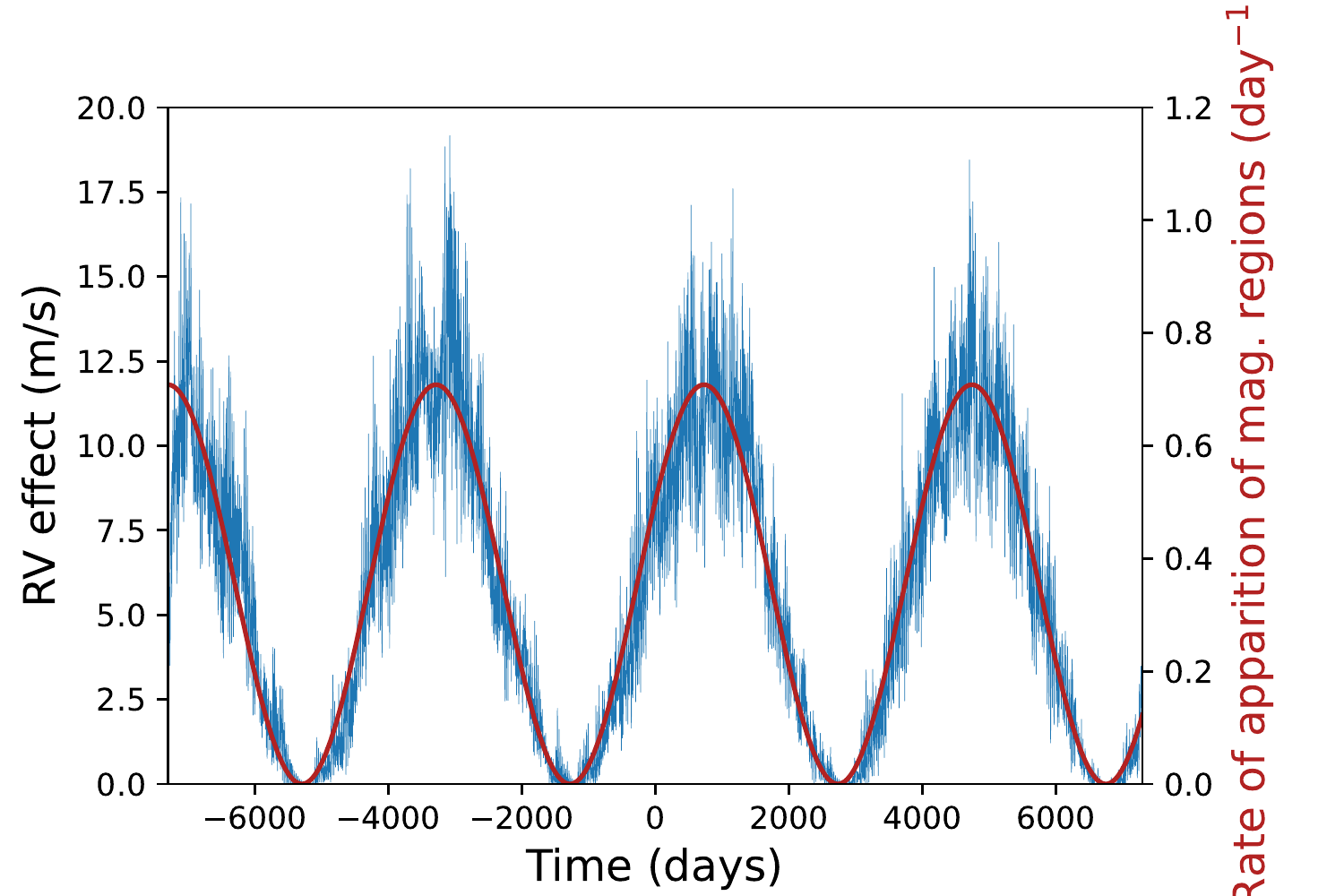}
    \caption{Simulated RV as a function of time when the RV effect is due to a combination of inhibition of convective blueshift and a RV photometric effect. The average lifetime of the magnetic regions is  15 days. The rate varies as a $A/2(1 + \cos 2\pi t/P_{magc})$ where $P_{magc}=11$years and $A$ is such that there are on average $50o$ spots visible each day.}
    \label{fig:mag_cycle_wolfn500}
\end{figure}

The rate at which active regions appear on the surface evolves with the magnetic cycle. A well known phenomenon on the Sun, where the counting of active regions has been the subject of numerous work. In the present section, we consider that the rate of appearance of spots is  $\lambda(t) = A(1 + \cos{2\pi/P t})$ where $P=11$ years and $t$ is the time. In Fig. \ref{fig:mag_cycle}, the red curve shows the daily apparition of spots as a function of time. The blue curve represents simulated RV data with active regions appearing with rate $\lambda$, described in the following section. 
In this particular simulation, the only active regions considered are spot groups and their surrounding faculae. The constant $A$ is chosen to have an average of $\sim 5$ visible spot groups per day. In Fig. \ref{fig:mag_cycle_wolfn500}, the number of spot groups is $\sim 50$ visible spot groups per day. 

Another option is to consider $\lambda(t)$ as a Gaussian process, similarly to the approach taken in \cite{camacho2022}. This approach is described in Appendix \ref{app:variable_rate}.

\subsection{Distribution of spot and faculae parameters }
\label{sec:distributions}

\begin{figure}
    \centering
    \includegraphics[width=\linewidth]{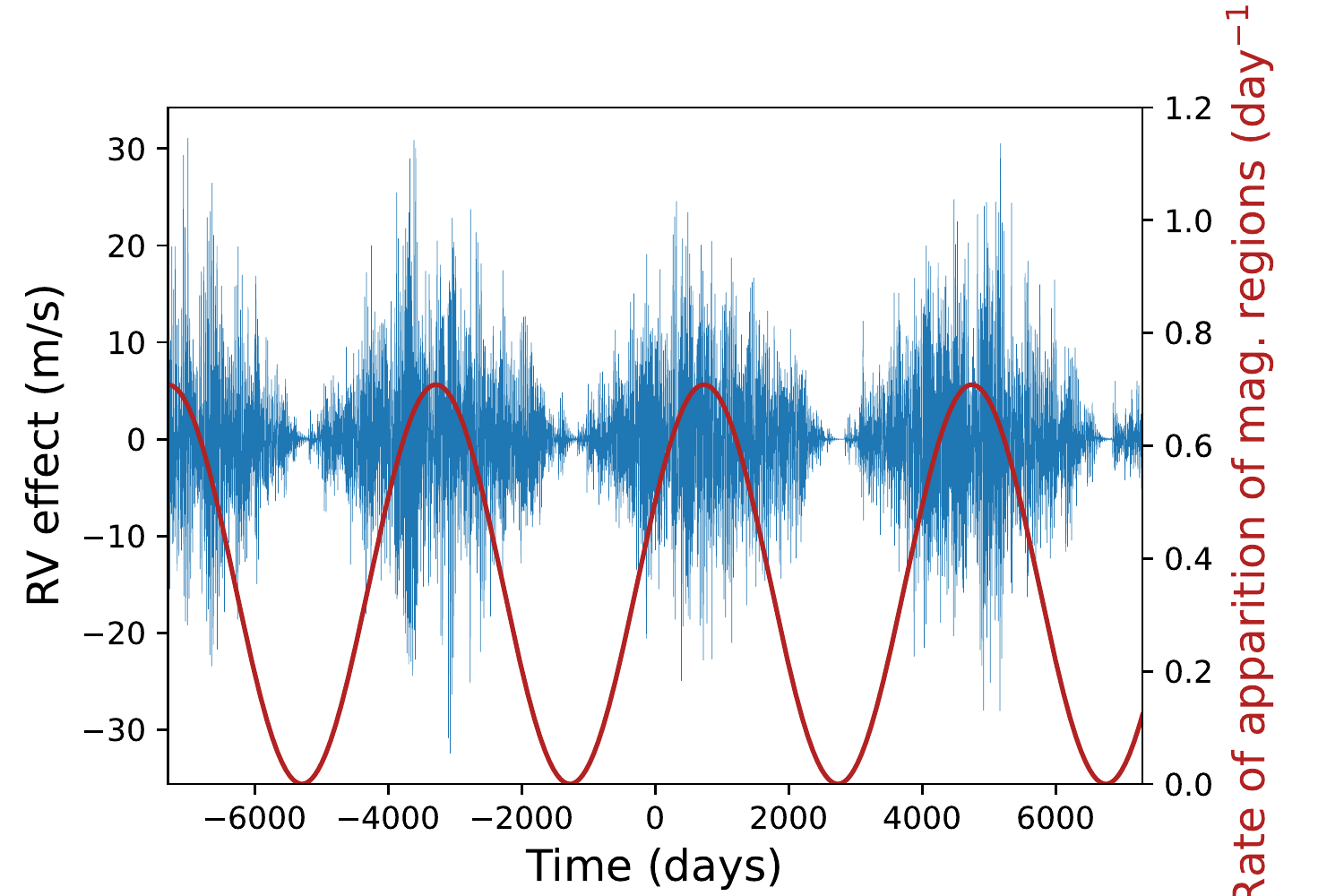}
    \caption{Simulated RV as a function of time when the RV effect is only due to the RV photometric effect. }
    \label{fig:mag_cycle_photom}
\end{figure}
%The covariance and higher order cumulants of a FENRIR process modelling stellar activity is given in Eqs. \eqref{eq:crosscova_body} and \eqref{eq:kappan}, in sections \ref{sec:spots}, \ref{sec:limbdarkening} and \ref{sec:group}  we specified the impulse responses $g(t, \gamma)$ and $h(t, \gamma)$. We now need to define two quantities. First, $p(\gamma \mid t, \eta)$, the statistical distribution of the spot and faculae parameters $\gamma$ knowing the time of appearance of the spot and $t$ and hyperparameters of the distribution $\eta$. Second, we need to choose a rate of appearance of spots and faculae $\lambda(t)$. 
%The covariance and higher order cumulants of a FENRIR process modelling stellar activity is given in Eqs. \eqref{eq:crosscova_body} and \eqref{eq:kappan}, and the 
%As we shall see, the integral over $\gamma$ can be evaluated numerically and the covariance interpolated. Thus, there is no need to specify a particular functional form for $p(\gamma \mid t, \gamma)$ for our formalism to operate in practice. Here, we discuss which choices seem reasonable, for this function but these are by no means a statement of what they should be. 

%We have 

In the model of Eqs. \eqref{eq:yph_main}, \eqref{eq:ycb_main} and \eqref{eq:z_main}, we chose to consider only the longitude as randomly drawn each time a new spot or faculae $\gamma = \delta$ is drawn. In the present work, we further assume that the longitude at which the feature reaches its maximal area is uniformly distributed on $[0, 2\pi]$. This assumption is partially untrue at least for the Sun, because active regions tend to appear close to existing active regions and on active longitudes, and will be refined in future work. 

We consider as hyperparameters of our process quantities which do not vary from one spot apparition to the other: the stellar rotation frequency $\omega$, the inclination $i$, the limb-darkening coefficients  $(a_k)_{k=1..d}$, the time-scale of the window function $\tau$ and the relative amplitude of a spot at the opposed longitude $\alpha$, and parameters describing the probability distribution of the latitude of stars. 

%The decay rates seem to be normally distributed \citep{martinezpillet1993}. 

%, isolated spots, with lower maximum areas and timescale, and sunspot in groups, which are larger and longer-lived , 

In the La Laguna classification (see \cite{martinezpillet1993}, there are twio types of spots, isolated ones (La Laguna type 3), and spot grous (La Laguna type 2). Within the La Laguna type 2, there is evidence for two population of sunspot groups \citep{munoz_jaramillo2015, nagovitsyn2016} with a separation in lifetime at $\tau = 5$ days. The lifetime of magnetic regions depends on their maximum area. The Gnevyshev-Waldmeier law states that the lifetime $\tau$ and maximum area $A_\mathrm{max}$ of a sunspot are proportionally related, $A_\mathrm{max} = a \tau$ with $a=10$ millionths of the solar
hemisphere (MSH) per day. However, there is a very large scatter about that law, as shown in Fig. 5 of \cite{henwood2010} and Fig 8. of \cite{forgacsdajka2021}.

The statistical distribution of spots and faculae should change with time. First, at minimum solar activity, sunspots appear on average at a latitudes $30-45^\circ$ away from the equator. As the activity level rises, the average latitude of sunspots move towards the equator. Second, the ratio of spots to facuale coverage varies with stellar activity \citep{nemec2022}, so that the parameters governing their effects should depend on time in a non stationary way. 

%Expressions \eqref{eq:yph_main}, \eqref{eq:ycb_main}, \eqref{eq:z_main} depend on the stellar inclination, latitude, longitude  stellar rotation frequency: $i, \delta, \phi$ and $\omega$. In the formalism of section \ref{sec:FENRIR}, we now turn to the distribution of these elements when a spot or facula appears at time $t$, depending on hyperparameters $\eta$, $p(\gamma \mid t, \gamma)$. Furt

%How to make large spots. 
%Differential rotation?

%Non stationarity due to magnetic cycles. 

\subsection{Refined model, extension to activity indicators}

  In the previous section, we assume that the measured RV is the  sum of the local stellar RVs weighted by their relative flux. However, this is a simplistic assumption. In Appendix \ref{app:ccf}, we derive the expressions of RV, as well as ancillary indicators considering that the measured cross correlation function (CCF) is a weighted sum of the local stellar CCFs, as is done in SOAP 2.0 \citep{dumusque2014}. This formalism takes into account in particular the inter-dependence of velocity, contrast and CCF width.

\subsection{Effects not included}

The angular velocity of the solar surface depends on the latitude, a phenomenon known as differential rotation. It means that our parameter $\omega$ is linked to $\delta$. Second, for very large spots, the assumption that they are pointwise breaks down. These features are not yet implemented in FENRIR. We note that from a preliminary analysis, we found that the first order effect of differential rotation is to reduce the time-scale of cohererence of the signal, in other words, to make the window function described in Section \ref{sec:windowfunction} shorter. So far, the correlation between the presence of existing magnetic regions and the location of new ones is only taken into account in the definition of $g$, to account for spots at opposed longitudes. We do not take into account other types of correlations.

  \section{Analysis of HARPS-N solar radial velocities and SORCE Total Solar Irradiance}
  
  \label{sec:sun}

\begin{figure}
    \centering
    \includegraphics[width=\columnwidth]{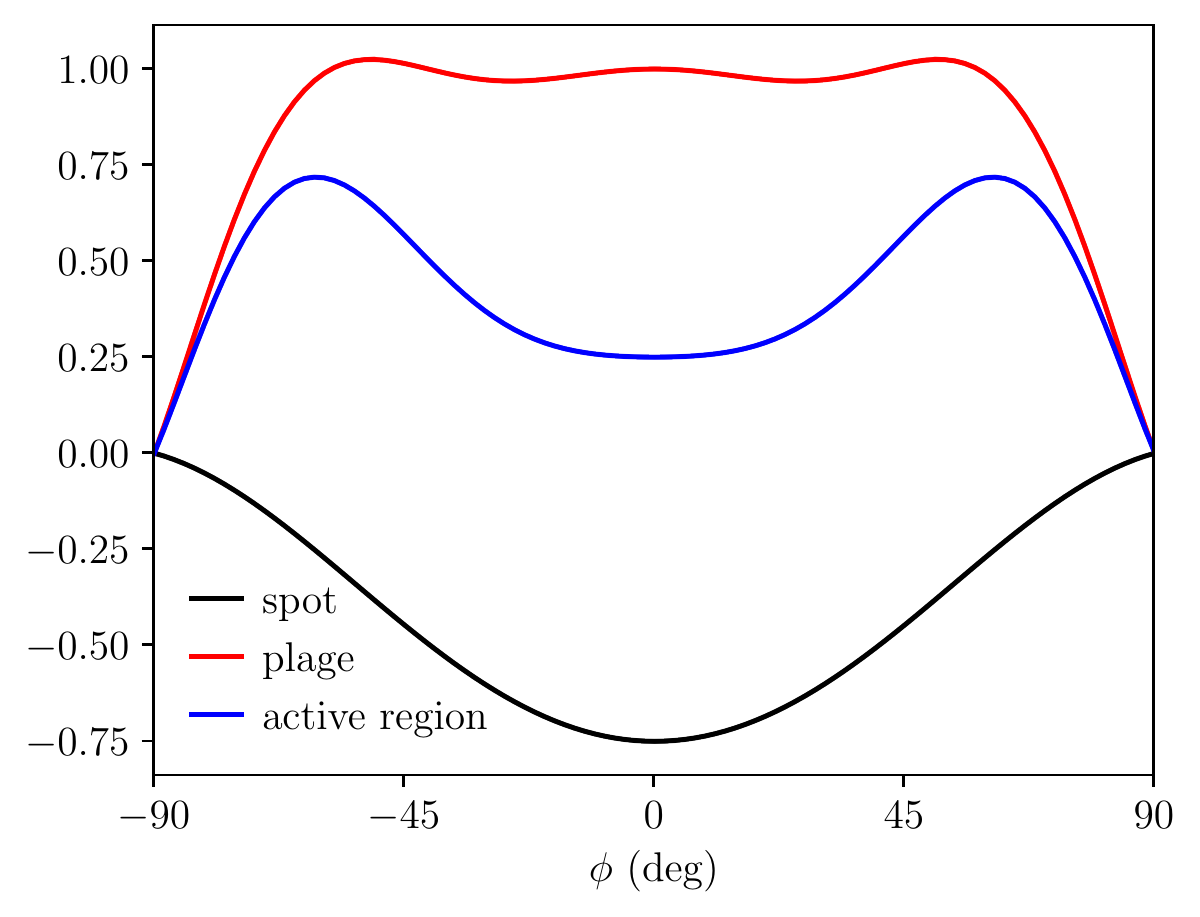}
    \caption{Illustration of the photometric effect of spots, plages, and global effect of an active region accounting for the limb-brightening effect, the limb-darkening effect, and the projection effect.}
    \label{fig:brightening}
\end{figure}

  To illustrate the use of FENRIR GP kernels, we analyzed the HARPS-N solar radial velocity time series provided by \citet{dumusque2020} together with the
  SORCE Total Solar Irradiance time series \citep{kopp2020}.
  We jointly modeled the radial velocities and the photometric time series using a FENRIR kernel.
  The SORCE photometry was binned by day, so we also binned the HARPS-N by day for consistency.
  The SORCE time series covers a time span of 17 years from February 2003 to February 2020,
  while the HARPS-N data cover 3 years from July 2015 to July 2018.
  We considered a population of spots/faculae following a symmetrical distribution of latitudes in the two hemispheres.
  The distribution was assumed to follow a mixture of two Gaussian distributions
  with means $\pm \mu_\delta$ and standard deviations $\sigma_\delta$.
  We used a quadratic limb-darkening law (see Eq.~(\ref{eq:ldlaw})) with coefficients $a_0=0.3$, $a_1=0.93$, $a_2=-0.23$ \citep[e.g.,][]{livingston2002}.
  Since faculae tend to appear in the vicinity of spots,
  we considered in our model the joint effect of spots and faculae in an active region.
  The faculae are affected by the limb-brightening effect \citep[e.g.][]{dumusque2014},
  which we model in the same fashion as the limb-darkening, using a quadratic law,
  with coefficients \citep[see][]{meunier2010a}:
  \begin{align}
    b_{\mathrm{plage},0} &= 1 - b_{\mathrm{plage},1} - b_{\mathrm{plage},2},\nonumber\\
    b_{\mathrm{plage},1} &= - \frac{407.7}{250.9 - 407.7 + 190.9},\nonumber\\
    b_{\mathrm{plage},2} &= \frac{190.9}{250.9 - 407.7 + 190.9}.
  \end{align}
  With this choice, the brightening is 1 when the faculae is at the center of the star.
  The contrast of a spot is on the contrary approximately constant (i.e., $b_{\mathrm{spot},0}=-1$, $b_{\mathrm{spot},1}=b_{\mathrm{spot},2}=0$).
  We now need to scale the contributions of the spots and faculae in an active region.
  While the flux decrease due to a spot is about 15 times stronger
  than the flux increase due to a plage of the same size,
  the surface covered by plages is about 20 times larger \citep[e.g.][]{meunier2010a}.
  We thus adopt the following brightening-law for an active region containing spots and faculae:
  \begin{align}
    b_0 &= \frac{1}{4} - b_1 - b_2,\nonumber\\
    b_1 &= b_{\mathrm{plage},1},\nonumber\\
    b_2 &= b_{\mathrm{plage},2}.
  \end{align}
  We illustrate in Fig.~\ref{fig:brightening} the photometric effect of a spot, a plage, and the global effect of an active region.
  We precomputed the five first Fourier coefficients (constant term, fundamental, and three first harmonics) of the periodic part of the kernel on a grid of values for $i$, $\mu_\delta$, and $\sigma_\delta$.
  
  We sampled $\bar{i}$ and $\mu_\delta$ on a regular grid of 51 values in the range $[0,\pi/2]$,
  and $\sigma_\delta$ on  a regular grid of 51 values in the range $[\pi/36, \pi/2]$.
  For each point in the grid, we sampled 101 values of $\delta$ in the range $[\bar{i}-\pi/2, \pi/2]$
  (latitudes that are visible from the observer)
  to integrate the Fourier coefficients over the population of spots/faculae (see Appendix~\ref{sec:spotpropdist}).
  We also compute the two coefficients
  (average photometeric effect and average RV convective-blueshift effect)
  of the terms due to the variations of the spot appeareance rate
  (i.e., magnetic cycle, see Appendix~\ref{app:variable_rate}).
  Finally we could then interpolate in the grid for any value of $\bar{i}, \mu_\delta, \sigma_\delta$,
  and renormalize the interpolated coefficients to set the amplitude of each effect as desired.
  We first include the effect of spots at the opposite longitude (parameter $\alpha_{180}$, see Eq.~(\ref{eq:g180})).
  We then multiply the two magnetic cycle amplitude by a factor $\gamma_\mathrm{mag.}$.
  This parameter measures the ratio between the amplitude of the magnetic cycle variations
  and the periodic variations due to stellar rotation.
  Finally we normalize all the coefficients such that
  the total amplitude (considering both the periodic and magnetic cycle components)
  in photometry is $\sigma_\mathrm{spot, phot.}$ ,
  the total amplitude
  of the RV convective blue-shift effect is $\sigma_\mathrm{spot, rv-cb}$,
  and the amplitude of the RV photometric effect is $\sigma_\mathrm{spot, rv-phot.}$.
  We note that in our model, there is no contribution of the RV photometric effect 
  in the magnetic cycle since this effect averages out.
  
  We used a single Matérn 1/2 kernel with timescale $\rho_\mathrm{spot}$ for the window function
  of the periodic component (i.e., sudden appearing and exponential decay of spots/faculae, see Appendix~\ref{sec:winfunc}),
  and we use another Matérn 1/2 kernel with timescale $\rho_\mathrm{mag}$ for the magnetic cycle.

  We modeled the super-granulation with a Matérn 1/2 kernel
  with timescale $\rho_\mathrm{gran.}$,
  amplitudes $\sigma_\mathrm{phot, gran.}$ in the photometry
  and $\sigma_\mathrm{rv,gran.}$ in the radial velocity,
  and no correlations between the two time series.
  We additionally included a jitter term (white noise) for both time series.
  We assumed that oscillations and short timescale granulation were
  captured by this jitter term since we used daily-binned data.
  Finally, we introduced an offset for each time series.

  Overall, this GP can be modeled with a \spleaf{} covariance matrix of rank $r=21$.
  The cost of likelihood evaluations of the model scales as $\mathcal{O}(r^2 n)$
  \citep[see][]{foremanmackey2016,delisle2022},
  where $n$ is the total number of measurements (including RV and photometry).
  
\begin{figure}
    \centering
    \includegraphics[width=\columnwidth]{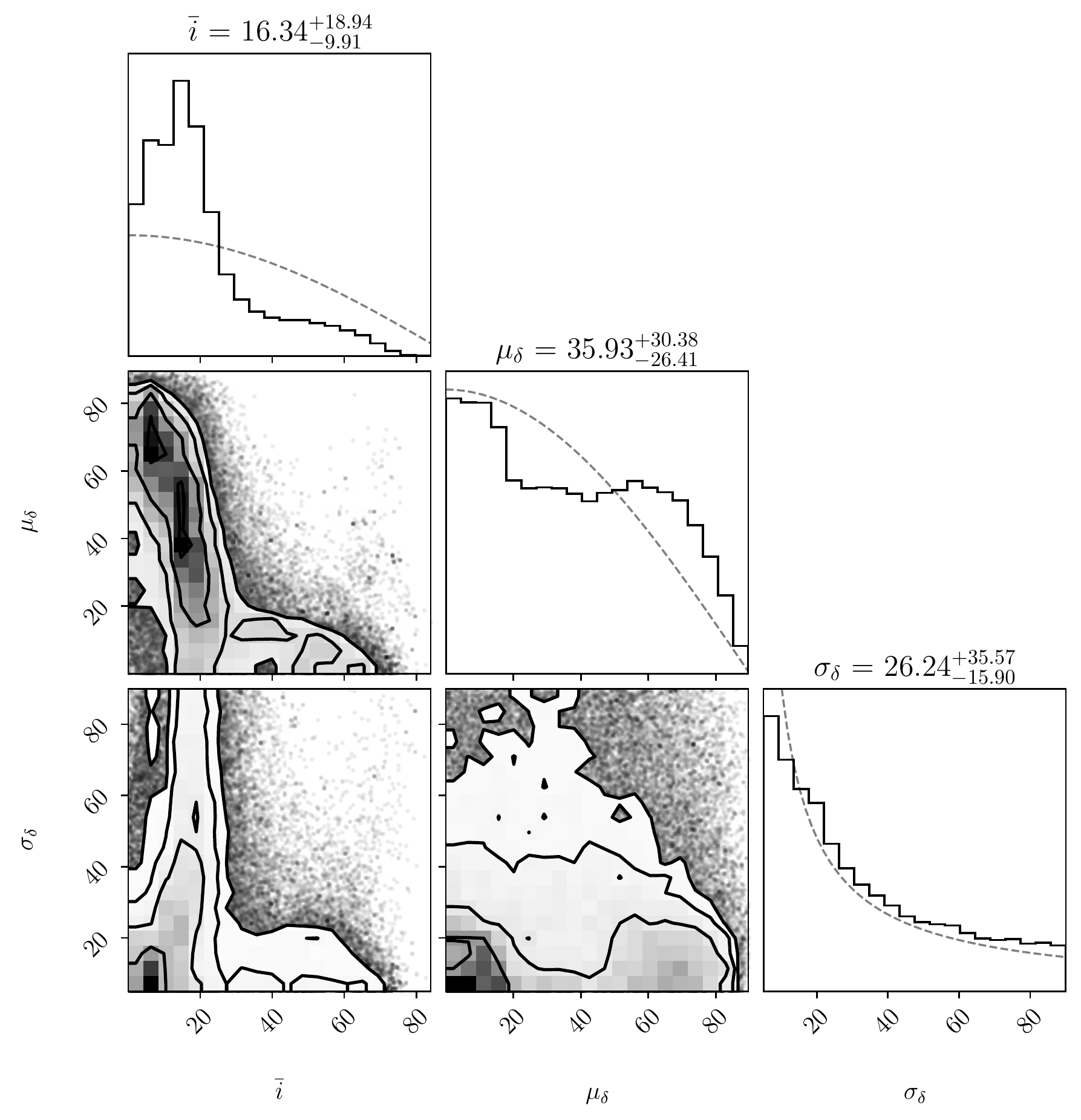}
    \caption{Corner plot of the MCMC samples for the Sun's inclination and the spots latitude distribution parameters.
    The dashed lines correspond to the priors.}
    \label{fig:mcmcmSun_corner}
\end{figure}

\begin{figure}
    \centering
    \includegraphics[width=\columnwidth]{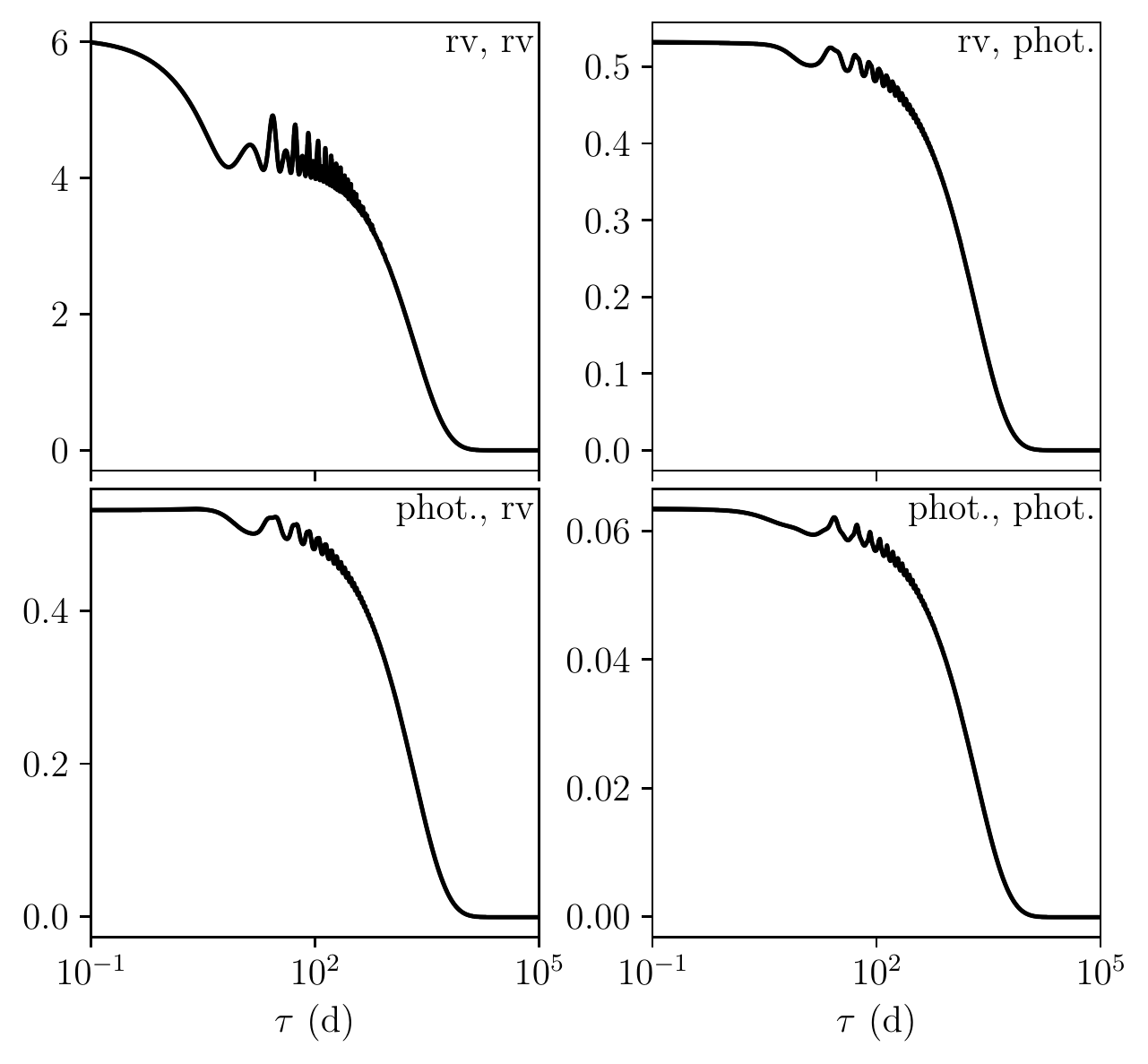}
    \caption{Kernel function using the maximum a posteriori parameters from the MCMC.}
    \label{fig:mcmcmSun_kernel}
\end{figure}

\begin{figure}
    \centering
    \includegraphics[width=\columnwidth]{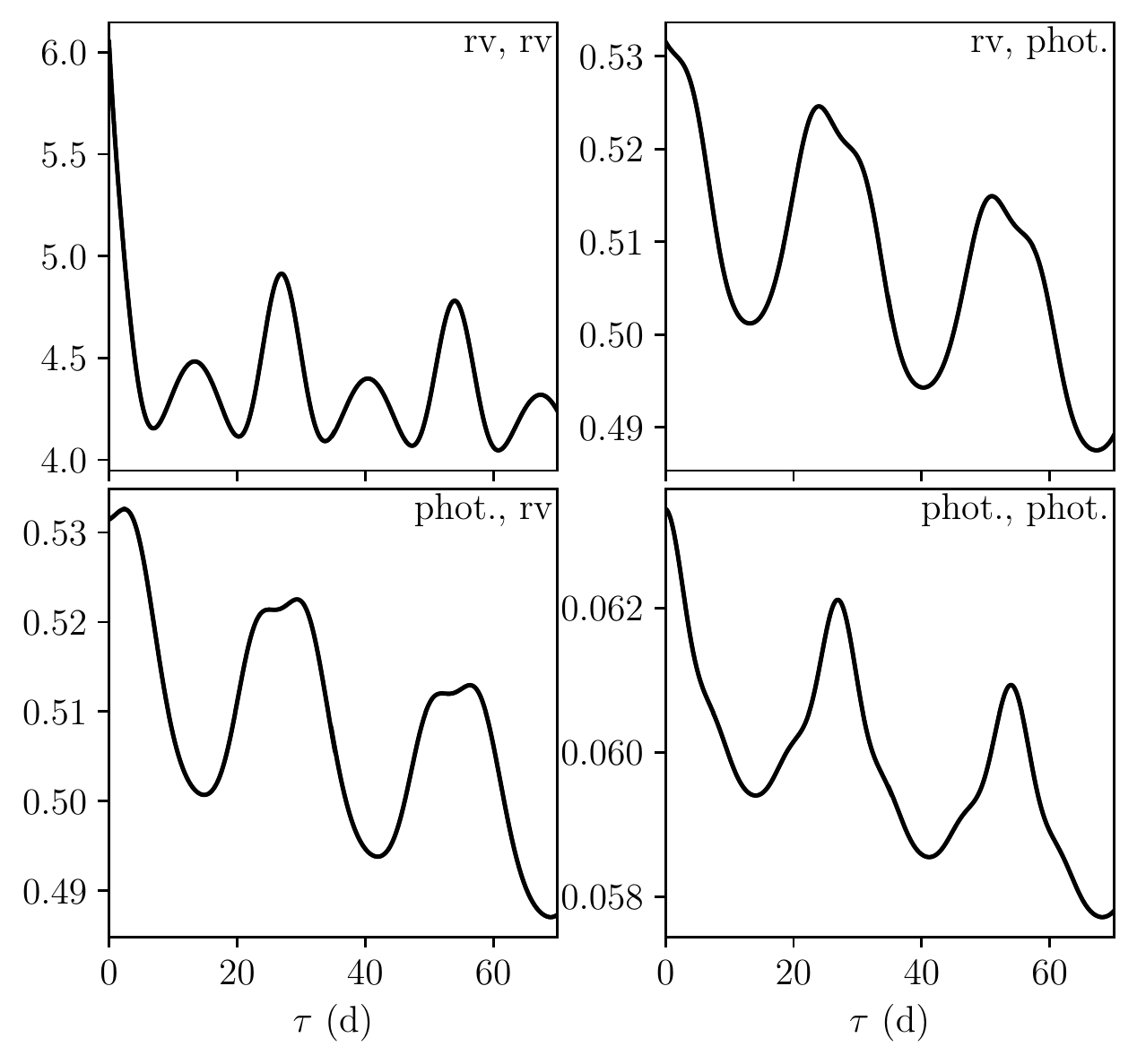}
    \caption{Zoom of Fig.~\ref{fig:mcmcmSun_kernel} in the range $\tau\in[0,70]$~d.}
    \label{fig:mcmcmSun_kernel_zoom}
\end{figure}

\begin{figure}
    \centering
    \includegraphics[width=\columnwidth]{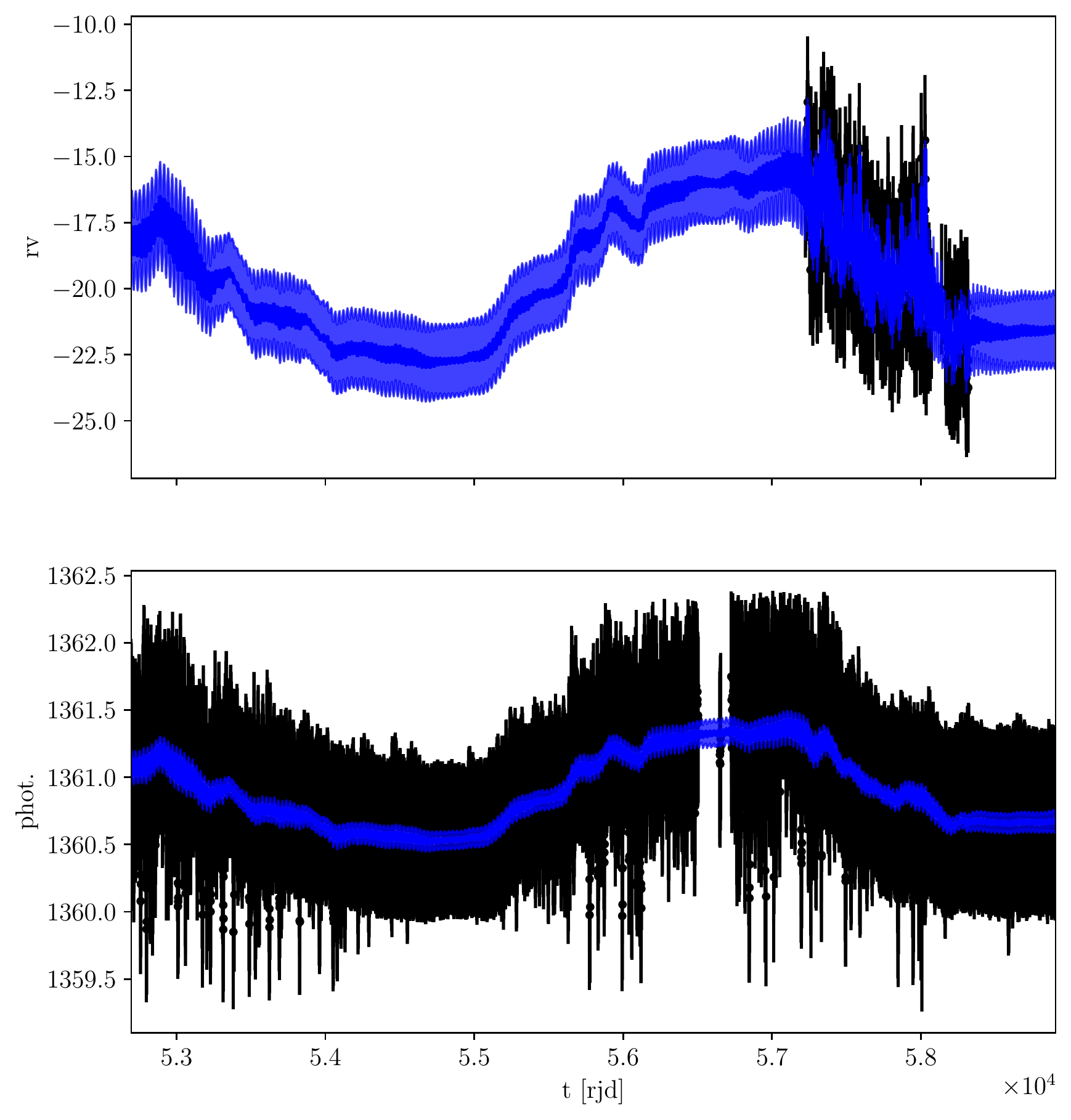}
    \caption{Maximum a posteriori solution superimposed over the RV (\textit{top})
    and photometric (\textit{bottom}) data.}
    \label{fig:mcmcmSun_timeseries}
\end{figure}

  We used the samsam MCMC sampler \citep[e.g.,][]{delisle2018} to explore the parameter space.
  We provide the priors and posteriors of all the explored parameters in Table~\ref{tab:mcmcSun}.
  In Fig.~\ref{fig:mcmcmSun_corner}, we show a corner plot of the stellar inclination $\bar{i}$ and the parameters of the distribution of spots in latitude.
  In Figs.~\ref{fig:mcmcmSun_kernel}~and~\ref{fig:mcmcmSun_kernel_zoom}, we show the kernel function corresponding to the maximum a posteriori set of parameters.
  Finally, Fig.~\ref{fig:mcmcmSun_timeseries} shows 
  the Gaussian process's conditional distribution
  corresponding to the maximum a posteriori set of parameters.

  From Fig.~\ref{fig:mcmcmSun_corner} we clearly see that our model is degenerate
  and that we cannot constrain well the distribution in latitude of spots.
  Since we observe the Sun from the Earth, the Sun's inclination is not constant
  and oscillates through the year ($\bar{i} \in [-7, 7]$~deg).
  We indeed find that small inclinations $\bar{i}$ are slightly preferred (see Fig.~\ref{fig:mcmcmSun_corner}),
  but this parameter remains poorly constrained in our model.
    
  \begin{table}[]
      \centering
\begin{tabular}{ccc}
\hline
\hline
Parameter & Prior & Posterior\\
\hline
\multicolumn{3}{c}{Offsets}\\
\hline
$\gamma_\mathrm{rv}$ & $\mathcal{U}(-10^5, 10^5)$ & $-20.20_{-8.85}^{+8.82}$\\
$\gamma_\mathrm{phot.}$ & $\mathcal{U}(-10^5, 10^5)$ & $1360.82_{-1.00}^{+0.99}$\\
\hline
\multicolumn{3}{c}{Jitter}\\
\hline
$\sigma_\mathrm{jit., rv}$ & $\mathrm{trunc}\mathcal{N}(0, 200)$ & $0.193_{-0.130}^{+0.147}$\\
$\sigma_\mathrm{jit., phot.}$ & $\mathrm{trunc}\mathcal{N}(0, 200)$ & $0.00505_{-0.00357}^{+0.00552}$\\
\hline
\multicolumn{3}{c}{Super granulation}\\
\hline
$\rho_\mathrm{gran.}$ & $\log\mathcal{U}(0.5, 10)$ & $1.667_{-0.276}^{+0.332}$\\
$\sigma_\mathrm{gran., rv}$ & $\mathrm{trunc}\mathcal{N}(0, 200)$ & $0.9767_{-0.0714}^{+0.0679}$\\
$\sigma_\mathrm{gran., phot.}$ & $\mathrm{trunc}\mathcal{N}(0, 200)$ & $0.00614_{-0.00434}^{+0.00677}$\\
\hline
\multicolumn{3}{c}{Spots/Faculae}\\
\hline
$\bar{i}$ & $\cos(\bar{i})\ (\bar{i}\in[0,90])$ & $16.3_{-10.0}^{+19.1}$\\
$\mu_\delta$ & $\cos(\mu_\delta)\ (\mu_\delta\in[0,90])$ & $36.0_{-26.5}^{+30.5}$\\
$\sigma_\delta$ & $\log\mathcal{U}(5, 90)$ & $26.2_{-15.9}^{+35.8}$\\
$P_\mathrm{rot.}$ & $\log\mathcal{U}(20, 60)$ & $26.979_{-0.108}^{+0.108}$\\
$\rho_\mathrm{spot}$ & $\log\mathcal{U}(10, 10^5)$ & $257_{-87}^{+135}$\\
$\sigma_\mathrm{spot, rv-phot.}$ & $\mathcal{N}(0, 200)$ & $-0.026_{-0.440}^{+0.453}$\\
$\sigma_\mathrm{spot, rv-cb}$ & $\mathcal{N}(0, 200)$ & $10.50_{-4.43}^{+3.99}$\\
$\sigma_\mathrm{spot, phot.}$ & $\mathcal{N}(0, 200)$ & $1.189_{-0.501}^{+0.425}$\\
$\alpha_\mathrm{180}$ & $\mathcal{U}(0, 1)$ & $0.092_{-0.066}^{+0.110}$\\
\hline
\multicolumn{3}{c}{Magnetic Cycle}\\
\hline
$\rho_\mathrm{mag.}$ & $\log\mathcal{U}(50, 10^5)$ & $46095_{-30714}^{+35264}$\\
$\gamma_\mathrm{mag.}$ & $\log\mathcal{U}(10^{-5}, 10^5)$ & $27.2_{-12.0}^{+13.5}$\\
\hline
\hline
\end{tabular}
      \caption{Set of parameters explored in the MCMC, with their priors and posteriors (median and 68.27\% interval).}
      \label{tab:mcmcSun}
  \end{table}


 \section{Discussion}
\label{sec:discussion}

\subsection{Granulation}
\label{sec:granulation}

We mentioned in Section \ref{sec:spots} that there is a convection pattern at the surface of the star. This creates a so-called granulation pattern. The surface of the star is composed of contiguous granules, such that the rising hot gas is at the center of the granule, cools down, and goes downwards at its periphery.  Besides the granulation phenomena, the Sun exhibits a so called super-granulation pattern. Large regions of the Sun have on average an upwards motion. This phenommenon occurs with a larger 1.8 days time scale. The existence of an intermediate scale meso-granulation effect is still debated.

Because regular, meso and super granulation proceed in a similar way and seem empirically to have comparable effects, we model them in the same way. The effect of a single granule, meso granule or super granule as a function of time can be modelled by
\begin{align}
     g'(t,\gamma) &= P(t, \gamma) W(t, \gamma)
\end{align}
where $W(t, \gamma)$ models the combined effect of the variation of brightness and area of the granule as a function of time, and $P(t, \gamma)$ is the periodic part due to stellar rotation and projection effect. 

On the Sun, Granules typically live for 15 minutes, however, averaging images of the Sun over an hour does not make the surface uniform because granules tend to appear at the same location \citep{cegla2013, cegla2018}. To take this into account, we assume that when a granule appears at time $t_0$, $N$ granules with the same properties (position, maximal amplitude, lifetime) will appear following at times $t_i$. These granules appear between time $t_0$ and $t_0+T$ with a rate $\lambda$, and thus their number $N$ follows a Poisson distribution of parameter $\lambda T$.  As a result, the impulse response for granulation is 
%g'(t,\gamma) &= W^\tau(t) l_{cb}^a(J(i,\delta,\phi))  \Delta V_{cb} J(i,\delta, \phi)^2 \approx W^\tau(t).
\begin{align}
     g_{gran}(t,\gamma) &= \sum\limits_{i=0}^N  P(t, \gamma) W(t-t_i) .
\end{align}
The periodic part is the same for all granules since they stay at the same position in the stellar frame, but they will grow and vanish randomly. We call the ensemble of $N$ granules appearing at the same position a granule packet. 

%In Section~\ref{sec:granulation}, we suggested a model of granulation where packets of granules can appear anywhere on the stellar granules. 

If a packet appears at time $t_0$, granules appear between time $t_0$ and $t_0+ T$ following a Poisson process with window function $W(t)$, which is approximately the profile of evolution of velocity times flux of a granule. In Appendix \ref{app:granulation}, we establish the form of the kernel corresponding to this process.  In the present section, to simplify the discussion, we assume that $T$ is sufficiently small compared to the stellar rotation period so that the star can be considered static during the packet lifetime. Then the granulation kernel is 
\begin{align}
    k_{gran}(\tau,\gamma) \propto \lambda T \int_{-\infty}^{\infty} W(t)   W(t + \tau)   \dd t  + (\lambda T)^2 \int_{-\infty}^{\infty}  w(t) w(t+\tau) \dd t, \label{eq:kgranul_body}
\end{align}
where
\begin{align}
    w(t) = \frac{1}{T} \int_{0}^T W(t-u) \dd u.
\end{align}
In Eq. \eqref{eq:kgranul_body}, there are two terms: the cross-correlation of the profile of evolution of velocity times flux of the granule, and the cross-correlation of $w(t)$. This one has a time-scale of order $T$, while the granule has a lifetime $1/\lambda$. This term appears if we take into account that granules appear consistently at the same location for a time $T$, this should then create correlation at a longer time-scale than the granule lifetime. In VIRGO data, it appears that the power spectral density of granulation decays with different rate between timescales of 30 min and 8 min, and beyond 4 min \citep{sulis2020}, the power spectrum between 4 and 8 min is dominated by asteroseismic oscillations.  The different slopes might be due to the time-scale of single granules and the time-scale of correlation of the granule locations. %This would mean that, at frequencies $\gtrssim$ 1 cycle/min the decrease of the PSD could be dominated by a $1/\nu^2$ term and 
%t could be responsible for the fact that there seems to be 

This model is only approximate on several accounts. First, we assume that the granule packets can appear anywhere. However, if a granule is present, it should be forbidden for a granule packet to appear at the same place. In other words, our model is similar to an urn model with replacement (the same position can be drawn twice), while it should be without replacement. Another option to represent granulation consists in modelling several granules simultaneously. Second, within a granule packet, we should forbid two granules to overlap. If we model collectively several granules, and assume that the impulse response is coherent over the timescale where granules appear at the same location, we should expect this as a single time-scale, and a covariance with a simple power decrease.

The power spectrum of the granulation effect on photometry and RV seems to be captured  by a so-called super-Lorentzian function with a power spectrum $P(\omega) = S_0/(1 + \omega^{a}/ \omega_0^{a})$ where $S_0,\omega_0$ and $a$ are free parameters \citep{harvey1985, dumusque2011i, kallinger2014, cegla2018, guo2022}. A value of $a \sim 4$, and a sum of at least two such processes seems to be favoured by photometric and RV observation \citep[][respectively]{kallinger2014, guo2022, luhn2022}, although there can be a higher discrepancy~\citep{dumusque2011i, cegla2018}.
 So called super and meso granulation seem also to be well approximated as a stochastic process with a super-Lorentzian power spectrum density with $a=4$, but different time scales and amplitude. In the formula \ref{eq:kgranul_body},  the kernel is dictated by the form of the granulation profile. We note that if $W(t)$ is a one sided exponential, that means the granules appear brutally and their RV signature wanes exponentially, then the kernel is a Matérn-3/2. The associated power spectral density of such kernels decreases asymptotically as $1/\omega^4$, like the super-Lorentzian profile with $a \sim 4$. A detailed discussion of granulation is left for future work.

%\subsection{Unsigned magnetic field}

	\subsection{Are stellar signals Gaussian?}
	\label{sec:testing_gaussianity}
	
	\subsubsection{Describing non Gaussian processes}
	
	  A stochastic process $X(t)$ is said to be a Gaussian process if for every finite collection of times $t_1,...,t_n$, then  $X(t_1),...,X(t_n)$ has a Gaussian multivariate distribution. A stationary process is such that the distribution of $X(t + \tau_1),...,X(t + \tau_n)$ does not depend on $t$. When a process is both Gaussian and stationary, it is fully characterized by its mean function $\mu(t) = \mathbb{E}\{X(t)\}$ and kernel, $k(\tau) = \mathbb{E}\{ ( X(t)- \mu(t)) ( X(t + \tau)- \mu(t + \tau)) \}$, which does not depend on $t$.  As we have seen, stellar activity is often assumed to be a stationary Gaussian process. 	In section~\ref{sec:usingFENRIR}, we established a covariance function for our stellar activity model, which can be used to parametrise a Gaussian process, but our calculation does not guarantee that for any choice of finite times $(t_i)_{i=1..N}$ the distribution of $y(t_1), ...y(t_N)$ should be Gaussian. 
  %In the present section, we suggest ways to study the non Gaussianity of stellar signals. This might be important both to better correct stellar activity signals, and to gain more information on the star from the data. 
  
Stellar activity signals cannot be strictly Gaussian. Indeed, if $X(t)$ is a Gaussian process of vanishing mean, $X(t)$ and $-X(t)$ have exactly the same covariance, and thus the same probability. However, the signature of the inhibition of the convective blueshift on radial velocity is not symmetrical, it always manifests as a net redshift (see Fig. \ref{fig:gs_example} (b2)). Furthermore, as soon as there is an excess of the effect of regions brighter or darker than the continuum, the flux effect, and the RV photometric effect do not have a symmetric distribution.

% We have seen in section \ref{sec:lambda} that because of magnetic cycles, the rate of appearance and properties of spots and faculae changes with time. As a result the assumption of stationarity over timescales of magnetic cycles ( $\sim 10$ years) is incorrect.

%As a result, the phase obtained on the first and second time interval are statistically independent. 

 %Let us consider a stochastic process $X(t)$ and a time span from times $t_0$ to $t_0+\Delta t$ where $\Delta t$ is fixed but $t_0$ might vary. %, and the second between times $t_3$ and $t_4$. %We choose $t2$ and $t_3$ such that $t_3-t_2$ is much greater than the process correlation.  
 %A stationary Gaussian process is fully characterized by its kernel, or equivalently by its Fourier transform, the power spectrum density. 
 
 The Fourier transform of the kernel is called the power spectral density, and has has also another interpretation: it is the expectancy of the squared modulus of the Fourier transform of the process.  Let us consider time span from times $t_0$ to $t_0+\Delta t$. If a stochastic process is Gaussian and stationary, its local Fourier transform on the timespan has the same expected modulus for all $t_0$, but we have no information on the phase: at all frequencies, the phase is distributed uniformly. If the process is Gaussian but non stationary, then the expected modulus of the Fourier transform computed on $t_0$ to $t_0+\Delta t$ depends on $t_0$, but the phases are completely random. In particular, knowing the phase at a certain frequency does not give information on the phase at other frequencies.    When we loose the assumption of Gaussianity, it might mean in particular that the phases of the process are linked with one another. To explore this aspect, we use the notion of cumulants and their Fourier transform, the polyspectra.

 The notion of cumulant is briefly presented here, and we refer the reader to \cite{mendel1991} for a more in-depth introduction. Suppose you have a stochastic process $X(t)$. Given $n$ time stamps $(t_i)_{i=1,..,n}$ the cumulant generating function of $X(t_1), ..., X(t_n)$ is 
\begin{align}
   K^n_{(t_i)_{i=1,..,n}}(\alpha_1,...,\alpha_n) = \ln \mathbb{E} \{ \e^{\alpha_1 X(t_1) +...+ \alpha_n X(t_n) }\}
\label{eq:cumulant_genfunction}
\end{align}
In the Taylor expansion of $K$ as a function of $\alpha_i$s, the cumulant of order $n$, $\kappa_n$, is the coefficient of their products, $\alpha_1\alpha_2...\alpha_n$. One of the properties of stationary Gaussian processes is that their cumulants of order equal or greater than 3 are equal to zero. Cumulants are thus used as Gaussianity tests. 
Determining in which extent stellar activity departs from a Gaussian behaviour is beyond the scope of this work. We here only discuss whether the FENRIR process adopted in Section \ref{sec:physmodel} is Gaussian, and show its behaviour is compatible with the observed asymmetry of radial velocities observed on the Sun. 

When the FENRIR process is stationary, its $n$-point correlation function, or cumulant of order $n$ is expressed as a function of time lags between $n-1$ points and a reference one,
 \begin{align}
   C(\tau_1,...,\tau_{n-1} ) = \lambda \iint\limits_{-\infty}^{\infty} g(t, \gamma )g(t+ \tau_{1}, \gamma )... g(t + \tau_{n-1}, \gamma ) p( \gamma \mid  \eta)   \dd t \dd \gamma .
   \label{eq:ncorr_function}
\end{align}
In this expression, $\lambda$ is the rate of apparition of stellar features, $g(t, \gamma(t))$ is the impulse response: the effect at time $t$ of a feature of parameters $\gamma$, and $p( \gamma \mid \eta)$ is the distribution of feature parameters depending on hyperparameters $\eta$, such as the stellar inclination and mean latitude of the features.
 Applying the Fourier transform to Eq. \eqref{eq:ncorr_function}, we obtain the polyspectrum of our FENRIR process
\small
\begin{align}
    S_n(\omega_1, ..., \omega_{n-1};\eta) = \lambda \iint \hat{g}(\omega_1, \gamma)  ... \hat{g}(\omega_{n-1}, \gamma) \hat{g}\left(-\sum_{i=1}^{n-1} \omega_{i}, \gamma  \right) p(\gamma\mid \eta) \dd \gamma,
    \label{eq:polyspectrum_body}
\end{align}
\normalsize
where $\hat{g}$ is the Fourier transform of $g$. 

\subsubsection{Poisson rate and asymmetry }

Based on expressions \eqref{eq:ncorr_function} and \eqref{eq:polyspectrum_body}, we can see right away that increasing the rate $\lambda$ tends to make the signal more Gaussian. Indeed, suppose we have a Gaussian process (GP) with the same mean and covariance as our FENRIR process (or same order 1 and 2 cumulants). Since it is stationary, from Eq. \eqref{eq:ncorr_function}, its variance at each time would be constant, equal to $\lambda \bar{g^2}$, where $E_{g^2}$ is the expectancy of $g^2$ taken over the parameters $\gamma$ and integrated on $t$.  If we were to estimate empirically the cumulants of this GP, the would not be exactly zero because of statistical fluctuation, but would have a certain standard deviation. The standard deviation of cumulant $n$ of a Gaussian process is proportiontal to $\sigma^n$, and in turn to $\lambda^{n/2}$. So, the ratio of the cumulants of order $n\geqslant2$ of the FENRIR process and its GP approximation is proportional to $\lambda/\lambda^{n/2}$, which tends to zero when $\lambda$ increases for all $n\geqslant 3$. Intuitively, we can see that if the rate is very low, there is only one feature at a time. If we see the beginning of the feature, the phase and amplitude of the signal can be predicted with infinite accuracy. As the rate increases, it becomes more difficult to predict the phase of the signal because we see at any time a superposition of several features. 

\begin{figure}
    \includegraphics[width = 0.94\linewidth]{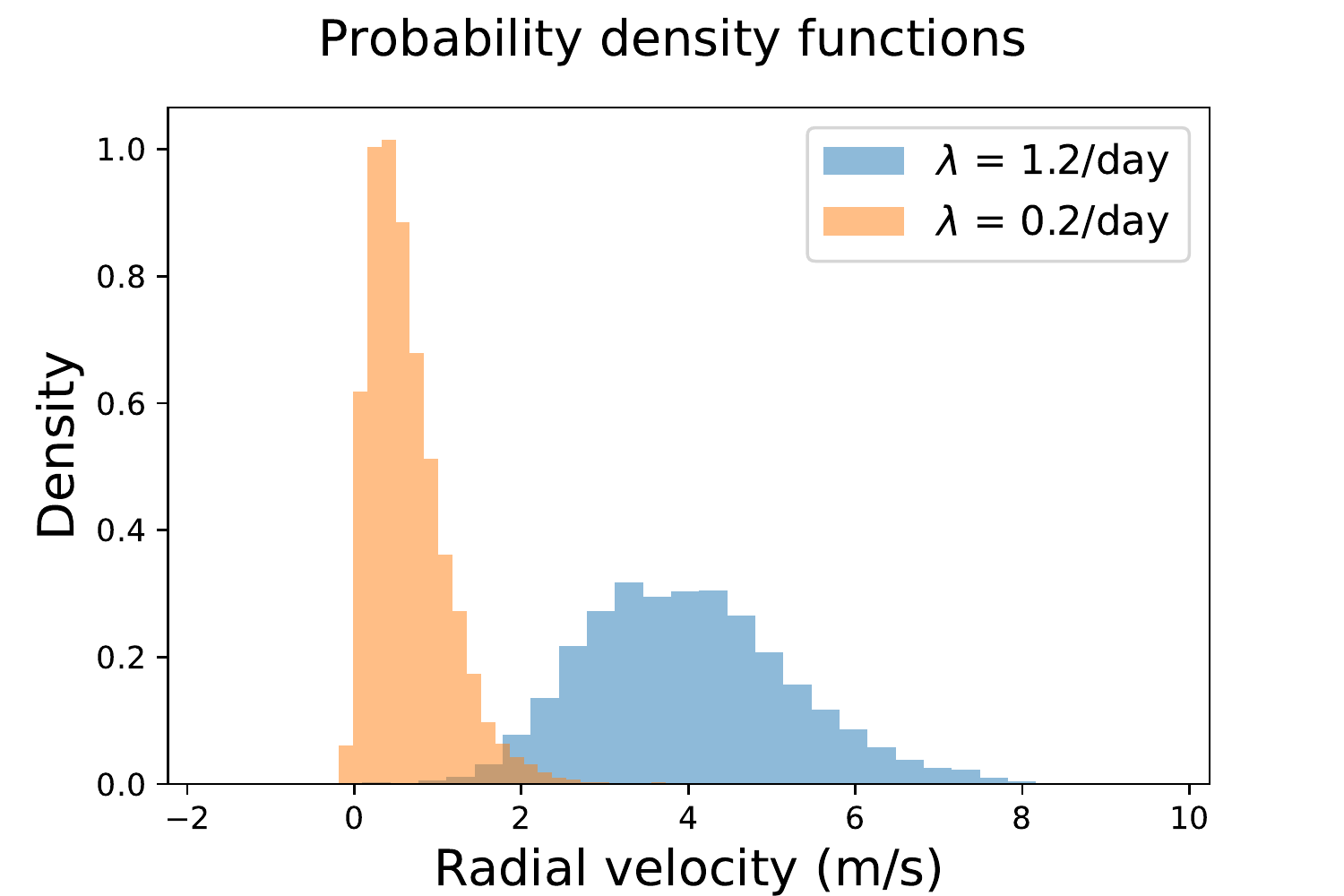}
    \caption{ Histogram of RVs simulated with a FENRIR process, with a rate of $\lambda$ = 1.2 and  $\lambda$ = 0.2 magnetic region appearing per day (respectively the blue and orange histograms).  
    }
    \label{fig:histFENRIR}
\end{figure}
\begin{figure}
    \includegraphics[width = 0.94\linewidth]{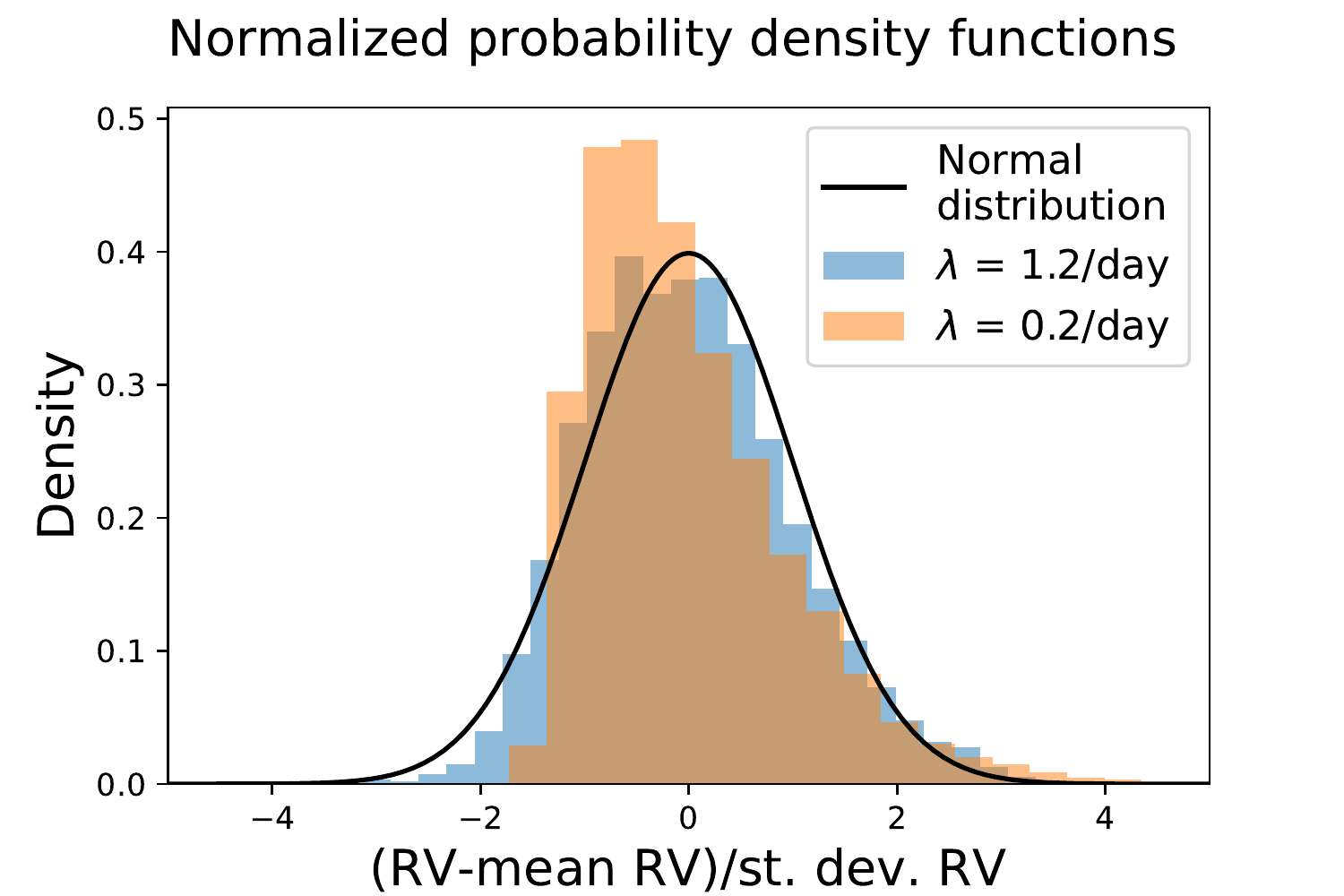}
    \caption{  Histogram of normalized RVs simulated with a FENRIR process, with a rate of $\lambda$ = 1.2 and  $\lambda$ = 0.2 magnetic region appearing per day (respectively the blue and orange histograms). RVs are normalized by subtracting the mean of the time-series and dividing by its empirical standard deviation. The black line represents a Gaussian distribution with vanishing mean and variance 1. 
    }
    \label{fig:histFENRIR_normalized}
\end{figure}
\begin{figure}
    \includegraphics[width = 0.94\linewidth]{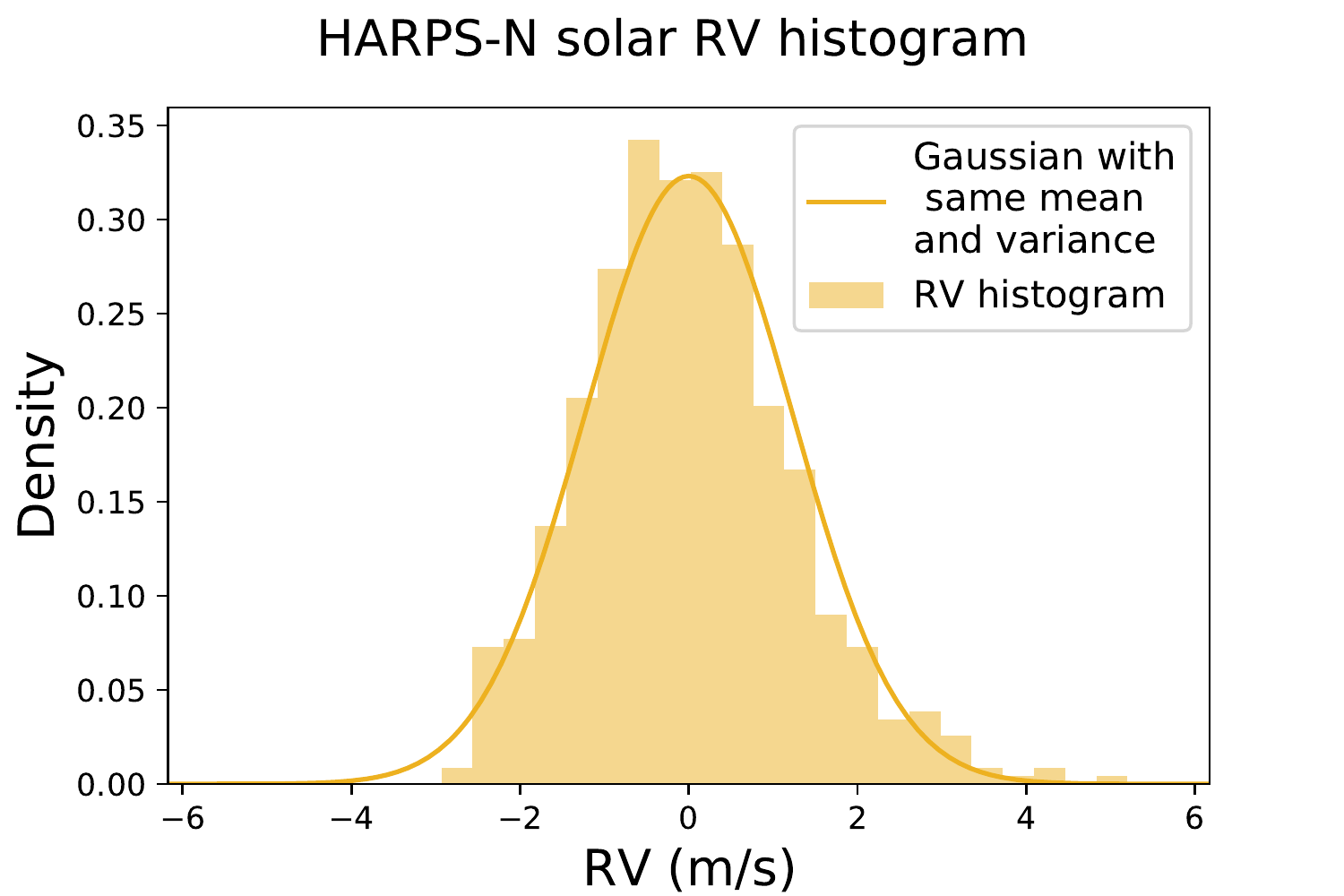}
    \caption{Yellow histogram: Histogram of RVs measured on the Sun \citep[public data][]{dumusque2020} by HARPS-N after binning by 15 hours, and fitting and subtracting a 9-th order polynomial. The yellow solid curve is a Gaussian ditribution with the same mean and variance as the data.  
    }
    \label{fig:histSun}
\end{figure}

As an illustration, we generate FENRIR processes with the same properties as in Section \ref{sec:rate} with a Poisson rate $\lambda$ of 1.2 magnetic regions appearing per day and 0.2 magnetic region per day, and a sampling of 1 point per day over 20 years. The histogram of the simulated RVs are shown in Fig. \ref{fig:histFENRIR}. As expected, the variance of the variance grows with $\lambda$. Furthermore, it seems that the RVs of the simulation with $\lambda=1.2$/day is closer to a Gaussian than the RVs obtained with a lower rate $\lambda=0.2$/day. To show it more clearly, we subtract the mean of each simulated RV time series, and normalize them by their means. In Fig. \ref{fig:histFENRIR_normalized}, we show the histogram of these normalized RVs. As expected, the distribution of $\lambda=1.2$/day RVs is closer to a normal distribution. Interestingly, both distribution seem to have a mode slightly shifted towards negative RV, but a heavier tail at positive RV and a sharp cutoff at negative RVs.

The FENRIR simulation is made to roughly reproduce characteristics of magnetic regions of the Sun.  
We consider the public three-year time series of RVs taken by HARPS \citep{dumusque2020}, bin them by 15 hours, subtract a 9th order polynomial to remove frequencies lower than $\approx 1$  year and further isolate the contribution of stellar rotation. 
In Fig. \ref{fig:histSun}, we show the histogram of the residual RVs, which show a similar behaviour as FENRIR processes: a sharp cut-off at negative RVs, heavier tails at positive RVs, and a maximum marginally shifted towards negative RVs. This behaviour does not depend on the degree of the polynomial fitted, nor the binning strategy. However, as one might expect, when the binning is made on shorter time intervals, high frequency noise has a tendency to make the distribution slightly more Gaussian. 

The $\lambda = 1.2$/day simulation has roughly the same standard deviation as the solar RVs (1.237 m/s and 1.234 m/s respectively). Interestingly, they have very similar empirical 3rd order moment: 1.02 $m^3/s^3$ and 0.91 $m^3/s^3$ for the FENRIR and solar RVs, respectively. If one generates a Gaussian white noise with the same standard deviation and number of observations as the Sun 15h-binned RVs, the standard deviation of the 3rd order moment is 0.18 $m^3/s^3$, so that the 3rd order moment of Sun RVs is 5 sigma significantly non zero. This must be tempered by the fact that solar RVs are time-correlated, which increases the dispersion of third order moments. A precise estimate of non Gaussianity is left for future work.

%The polyspectrum of order 3 is called the bispectrum, and should be identically equal to zero if the process is Gaussian. Polyspectra are used in particular to test the Gaussianity of the cosmic wave background \citep[e. g.][]{planck2020}.For stationary Gaussian processes, for all $n \geqslant 3$, Eq. \eqref{eq:polyspectrum_body} should be equal to zero for all choice of frequencies $\omega_1, ..., \omega_{n-1}$. Conversely, if a polyspectrum with order $n \geqslant 3$ is significantly different from zero, it means the data is not Gaussian. 

%. This one is 

\subsubsection{Bispectrum}

\begin{figure}
    \includegraphics[width = 1\linewidth]{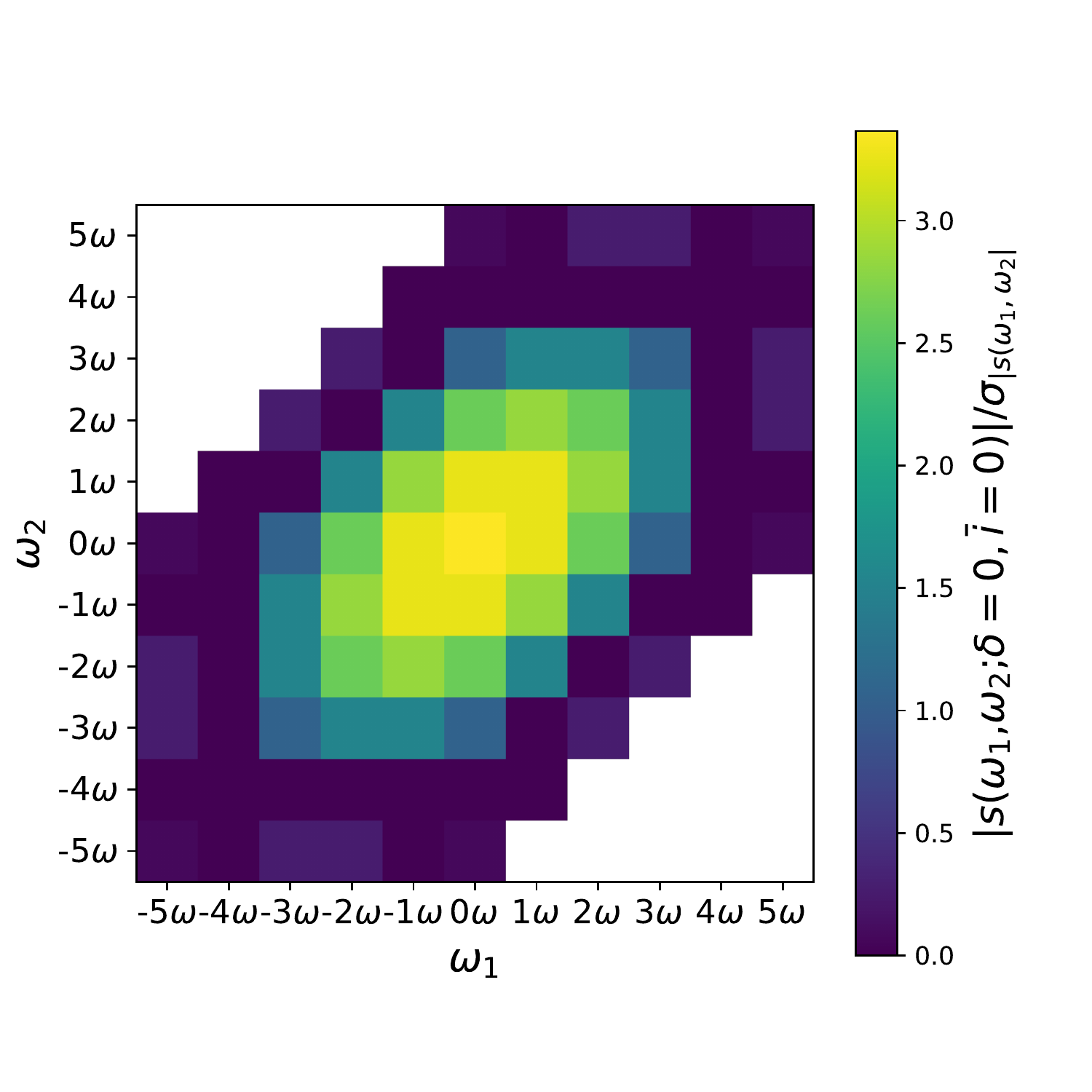}
    \caption{ Absolute value of the bispectrum divided by an upper bound on its mean squared error, for a purely periodic, equatorial spot signal with a contribution of the convective blueshift inhibition twice that of the RV photometric effect, an inclination $\bar{i}=0$.
    }
    \label{fig:bispec}
\end{figure}

The polyspectrum of order 2, called the bispectrum, is often used to test for non Gaussianity in time series. From \ref{eq:polyspectrum_body}, we obtain its expression for a FENRIR process:
\begin{align}
    S_3(\omega_1,\omega_{2};\eta) &= \lambda \int s(\omega_1, \omega_2, \gamma)  p(\gamma\mid \eta) \dd \gamma, \label{eq:bipectrum_body} \\
    s(\omega_1, \omega_2, \gamma) &= \hat{g}(\omega_1, \gamma) \hat{g}(\omega_{2}, \gamma) \hat{g}(-\omega_{1}-\omega_{2}, \gamma).
    \label{eq:integrand}
\end{align}
The bispectrum is a function of the Fourier transform of the impulse response $g$, which can be expressed as a product of three terms: the effect of a magnetic region when it is visible, an indicator function equal to one when the region is visible and 0 otherwise, and and what we called a window function, modulating the amplitude of the signal as the region grows and decays in size (see Section \ref{sec:windowfunction}). We first assume that there is no decay. The product of the first two terms is periodic, and can be expressed as a Fourier series truncated to arbitrary order
\begin{align}
    g(t) = \sum_{k=-k_{max}}^{k=k_{max}} a_k \e^{\ii \omega t} .
\end{align}
This means that the Fourier transform of $g$ is only non zero at frequencies $k\omega$, $k=-k_{max},...,k_{max}$, and the only 
non-zero products in the integrand of Eq. \eqref{eq:bipectrum_body} are such that $\omega_1 = k_1 \omega, k_2\omega_2$ such that $ k_1,  k_2$ and $-k_1-k_2$ are all integers between $-k_{max}$ and $k_{max}$. In practice, $\hat{g}$ is only estimated. From an uncertainty on its amplitude we can compute the mean squared error on $s(\omega_1, \omega_2, \gamma)$ defined in Eq. \eqref{eq:integrand}. In Fig. \ref{fig:bispec}, we show the ratio of $s$ and an upper bound on its mean squared error assuming an inclination $\bar{i}=0$, a latitude $\delta=0$ and ratio between the photometric RV and convective blueshift inhibition effects $B=1$.  In this case we assume that $\hat{g}(\omega)$ is known with a 15\% accuracy for all $\omega$, and the uncertainties on $\hat{g}(\omega)$, $\hat{g}(\omega')$ are independent. It appears that several bispectrum coefficients are more than three times greater than their mean squared error.

\begin{figure}
    \includegraphics[width = 1\linewidth]{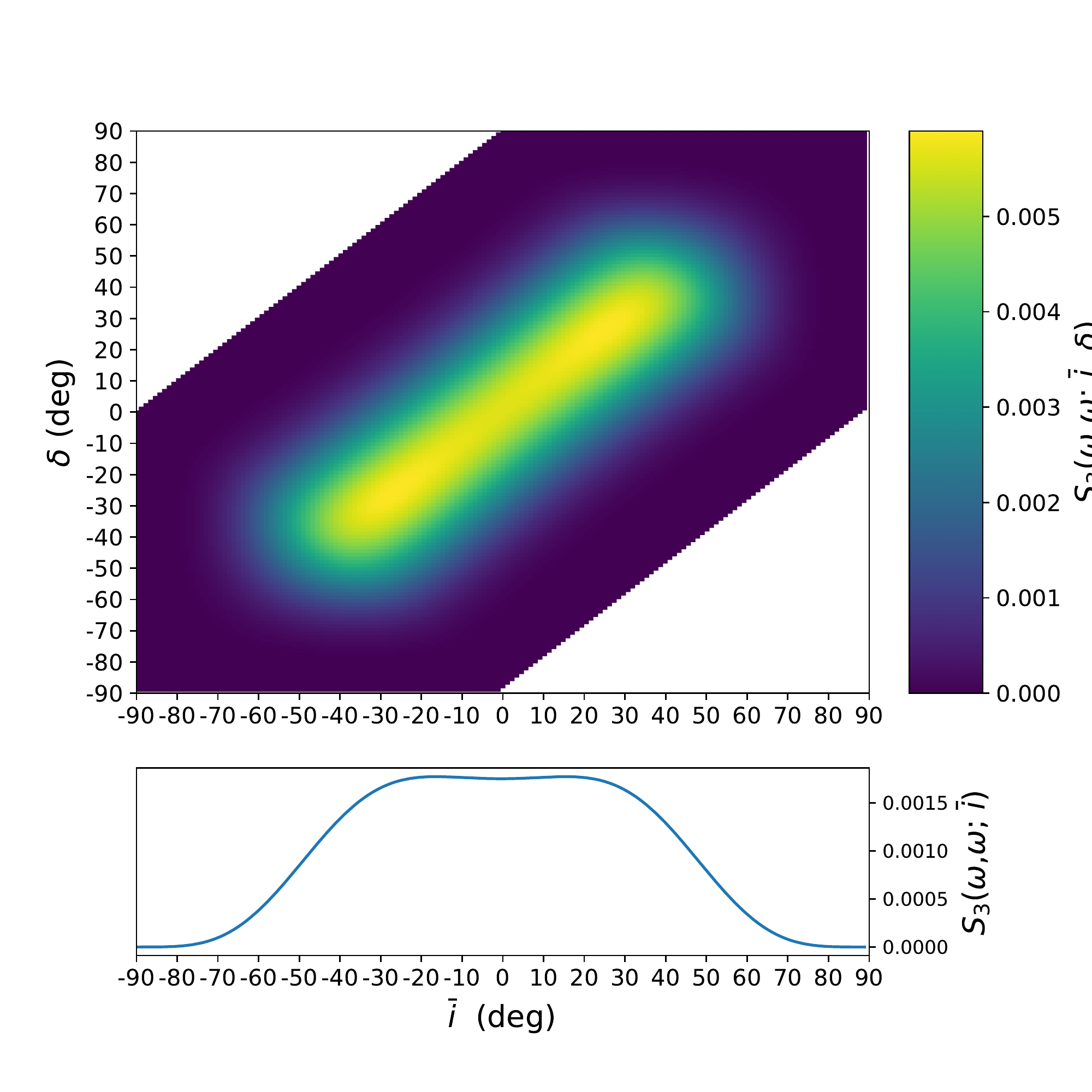}
    \caption{Top: real part of the $\omega, \omega$ coefficient of the bispectrum as a function of stellar inclination $\bar{i}$ and latitude of the spot $\delta$. The bottom plot shows the value of the bispectrum averaged along $\delta$.
    }
    \label{fig:bispecdeltai}
\end{figure}

The bispectrum shown in Fig. \ref{fig:bispecdeltai} concerns a single magnetic region with fixed latitude and for a given stellar inclination. In Fig. \ref{fig:bispecdeltai}, we show the real part of Eq. \eqref{eq:integrand} for $\omega_1=\omega_2 = \omega$ as a function of inclination $\bar{i}$ and latitude $\delta$. $\bar{i} = -90^\circ$ means that the one pole of the star is pointing to the observer, and $\bar{i} = 90^\circ$, the other pole, $\bar{i} = 0^\circ$ means the stellar rotation axis is perpendicular to the line of sight. For each inclination, only regions satisfying $-\tan \bar{i} \tan{\delta} \leqslant 1$ are visible (see Appendix \ref{app:spoteffect}). The bispectrum (Eq. \eqref{eq:bipectrum_body}) requires to integrate Eq. \eqref{eq:integrand} over the distribution of parameters of appearing magnetic regions, and in the bottom of  Fig. \ref{fig:bispecdeltai} we show the value of $s$ averaged over latitude, assuming that all latitudes are equally probable. It is apparent that the real part of the bispectrum does not average out, regardless of which exact distribution is chosen for $\delta$. 

In the case where the window function (growth and decay of the spot area) is taken as non constant, hanks to the convolution theorem, the Fourier transform of $g$ is
\begin{align}
    \hat{g}(\nu) = \sum_{k=-k_{max}}^{k=k_{max}} a_k \hat{W}(\nu-k\omega).
\end{align}
Taking into account the window function would give a smeared version of Fig. \ref{fig:bispec}.

%To evaluate if it is significantly non-zero, we can compare it to the mean sq

%This model can be expressed as a product of a periodic signal due to the rotation of the star $P(t)$, itself the prod and a window function traducing the effect of the growing and decaying size of the region $W(t)$. To determine whether our FENRIR model is close to Gaussian or not, we first recall that 

%several parameters: the inclination of the star, the latitude of the region, the limb-darkening law and ratio of photometric and inhibition of the convective blueshift effect, shape of the window function, etc. To determine whether our FENRIR model is close to Gaussian or not, we first recall that 

\subsection{Phase shifts might relate to the ratio of spots to faculae}

\label{sec:phaseshifts}. 

%\citet{aigrain2012} argue that the RV effect of stellar spots could be predicted based on measurements of their photometric effect. They model the flux of light as a function of time $t$ as $\Psi_0\{1-F(t)\}$, where $\Psi_0$ is a constant with dimensions of flux and $F(t)$ models the time variability of the photometric signal,  and model the  RV variation contamination as  $-\alpha \dot{F} F + \beta F(t)^2$ with $\alpha, \beta >0$ estimated from the light flux. They smooth the photometric time-series with a Gaussian process and estimate the RV as a linear combination of its square and derivative. 

Using the definition of $J$ in \eqref{eq:J_body}, assuming a constant limb darkening, as noted in \cite{aigrain2012}, the RV effect due to the photometric effect (Eq. \eqref{eq:yph_main}) is proportional to the  the derivative of the inhibition of convective blueshift effect (Eq. \eqref{eq:ycb_main}). 
We can write that RV impulse response $g$, as   $g(t+ \Delta t)  = g_{cb}(t) +  g_{cb}'(t) \Delta t $  and $g_{cb}'(t) = \alpha g_{ph}(t) $.  This is just another way to rewrite the $FF'$ approximation of \cite{aigrain2012}, and this is valid only if the limb-darkening effect is the same for the RV photometric effect, and the convective blueshift inhibition RV effect. This is incorrect, in particular because plages have a limb-brightening effect \citep{meunier2010b}. Nonetheless, we lay out the reasoning with the assumption that the limb darkening is constant as a starting point for a more realistic model. 

The opposite of the flux impulse response in Eq. \eqref{eq:z_main}, as apparent in Fig. \ref{fig:gs_example} is in phase with the inhibition of the convective blueshift. As long as the RV photometric effect is smaller than the inhibition of convective blueshift by a factor $3-4$,  we can make the approximation 
\begin{align}
\Delta t \approx -\frac{\Delta f R^\star}{2 ( \Delta f V_{cb} + \Delta V_{cb}(f + \Delta f))} .
%    \Delta t \approx -\frac{\Delta f R^\star}{2 \Delta V_{cb}} .
\end{align} 
The absolute value of the phase shift depends on several parameters. Its sign is more robustly defined by the ratio of spots to faculae through the sign of $\Delta f$. In any case, $\Delta V_{cb}$ is positive because redshifts are positive, and $\Delta f$ is small compared to $f$. If $\Delta f$ is negative, then $\Delta t$ is positive. 

%the flux difference of the magnetic region compared to the continuum.  

We expect that the $\log R'_{HK}$ effect in RV is approximately proportional to the projected area of the magnetic region. As a result, it behaves approximately proportionally to the flux. If the magnetic region is bright, then RV should be late compared to photometry and if the region is dark, RV is in advance compared to the $\log R'_{HK}$.  In \cite{hara2021b}, we showed that in the HARPS-N RV observations of the Sun \citep{dumusque2020}, RV is consistently in advance of 20 $\pm$ 6$^\circ$, which is consistent with the fact that the RV photometric effect is dominated by spots on the Sun. With spot and faculae models such as \cite[]{dumusque2014}, this could be linked in turn to whether the star is spot-dominated or facula dominated \citep[e.g.][]{nemec2022}. 

Again, we stress that the model of the present section relies on several assumptions which are incorrect, but the present argument serves as a starting point to link the phase shifts to whether the star is in a plage-dominated or spot-dominated regime. 

\subsection{Building relevant indicators}

\label{sec:closed_loop}

\begin{figure*}
    \includegraphics[width = 0.49\linewidth]{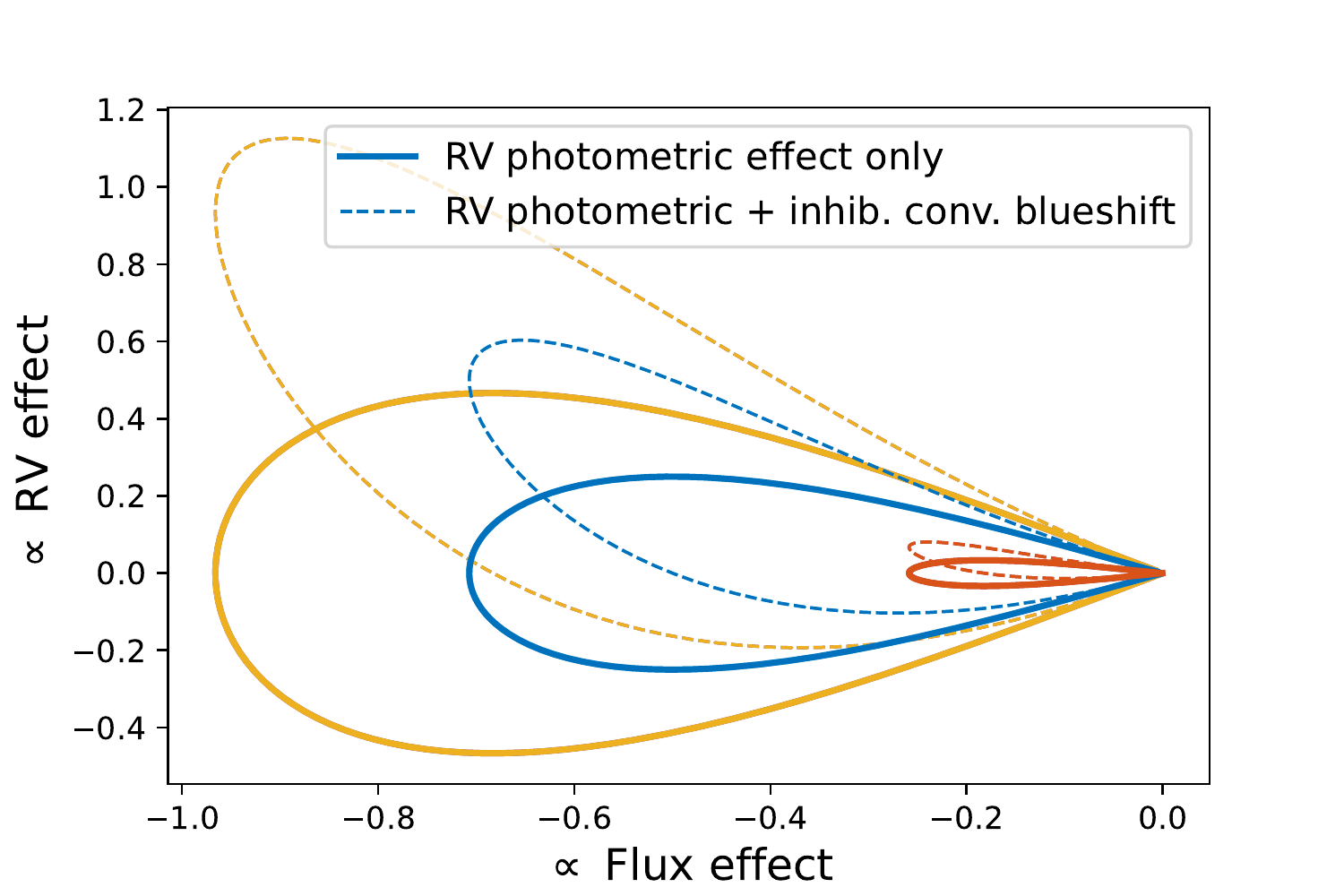}
    \includegraphics[width = 0.49\linewidth]{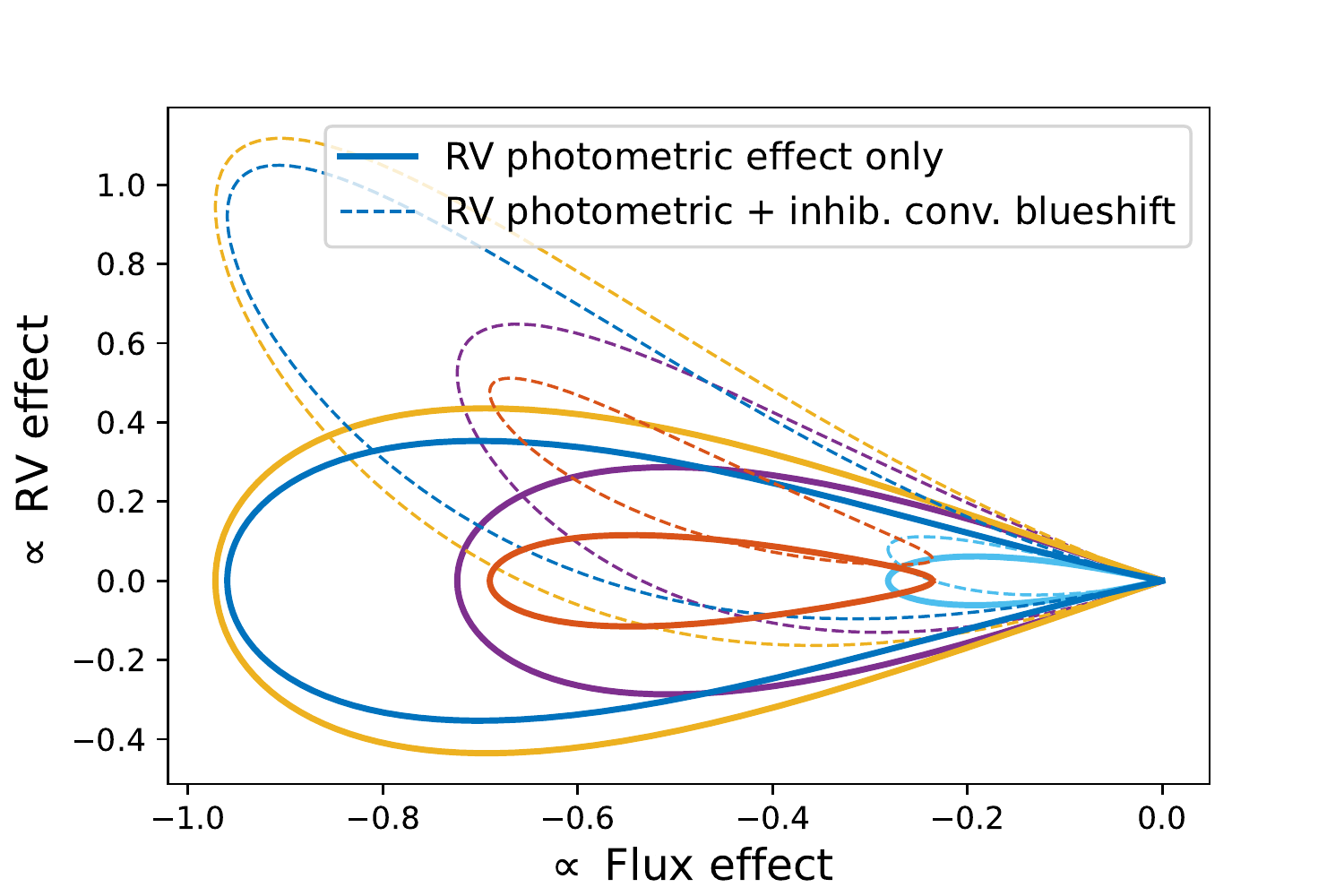}

    \caption{RV effect as a function of the flux effect. Plain lines correspond to a RV effect only due to the inhibition of the convective blueshift and dashed line to an RV signal with equal contribution from convective blueshift inhibition and photometric effects. Left: $i=0$ and right $i=0.5$ rad. 
    Colors are consistent with Fig. \ref{fig:gs_example} 
    }
    \label{fig:closedloop}
\end{figure*}

As mentioned in Sectin \ref{sec:usingFENRIR}, we could try to find a spectral variability indicator such that the impulse response of a stellar feature $h(t,\gamma)$ in this indicator is such that the effect of the feature on the signal of interest (photometry or radial velocity), $g(t, \gamma)$, is proportional to $h(t,\gamma)$ for all $\gamma$. However, in most cases indicators do not exhibit a linear relationship. 
It is known that RV and photometry, or RV and indicators might exhibit a so-called closed-loop relations \citep{bonfils2007,forveille2009,  santerne2015,lanza2018, colliercameron2019}. This means that when one is plotted against the other, the figure seems to close on itself. In Fig. \ref{fig:closedloop}, we show the behaviour of RV vs photometry when they are modelled as in \ref{sec:overall_modell}. 

%We suggest here an interpretation based on the following observation. If two observed channels $y(t)$ and $z(t)$ can be modelled with FENRIR processes of impulse responses $g(t, \gamma)$ and $h(t, \gamma)$,  if  $g(t, \gamma) \approx \alpha g(t +\Delta t, \gamma)$ for some $\alpha$ and $\Delta t$, then there should be a closed-loop relationship between $y(t)$ and $z(t)$. In that case, the phase-shift between photometry and RV as a function of the ratio of spots to faculae on the stellar surface. 

Another approach would be to build several indicators, such that the impulse response of stellar features in the signal of interest is a linear combination of the responses in the different channels. In \cite{haywood2022}, it is argued that the unsigned magnetic flux is a good proxy for the RV component due to the inhibition of the convective blueshift. Neglecting the limb-darkening effect, the radial magnetic field projected onto the line of sight is proportional to RV convective blueshift effect. Since this one dominates on the Sun, we expect it to be a good indicator. However, the limb-darkening law might differ for the velocity and the magnetic field might differ, so we do not expect the correlation to be exact.

\subsection{Link with Doppler imaging}

\label{sec:doppler_im}
%It has been found in \cite{gilbertson2020} that for Sun-like stars, it might be more appropriate to 

 %They describes the observational channels by expressing them as a linear transformation of the stellar surface brightness, itself considered as a random field. In our  we consider that each observational channel has a certain response to stellar features, with a 

Our work has a similar purpose as \cite{luger2021_mappingI,luger2021_mappingII,luger2021_mappingIII}, which is to build a statistical model of the observational challenge rooted in physical quantities. In particular \cite{luger2021_mappingIII} focuses on the Doppler imaging problem.

Doppler imaging consists in considering a time series of line spectral decompositions (LSD), similar to a time series of CCFs. The temporal variations of the LSD are mapped to temporal variations of temperature and magnetic field at the surface of the star. This method can be applied on the intensity spectrum \citep[Stokes I profile][]{deutsch1958, khokhlova1976, goncharskii1977, goncharskii1982, vogt1983, vogt1987} or the polarized light spectrum \citep[Stokes I profile][]{donati1997}. Given a certain profile of temperature, chemical composition expressed in spherical harmonics, and magnetic properties of the star, the shape of the LSD is forward-modeled and compared to the measurements. The inverse problem: finding the stellar surface by minimizing the squared difference between the measured LSD profile and the forward model is degenerate. The classical approach to Doppler introduces an entropy regularization term to the least square minimization \citep[e.g.][]{petit2015,yu2019}. 

As noted in \cite{luger2021_mappingIII}, the maximum entropy procedure gives a point estimates, and does not allow to measure precise uncertainties on the stellar surface profile. Their approach consists in building a Gaussian process representation of the stellar surface properties, translating to a Gaussian process representation of the forward modeled spectra, CCF or LSD time series and they compute the posterior distribution of the stellar surface properties. As such, it gives an instantaneous map of surface brightness.

However, our framework could be extended to perform more classical Doppler imaging.  \cite{luger2021_mappingIII} maps linearly the stellar surface brightness developed in spherical harmonics to the spectrum. In \cite{lehmann2022}, the authors the time series of LSD onto principal components, which can be seen as a data-driven way to retrieve the spherical harmonics.
By defining observation channels as the spectrum projected on a basis corresponding to the spherical harmonics, we could obtain an estimate of the evolution of their coefficient with time.

In its current form, our work is rather oriented towards two goals: to analyze data to detect exoplanets, and retrieving statistical properties of the stellar surface, not instantaneous ones. We do not model the full spectrum nor the full stellar surface, but we offer a flexible framework to test different hypotheses on the effect of magnetic regions, including inhibition of convective blueshift, and express our framework in the S+LEAF form to further improve the run speed. 
As shown in Fig. \ref{sec:sun}, from the statistical properties of the spot we can retrieve an estimate of the inclination between the stellar rotation axis and the sky plane.

%In an upcoming article, we will explain how to use practically our formalism to model RV, photometry and indicators to detect exoplanets, and to perform Doppler imaging. 

%Why Langellier+ results are crap.

%La variation des vitesses HARPS-N solaires est beaucoup plus petite que celles mesurées par Haywood+ 2016 sur Vesta (7-10 m/s), comment ça se fait ?

%\subsection{Linear models}
	
	%When using linear activity indicator models the uncertainty of the indicators must be propagated. $\alpha \vec q(t)$ follows a law of mean $\alpha \vec \hat{q}(t)$ and variance $\alpha^2 \mat Q$ where $\mat Q $ is the covariance of $q(t)$.

	%\subsection{Inclusion in Bayesian inference}
	
	%Let us suppose that the variations of $\lambda$ are parametrised by $\eta$. For instance, $\lambda$ can be chosen as a Gaussian process whose hyperparameters are $\eta$. We further assume that the distribution of $\gamma$ as a function of $t$ is only a function of $\lambda(t,\gamma$. We can envision a simple  model
	%\begin{align}
	%    RV(t) = a \lambda(t,\eta) + \epsilon + \mathrm{noise} + \mathrm{Keplerians}\\
	%    \log R'_{HK} = b \lambda(t,\eta) + \epsilon'
	%\end{align}
	%where $\epsilon$ is a Gaussian process whose kernel is given by $\ref{eq:nonstaationary}$. To furt
%		\begin{figure}
%	\includegraphics[width=9cm]{stellar_activity_1P_diam.pdf}
%	\caption{Effect of convective blueshift inhibition (blue) and photometric effect (orange) for two diametrically opposed spots.}
%	\label{fig:stellar_activity_diam}%
%	\end{figure}
	
	\section{Conclusion}
	\label{sec:conclusion}
	
	Our initial aim was to build a representation of stellar activity whose form is dictated by physical considerations. We suppose that we have several observation channels such as RVs, photometry, activity indicators, or even a time-series of spectra. The Finite ENergy Random Impulse Response (FENRIR) model we introduce represents them through three ingredients: First, the effect of a given stellar feature (stellar spot, plage or combination of the above, granulation cell), as a function of its parameters (size, latitude...), which we call the impulse . Second, the statistical distribution of the feature parameters knowing some hyperparameters $\eta$ such as the mean spot latitude and stellar inclination. The third ingredient is the rate at which features appear, which might vary over the magnetic cycle. 
	
	The FENRIR model gives a Gaussian process representation of the different channels with physical hyperparameters $\eta$. We express our formalism in the S+LEAF framework \citep{delisle2019b, delisle2022}, so that likelihood evaluation have a cost linear in the number of observations, so that our algorithm is applicable to a wide range of datasets. This includes observations of the Sun, essential to calibrate our models. Furthermore, the FENRIR framework allows to interpret the non Gaussianity present in the Solar data, and should in the long term, allow to go beyond the Gaussian process framework. 
	
	If we are able, from simulations and analysis of existing datasets, to constrain precisely these three ingredients based on the type of star, this will give a principled, efficient model of stellar activity with two main advantages: a greater ability to correct stellar signal and find smaller planets, and the possibility to perform ``statistical Doppler imaging'', that is retrieving the statistical properties of spots rather than their instantaneous values. We tested the ability of our model to retrieve the inclination of the Sun based on HARPS-N radial velocity observations and SORCE photometry, and obtain a constrained value. However, we note that the results are still dependent on the exact choices made in the parametric form of the effect of magnetic regions, and further work is needed to make our inclination of estimation robust. 
	
	In Section \ref{sec:discussion}, we suggested several avenues for reflection: statistical models of granulations based on the effect of single granules or super-granules, as mentioned above, exploiting non-Gaussianity in the signal, interpreting phase shifts between RV and photometry or $\log R'_{HK}$ as a signature of the ratio of spots and plages filling factors, how to build relevant activity indicators and discussed the extension of our work to Doppler imaging. All these aspects will require further work to yield their full potential. 
	
	%Hypotheses of the physical model. 	Need to isolate the effects that appear independently of each other (an ``atom'' of stellar activity). 
	
	%Hypotheses to have a spleaf representation: form of the impulse response window times cosines, random phase. 
	
	\begin{acknowledgements}
	The authors warmly thank Vincent Bourrier and Xavier Dumusque for their helpful suggestions. 
\end{acknowledgements}

\bibliographystyle{aa.bst}
\bibliography{biblio}

\appendix

\section{Covariance and cumulants}

\subsection{Covariance in the general nonstationary case}
\label{app:nonstationary}

In this appendix, we compute the expression giving the mean and autocovariance of process $y(t)$, and the covariance of $y(t)$ and $z(t)$. These expressions are particular case of the general formula for cumulants of order $n$ derived in section \ref{app:cumulants}. Since the manipulation of cumulants might be unfamiliar to our reader we here give a calculation with more familiar expressions.  

Let us consider that the spots appear independently on a time interval $[-L/2, L/2]$ with rate $\lambda(t)$. Then the times of spot appearance follow the distribution $p(t) = \lambda(t) / \int_{-L/2}^{L/2} \lambda(t) \dd t $. The signal is modelled as 
\begin{equation}
	y^L(t) = \sum_{k=1}^n g(t-t_k, \gamma_k)
\end{equation}
and we denote by $y(t)$ the limit in probability of $y^L(t)$ as $L$ tends to infinity ($y^L(t)$ is a random variable). Here $n$ is the number of spots such that $t_k \in [-L/2, L/2]$, and follows a Poisson distribution of parameter $I_\lambda^L := \int_{-L/2}^{L/2} \lambda(t) \dd t $. To compute the covariance of $y$, which by definition is 
\begin{align}
	\mathrm{Cov}(y(t_a), y(t_b)) = \mathbb{E}\{y(t_a) y(t_b)\}  - \mathbb{E}\{y(t_a)\}\mathbb{E}\{y(t_b)\}
\end{align}
we first compute the average value of $y$ at $t$, $\mathbb{E}\{y(t_a)\}$, then the product expectancy, $\mathbb{E}\{y(t_a) y(t_b)\}$. In both case, we do the calculation for $y^L$ and let $L$ tend to infinity.

The average value of the $y$ process is
\begin{align}
	\mathbb{E}\{y^L(t)\} = \e^{-I_\lambda^L}\sum\limits_{n=1}^{+\infty}  \frac{(I_\lambda^L)^n}{n!} \iint \sum\limits_{k=1}^n g(t-t_k, \gamma_k) \frac{\lambda(t_k)}{I_\lambda^L} p(\gamma_k) \dd (t) \dd (\gamma)  
\end{align}
%Provided the infinite sum and integral can be inverted
And we have 
\begin{align}
 \iint \sum\limits_{k=1}^n g(t-t_k, \gamma_k) \frac{\lambda(t_k)}{I_\lambda^L} p(\gamma_k) \dd (t) \dd (\gamma)  
  = \\ \frac{n}{I_\lambda^L} \iint  g(t-t_0, \gamma) p(\gamma \mid t_0)  \lambda(t_0)\dd t_0 \dd \gamma. 
\end{align}
So 
\begin{align}
	\mathbb{E}\{ y^L(t) \} &= \e^{-I_\lambda^L} \sum\limits_{n=1}^{+\infty}  \frac{(I_\lambda^L)^{n-1}}{(n-1)!} \iint_{-L/2}^{L/2} g(t-t_0, \gamma) p(\gamma \mid t_0) \lambda(t_0)\dd t_0  \dd \gamma \\
	& = \iint_{-L/2}^{L/2} g(t-t_0, \gamma_k) \lambda(t_0) p(\gamma \mid t_0) \dd t_0 \dd \gamma .
\end{align}
Provided this integral converges when $L$ tends to infinity, 
\begin{align}
	\mathbb{E}\{y(t)\}  = \iint g(t-t_0, \gamma) p(\gamma \mid t_0)  \lambda(t_0)\dd t_0 \dd \gamma. 
\end{align}

We now compute the expectancy of the product $y^L(t_a) y^L(t_b) $. Here also, provided the infinite sum and integral can be inverted, since for $k \neq j$, $t_k$ and $t_j$ are statistically independent, 
\tiny
\begin{align} 
	\mathbb{E}\{y^L(t_a) y^L(t_b) \} &= \e^{-I_\lambda^L} \sum\limits_{n=1}^{+\infty} \frac{(I_\lambda^L)^n}{n!} \times \\ \sum\limits_{k,j=1, k\neq j}^{n}  \iint_{-L/2}^{L/2} &  g(t_a  - t_k, \gamma_k)\frac{\lambda(t_k) }{I_\lambda^L} p(\gamma_k) \dd t_k \dd \gamma_k \iint_{-L/2}^{L/2} g(t_b  - t_j, \gamma_j)\frac{\lambda(t_j) }{I_\lambda^L}  p(\gamma_j) \dd t_j \dd \gamma_j  \label{eq:crossterm}\\  + \sum\limits_{k=1}^n \iint_{-L/2}^{L/2} & g(t_a  - t_k, \gamma_k)g(t_b  - t_k, \gamma_k)   \frac{\lambda(t_k)}{(I_\lambda^L)^2} p(\gamma_k) \dd t_k \dd \gamma_k \label{eq:diagterm} .
\end{align}
\normalsize
Distributing the product on the terms~\eqref{eq:diagterm} and~\eqref{eq:crossterm}, since there are $n(n-1)$ pairs of $k,j$ where $k \neq j$, $\mathbb{E}\{y^L(t_a) y^L(t_b) \}$ can be written as the sum of two terms $C^L$ and $D^L$, 
\tiny
\begin{align} 
	 C^L(t_a,t_b) &= \e^{-I_\lambda^L} \sum\limits_{n=2}^{+\infty} \frac{(I_\lambda^L)^n}{n!} \frac{n(n-1)}{(I_\lambda^L)^2}  \\ & \times \iint_{-L/2}^{L/2}  g(t_a  - t_0, \gamma) \lambda(t_0)  p(\gamma) \dd t_0 \dd \gamma \iint_{-L/2}^{L/2} g(t_b  - t_0) \lambda(t_0)  p(\gamma) \dd t_0 \dd \gamma \\ 
	% & = \iint_{-L/2}^{L/2}  g(t_a  - t_0, \gamma) \lambda(t_0)  p(\gamma|t_0) \dd t_0 \dd \gamma \iint_{-L/2}^{L/2} g(t_b  - t_0, \gamma) \lambda(t_0)  p(\gamma) \dd t_0 \dd \gamma 
\end{align}
\normalsize
and
\begin{align} 
	 D^L(t_a,t_b) =  \iint_{-L/2}^{L/2}  g(t_a  - t_0) g(t_b  - t_0) \lambda(t_0)  p(\gamma|t_0) \dd t_0 \dd \gamma.
\end{align}
Let us remark that as $L$ tends to infinity, provided the integral over $t_k$ converges, $C^L(t_a,t_b) $ is the product of $\mathbb{E}\{y(t_a)\}$ and $\mathbb{E}\{y(t_b)\}$. If $D^L(t_a,t_b)$ admits a limit as $L$ tends to infinity, the covariance of $y$ is 
 \begin{align}
	\mathrm{Cov}(y(t_a), y(t_b)) = \iint  g(t_a  - t_0, \gamma) g(t_b  - t_0, \gamma) \lambda(t_0)  p(\gamma|t_0) \dd t_0 \dd \gamma
\end{align}
In this formula, the distribution of $\gamma$ can depend on $t_0$. 
 
 Note that if we replace $y(t_b)$ by $z(t_b)$ where $z$ models the effect of stellar activity on another ancillary indicator,
 \begin{align}
    z(t) = \sum\limits_{k=-\infty}^{+\infty}  h(t-t_k, \gamma(t_k))
\end{align}
the reasoning is unchanged, and provided all the limits when $L$ tends to infinity converge, 
 \begin{align}
	\mathrm{Cov}(y(t_a), z(t_b)) = \iint  g(t_a  - t, \gamma) h(t_b  - t, \gamma) \lambda(t)  p(\gamma|t) \dd t \dd \gamma \dd \gamma
	\label{eq:crosscovar}
\end{align}

\subsection{Cumulants and polyspectra}
\label{app:cumulants}

\subsubsection{Purpose}
In this section, we compute the analytical formula of the cumulants and polyspectra of the FENRIR process $y(t)$, defined as
\begin{align}
    y(t) & = \sum\limits_{k=-\infty}^{+\infty}  g(t-t_k, \gamma(t_k))  . 
    \label{eq:ytf_app} 
\end{align}
We have seen that for a stationary FENRIR process, the autocovariance is of the form $k(\tau) = \int_{-\infty}^{+\infty}g(t,\gamma)g(t+\tau,\gamma) p(\gamma) \dd t$. Even if the $\gamma$ can take only one value, the knowledge of $k(\tau)$ does not give unequivocally $g$: several functions can have the same autocorrelation. If $y(t)$ is Gaussian, based on it the best we can do is to estimate $k(\tau)$. If $y(t)$ is non Gaussian we can characterize $g$ more precisely, thus have not only less biases but also leverage more information. The purpose of cumulants is to reveal non Gaussian behaviours.

The cumulant generating function of $y(t_1), y(t_1) ..., y(t_n)$ is, regardless of $t$, 
\begin{align}
   K^n_{t_1,t_2,...,t_n}(\alpha_1,...,\alpha_n) = \ln \mathbb{E} \{ \e^{\alpha_1 y(t_1) +...+ \alpha_n y(t_n) }\}
\label{eq:cumulant_genfunction}
\end{align}
In the Taylor expansion of $K$ as a function of $\alpha_i$s, the cumulant of order $n$, $\kappa_n$, is the coefficient of their products, $\alpha_1\alpha_2...\alpha_n$. Considered as a function of $t_1,t_2,...,t_n$ it is sometimes called the $n$ point correlation function.  For a Gaussian process, cumulants of order 3 and 4 are zero. Our main result is that
\begin{theorem}
\begin{align}
\begin{split}
   & \kappa_n(y(t_1),y(t_2),...,y(t_n)) = \\ & \iint g(t-t_1, \gamma) g(t-t_2, \gamma)   ... g(t-t_n, \gamma) \lambda(t) p(\gamma,t) \dd t \dd \gamma. 
    \label{eq:kappan_app}
    \end{split}
\end{align}
\end{theorem}
The covariance is the 2-point correlation function.  When Gaussian processes they are stationary, we have $\kappa_n(t_1,t_2) =\kappa_n(|t_1-t_2|)$ they are equivalently represented by the Fourier transform of the kernel $\kappa(\tau)$: their power spectrum density (PSD). Similarly, under the hypothesis of stationarity of the process, the quantity  $\kappa_n(t,t + \tau_1,...,t + \tau_{n-1})$ does not depend on $t$. Considering the $n$ point correlation function as a function of the $n-1$ variables $\tau_{1},...,\tau_{n-1}$, we can define the polyspectra as the $n-1$ dimensional Fourier transform of $\kappa_n(t,t + \tau_1,...,t + \tau_{n-1})$. 

The proof of Eq. \eqref{eq:kappan_app} can be modified straightforwardly to prove a more general result, 
\begin{align}
\begin{split}
   & \kappa_n(x_1(x_1),x_2(t_2),...,x_n(t_n))  = \\ & \iint f_1(t-t_1, \gamma) f_2(t-t_2, \gamma)   ... f_n(t-t_n, \gamma) \lambda(t) p(\gamma,t) \dd t \dd \gamma. 
    \label{eq:kappan_app}
    \end{split}
\end{align}
where $x_i(t) = y(t)$ or  $x_i(t) = z(t) $. If $x_i(t) = y(t)$, then $f_i(t) = g(t=$ and if $x_i(t) = z(t)$, then $f_i(t) = h(t)$. This expression generalises eq \eqref{eq:crosscovar}. This expression has the advantage to be valid if the process is non stationary, and can explore non Gaussian dependencies across channel. However, it is unlikely that all these aspects could be explored simultaneously in a practical way.   

In the case where patterns appear with a constant Poisson rate, and the distribution followed by $\gamma(t_k)$ does not depend on time, $\gamma(t_k) \sim p(\gamma)$, the stochastic process $y(t)$ is stationary. 
%, which means that the joint  distribution of $y(t_1), y(t_2), ..., y(t_n)$ is equal to $y(t), y(t +\tau_1) ..., y(t_n+\tau_{n-1})$ for any $t, t_2...,t_n$, as long as $\tau_i = t_{i+1}-t_{i}$. 
We can write Eq. \eqref{eq:kappan_app} as a function of time differences $\tau_i = t_i - t_1$. With the change of variable $t \leftarrow t - t_1$,
\small
\begin{align}
    \kappa_n(\tau_1, ..., \tau_{n-1};\eta) = \lambda \iint g(t, \gamma) g(t+\tau_1, \gamma)  ... g(t+\tau_{n-1}, \gamma) p(\gamma \mid \eta) \dd t \dd \gamma,
    \label{eq:kappan_stationary_app}
\end{align}
\normalsize
In that case, the Fourier transform of $\kappa_n$ as a function of $\tau_1, ..., \tau_{n-1}$ is called the polyspectrum of order $n$. Here, provided the integrals on $\gamma$ and $\tau_1, ..., \tau_{n-1}$ can be inverted, it is equal to 
\small
\begin{align}
    S_n(\omega_1, ..., \omega_{n-1};\eta) = \lambda \int \hat{g}(\omega_1, \gamma)  ... \hat{g}(\omega_{n-1}, \gamma) \hat{g}\left(-\sum_{i=1}^{n-1} \omega_{i}, \gamma  \right) p(\gamma \mid \eta) \dd \gamma.
    \label{eq:polyspectrum_app}
\end{align}
\normalsize
To understand the purpose of polyspectra, let us go back to the Gaussian process case. The power spectral density has also another interpretation than the Fourier transform of the kernel: it is the expectancy of the squared modulus of the Fourier transform of the process, stationary Gaussian processes thus have completely random phase. Suppose we compute the phase of a stationary Gaussian process between times $t_1$ and $t_2$ on the one hand and between times $t_3$ and $t_4$ on the other hand, such that $t_3-t_2$ is much greater than the process correlation, the phase obtained on the first and second time interval are statistically independent. For stationary Gaussian processes, for all $n \geqslant 3$, Eq. \eqref{eq:polyspectrum_app} should be equal to zero for all choice of frequencies $\omega_1, ..., \omega_{n-1}$. As a consequence, non zero polyspectra traduce non Gaussianity.  Let us consider a quantity $\nu$ and all combinations of $n-1$ frequencies such that $\sum_{i=1}^{n-1} \omega_{i} = \nu$. In Eq. \eqref{eq:polyspectrum_app}, the phase of the Fourier transform evaluated at $\nu$ can be seen as a point of reference. If the frequencies $\omega_1, ... \omega_{n-1}$ and $\nu$ are not independent, the poly spectrum is non zero.

%can be seen as taking as a point of reference the phase at frequency $ -\sum_{i=1}^{n-1} \omega_{i}$,

%Now considering the random process $\cos(\omega t + \phi)$ where $\phi$ is random, it is clear that  

%in the case where patterns appear with a constant Poisson rate, and the distribution followed by $\gamma(t_k)$ does not depend on time, $\gamma(t_k) \sim p(\gamma)$. In this case, the $y(t)$ is stationary, which means that the joint  distribution of $y(t_1), y(t_2), ..., y(t_n)$ is equal to $y(t), y(t +\tau_1) ..., y(t_n+\tau_{n-1})$ for any $t$, $t_1,...,t_n$, as long as $\tau_i = t_{i+1}-t_{i}$.

%The PSD has also another interpretation: it is the expectancy of the squared modulus of the Fourier transform of the process, stationary Gaussian processes thus have completely random phase. Suppose we compute the phase of a stationary Gaussian process between times $t_1$ and $t_2$ on the one hand and between times $t_3$ and $t_4$ on the other hand, such that $t_3-t_2$ is much greater than the process correlation, the phase obtained on the first and second time interval are statistically independent. 

\subsubsection{Proof}

To establish Eq. \eqref{eq:kappan_app}, we will use several properties of cumulants, explicited below. 
\begin{theorem}
Let us consider $n$ random variables $Y_1,Y_2,...,Y_n$ (defined on the same $\sigma$ algebra).
 (i) Cumulants are multilinear, for all $i$, for two real numbers $a$ and $b$, $\kappa(Y_1, ..., aY_i+ bY_i', ..., Y_n) = a\kappa(Y_1, ..., Y_i, ..., X_n) + b\kappa(Y_1, ..., Y_i', ..., Y_n)$.
 
 (ii) If there is at least two indices $i\neq j$ such that $Y_i$ and  $Y_j$ are independent, $\kappa(Y_1, ..., Y_i, ..., Y_j,...,Y_n)=0$.
 
 (iii) If $N$ is a random variable following a Poisson distribution of parameter $\Lambda$, $\kappa(N,N...,N) = \Lambda$ regardless of the number of times $N$ is repeated. 
 
 (iv) Law of total cumulance. Suppose that we have a random variable $Y$ and we can define the conditional distribution of $Y_i$s knowing $X$. Then 
 \begin{align}
    \kappa(Y_1,...,Y_n) = \sum\limits_{\pi \in \mathbb{P}(n)} \kappa(\kappa(Y_i: i\in B \mid X) : B \in \pi)
    \label{eq:law_total_cumulance}
 \end{align}
where $\pi \in \mathbb{P}(n)$ means that $\pi$ runs through all the possible partitions of indices of indices $\{1,...,n\}$. $B \in \pi$ means that $B$ runs through the blocks of permutation $\pi$ and $\kappa(Y_i: i\in B \mid X)$ is the cumulant of $Y_i: i\in B $ knowing $X$, which is a random variable as a function of the random variable $X$. For instance, suppose we want to compute $\kappa(Y_1,Y_2,Y_3)$. The partitions of our indices $\{1,2,3\}$ are $\{\{1\},\{2\},\{3\}\},\{\{1\},\{2,3\}\},\{\{2\},\{1,3\}\},  \{\{3\},\{1,2\}\}, \{\{1,2,3\}\} $. In the partition $\{\{1\},\{2\},\{3\}\}$ we have three blocks, $\{1\},\{2\}$ and $\{3\}$. In the partition $\{\{1\},\{2,3\}\}$ we have two blocks, $\{1\}$ and $\{2,3\}$. For $n=3$, the law of total cumulance then writes
 \begin{align*}
   \kappa(Y_1,Y_2,Y_3)  &= \kappa( \kappa(Y_1,Y_2,Y_3 \mid X) )  \\ & +   \kappa(\kappa(Y_1,Y_2 \mid X), \kappa(Y_3 \mid X) )  \\ &+  \kappa(\kappa(Y_1,Y_3 \mid X), \kappa(Y_2 \mid X) ) \\ &  +  \kappa(\kappa(Y_2,Y_3 \mid X), \kappa(Y_1 \mid X) )  \\
  & +\kappa(\kappa(Y_1\mid X), \kappa(Y_2 \mid X), \kappa(Y_3 \mid X))
 \end{align*}
 where the quantities $\kappa(Y_i: i\in B \mid X) $ are random variables because they are functions of the random variable $X$.
 (v) With the same notations, the link between cumulants and non centred moments is the following
 \begin{align}
     \mathbb{E}\{Y_1Y_2...Y_3\} =  \sum\limits_{\pi \in \mathbb{P}(n)} \prod\limits_{ B \in \pi} \kappa(Y_i: i\in B \mid Y) . 
         \label{eq:moments_and_cumulants}
 \end{align}
 For instance
  \begin{align*}
   \mathbb{E}\{Y_1 Y_2 Y_3\}  &=  \kappa(Y_1,Y_2,Y_3 )   \\ & +   \kappa(Y_1,Y_2)\kappa(Y_3 ) \\ &+  \kappa(Y_1,Y_3 ) \kappa(Y_2 ) \\ &  + \kappa(Y_2,Y_3 ), \kappa(Y_1 )  \\
  & + \kappa(Y_1) \kappa(Y_2 ) \kappa(Y_3)
 \end{align*}
 \label{thm:cumulant_results}
 \end{theorem}
 Formulae \eqref{eq:law_total_cumulance} and \eqref{eq:moments_and_cumulants} are similar. In the first case one takes the cumulants over the different blocks of the distribution, in the second one simply takes the product. This will be useful in the proof of Eq. \eqref{eq:kappan_app}.

%To prove formula \eqref{eq:kappan_app}, we first consider \eqref{eq:ytf_app} restricted to a certain time interval. 
 Let us consider $n$ times ordered increasingly $T_1,...,T_n$. Let us suppose that $g$ has a finite support, meaning that for some $T$, for $x \in [-T/2, T/2]$, $g(x,\gamma) = 0$. Because $g$ has a finite support, computing the $n$ order cumulant of $y(T_1),...,y(T_n)$ for $y$ as defined in Eq. \eqref{eq:ytf_app} is equivalent to compute the cumulant of the variables $Y_i$
 \begin{align}
    Y_i & = \sum\limits_{k=1}^{N}  g(T_i-t_k, \gamma(t_k))  . 
    \label{eq:ytf_app} 
\end{align}
where features appear at $t_k$, following a Poisson process between $t_1 - T/2 $ and $t_n + T/2$ with a variable rate $\lambda(t)$. $N$ is the number of features in this interval and follows a Poisson distribution with parameter $\Lambda = \int_{t_1 - T/2}^{t_n + T/2} \lambda(t)$. A feature appearing at time $t_k$ has parameters drawn from the distribution $p(\gamma,t)$. 

Thanks to the multilinearity of cumulants (property $(i)$ in Theorem \ref{thm:cumulant_results}), we have 
\begin{align}
    \kappa(Y_1,...,Y_n \mid N) =  \sum\limits_{i_1=1}^N... \sum\limits_{i_n=1}^N \kappa(G(T_1 - t_{i_1}), ... ,G(T_n - t_{i_n} \mid N)
\end{align}
Measurable functions of independent variables are independent. Since $t_i$ and $t_j$ are independent for $i\neq j$, $G(T_1 - t_{i})$ and $G(T_1 - t_{j})$ are not independent if and only if $i=j$.
Now, because cumulants are zero if two or more variables are independent (property $(ii)$ in Theorem \ref{thm:cumulant_results}), 
\begin{align}
    \kappa(G(T_1 - t_{i_1}), ... ,G(T_n - t_{i_n}) \mid N) \neq 0
\end{align}
if and only if $i_1 = i_2 = ... i_n$, and 
\begin{align}
    \kappa(Y_1,...,Y_n \mid N) =  \sum\limits_{i=1}^N \kappa(G(T_1 - t_{i}), ... ,G(T_n - t_{i}) \mid N ) 
\end{align}
At each $i$, the value $G(T_1 - t_{i}), ... ,G(T_n - t_{i})$ is drawn from the same distribution. As a consequence, $ \kappa(G(T_1 - t_{i}), ... ,G(T_n - t_{i}) \mid N )  =  \kappa(G(T_1 - t_{j}), ... ,G(T_n - t_{j}) \mid N ) $ for all $i, j= 1..N$ and
\begin{align}
    \kappa(Y_1,...,Y_n \mid N) =  N \kappa(G(T_1 - t_{i}), ... ,G(T_n - t_{i})
    \label{eq:Ncumulant}
\end{align}

Thanks to the law of total cumulance (property $(iv)$ in Theorem \ref{thm:cumulant_results}), 
\begin{align}
    \kappa(Y_1,...,Y_n ) =  \sum\limits_{\pi \in \mathbb{P}(n)} \kappa(\kappa(Y_i: i\in B \mid N) : B \in \pi)
\end{align}
Injecting Eq. \eqref{eq:Ncumulant}, we have 
\begin{align}
    \kappa(Y_1,...,Y_n ) =  \sum\limits_{\pi \in \mathbb{P}(n)} \kappa(N\kappa(G_i: i\in B ) : B \in \pi)
\end{align}
Thanks to the multilinearity of cumulants, we have 
 \begin{align}
 \kappa(N\kappa(G_i: i\in B ) : B \in \pi) =   \prod\limits_{i\in B} \kappa(G_i: i\in B )  \kappa_{|B|}(N)
\end{align}
where $\kappa_{|B|}{N} = \kappa(N,...,N)$ where $N$ is repeated $|B|$ times, the number of block. For a Poisson distribution of parameter $\Lambda$, $\kappa(N,...,N) = \Lambda$ (property $(iii)$ in Theorem \ref{thm:cumulant_results}). We now have 
\begin{align}
    \kappa(Y_1,...,Y_n ) = \Lambda \sum\limits_{\pi \in \mathbb{P}(n)}  \prod\limits_{B \in \pi} \kappa(Y_i: i\in B \mid N) 
\end{align}
and thanks to the link between moments and cumulants (property $(v)$ in Theorem \ref{thm:cumulant_results}) we can write 
\begin{align}
\kappa(Y_1,...,Y_n ) = \Lambda \mathbb{E}\{G(T_1-t)...G(T_n-t)\}
% \sum\limits_{\pi \in \mathbb{P}(n)}  \prod\limits_{B \in \pi)} \kappa(Y_i: i\in B \mid N)  = \mathbb{E}\{G(T_1-t)...G(T_n-t)\}
\end{align}
Finally, by definition of  $\mathbb{E}\{G(T_1-t)...G(T_n-t)\}$,
\begin{align}
&\mathbb{E}\{G(T_1-t)...G(T_n-t)\} = \\ & \frac{1}{\Lambda} \int\limits_{t_1 - T/2}^{t_n + T/2} g(T_1 - t, \gamma(t) )... g(T_n - t, \gamma(t) ) \lambda(t) p( \gamma \mid t)   \dd t \dd \gamma \label{eq:almost_done}
\end{align}
simplifying by $\Lambda$, and recognizing that because of the finite support of $g$, the integral \eqref{eq:almost_done} is unchanged if we take as lower and upper bounds $-\infty$ and $+\infty$, and is true for any $T$, we can thus extend our results to functions with infinite support and  we have the desired result, 
\begin{align}
    \kappa(Y_1,...,Y_n ) =  \int\limits_{-\infty}^{\infty} g(T_1 - t, \gamma(t) )... g(T_n - t, \gamma(t) ) \lambda(t) p( \gamma \mid t)   \dd t \dd \gamma .
\end{align}

 %The rationale behind questioning the Gaussianity of stellar activity signals is two-fold. Estimating stellar parameters as hyperparameters of a Gaussian process can be done, but if the process is in fact non Gaussian the estimates might be skewed. Second, 

%For a stationary process, cumulants are written as a function of time  differences with $t_1$. We write $t_i = t_1 + \tau_i $. and the cumulant of order $n$ as $\kappa_n(\tau_1, ..., \tau_{n-1})$. 

	\section{Analytical approximation of the spot/facula RV effect}
	\label{app:spoteffect}

\subsection{Weighted RV}
 
In this appendix we approximate the RV and photometric effect as follows. We assume that the RV effect on the data of the quiet stellar surface is 
\begin{align}
    RV_{quiet}(t) = \frac{1}{F_0} \iint F(\vec x) \vec V(\vec x) \dd \vec x \cdot %\vec e_{obs}
\end{align}
where $\vec x$ is the position on the visible disk, $\vec V(\vec x)$ is the local velocity,  $F(\vec x)$ the local flux, $F_0$ the flux integrated on all the stellar surface and $ \vec e_{obs}$ the unit vector pointing from the observer to the star. Where 
\begin{align}
    \vec V(\vec x) = V_{cb} \vec e_r + \vec \omega \wedge \vec r
\end{align}
where  $V_{cb}$ is the velocity modulus of the convective blueshift effect, $\vec e_r$ is pointing radially outwards,  $\vec \omega $ is the rotation axis times the rotational velocity and $\vec r$ is the vector joining the stellar center and the point at position $\vec x$ on the stellar surface. 

 Denoting by $RV_{mag}(t)$ the RV when there a magnetic region is present, and by $V'(\vec x)$ and $F'(\vec x)$ the velocity and flux fields ,
\begin{align}
    RV_{mag}(t) = \frac{1}{F_0'} \iint F'(\vec x) \vec V'(\vec x) \dd \vec x \cdot \vec e_{obs}
\end{align}
where
\begin{align}
    \vec V'(\vec x) &= (V_{cb} +\Delta V_{cb})  \vec e_r + \vec \omega \wedge \vec r \\
    F'(\vec x)  &= F(\vec x) + \Delta F(\vec x).
\end{align}
We are interested in the difference between $RV_{mag}(t)$ and $RV_{quiet}(t)$. Developing $F'_0$ at first order in $\Delta F$, we have 
\begin{align}
  &\left(\frac{f'(\vec x)  \vec V'(\vec x)}{F_0'} -  \frac{f(\vec x)\vec V(\vec x)}{F_0} \right) \cdot \vec e_{obs}  = y_{cb} + y_{ph} \\  & y_{cb}  =  \left(\Delta F V_{cb} \left( 1 - \frac{F}{F_0} \right) + \Delta V_{cb}(F + \Delta F)  \right)\vec e_r  \cdot \vec e_{obs} \label{eq:rvcb}\\&   y_{ph} =  \Delta F \vec \omega \wedge \vec r \cdot \vec e_{obs} 
\end{align}
The term $y_{cb}$ and $y_{ph}$ are respectively called the inhibition of convective blueshift term RV photometric terms. 

Assuming that the magnetic region is small, the flux can be written as product of the flux per surface unit $f(\vec x)$ times the projected area of the magnetic region, $A(\vec x)  P(\vec x)$ where $A(\vec x)$ is the intrinsic area and $P(\vec x)$ the projection effect, times the limb-darkening effect $l(\vec x)$. We write
\begin{align}
    F(\vec x) = f(\vec x) A(\vec x)  P(\vec x) l(\vec x)  \\
    \Delta F(\vec x) = \Delta f(\vec x) A(\vec x)  P(\vec x) l(\vec x).
\end{align}
In the following, we establish the expression of the flux and RV as a function of the position of the magnetic region (longitude $\phi$ and latitude $\delta$) on an inclined star. For small spots, we can neglect the $F/F_0$ term in Eq. \eqref{eq:rvcb}.

We model the surface of the star by a sphere and consider that a spot or facula is an infenetesimal area of that sphere of unit radius. For the sake of simplicity, in the following we simply refer to a spot, but the reasoning for a facula is identical.
We consider a direct frame $x,y,z$ such that $x$ points in the direction of the observer and $y,z$ defines the sky plane. 

Let us first assume that the rotation axis of the star is aligned with $z$ denote by $\phi, \delta$ the spherical coordinate of the center of a spot, such that its position in the Cartesian frame is
\begin{align}
    x &= \cos\delta \cos\phi \\
    y &= \cos\delta \sin\phi \\
    z &= \sin\delta   
\end{align}
The local frame $(u,v,w)$ at  $(x,y,z)$ is such that 
\begin{align}
    u &= \cos\delta \cos\phi x + \cos\delta\sin\phi y + \sin\delta z \\
    v &= -\sin \phi x + \cos\phi y \\
    w &= -\sin\delta \cos\phi x  - \sin \delta \sin\phi y + \cos \delta z   
\end{align}
The effect of a spot on the RV depends on the inclination of the star with respect to the plane of the sky $i$. To compute the position of the spot and the projection of the local frame centered at the spot in the reference frame, we apply a rotation of axis $y$ and angle $i \in [0,\pi/2]$. We now have 
\begin{align}
    x &= \cos i\cos\delta \cos\phi + \sin i \sin \delta  \label{eq:x}\\
    y &= \cos\delta \sin\phi \\
    z &= -\sin i \cos\delta \cos\phi + \cos i\sin\delta   
\end{align}
and
\begin{align}
    u = & (\cos i \cos \delta \cos\phi + \sin i \sin \delta) x  \\
     & +\cos \delta \sin \phi y + (-\sin i \cos \delta \cos \phi + \cos i \sin \delta) z \\ 
    v = &-\cos i \sin \phi x + \cos \phi y+ \sin i \sin \phi y \label{eq:v}\\
    w =& -\cos i \sin \delta \cos \phi x  -\sin\delta\sin \phi y \\  & +  (\sin i\sin \delta \cos\phi + \cos i \cos \delta ) z   \label{eq:w}
\end{align}

 We assume that for an infinitesimal portion of the stellar surface, the effect of the spot is proportional to its projected area onto the sky plane times a velocity projected onto the $x$ axis (more precisely -$x$ axis, since the velocity is assumed to be positive in the direction observer - star) multiplied by a limb-darkening effect. In the case of the photometric effect, the velocity in question is the the velocity of the stellar surface due to the stellar rotation in the rest frame, it is therefore the $v$ component of the velocity. The convective blueshift inhibition effect is due to the motion of the gas from the center of the star to the stellar surface. We are therefore interested in the $u$ component of the velocity projected onto $x$.  The spot position on the stellar surface changes its projected surface onto the sky plane. The correction factor is equal to the Jacobian of the projection from the $(v,w)$ onto the $(y,z)$ plane, from Eq.~\eqref{eq:v} and Eq.~\eqref{eq:w}, the projected area of the spot is proportional to
 \begin{align}
 J(i,\delta, \phi)  := & \cos \phi (\sin i\sin \delta \cos \phi + \cos i\cos \delta) \\ &- (-\sin\delta \sin \phi)\sin i \sin \phi \\
    J(i,\delta, \phi) = &   \sin i \sin \delta + \cos i \cos \delta \cos \phi \label{eq:J}
 \end{align} 
 The spot is visible if its position is such that $x\geqslant 0$. From Eq.~\eqref{eq:x}, this translates to the condition
 \begin{align}
     \cos \phi \geqslant -\tan i \tan \delta \label{eq:seespot}
 \end{align}
 Overall, the velocity contribution of the photometric and inhibition of convective blueshift effects on RV ($y_{ph}$ and  $y_{cb}$, and its effect on photometry $z_{ph}$ is 0 when \ref{eq:seespot} is not satisfied, and when it is:
\begin{align}
    y_{ph}(i,\delta,\phi) =& A l(J(i,\delta,\phi)) \Delta f \omega R^\star  J(i,\delta, \phi) \cos i \sin \phi  \cos \delta \label{eq:yph}\\
    y_{cb}(i,\delta,\phi)  =&  A l(J(i,\delta,\phi))  (\Delta f V_{cb} + \Delta V_{cb}(f + \Delta f)  ) J(i,\delta, \phi)^2 \label{eq:ycb} \\
    z_{ph} (i,\delta,\phi) =  & A \Delta f l(J(i,\delta,\phi)) J(i,\delta,\phi)
    %&-\cos i\cos (\delta-i)\sin \phi \cos \phi \\
    % \propto  & -\cos i\cos (\delta-i)\sin 2\phi \\
   % y_{cb} \propto & -\cos i \cos (\delta-i) \sin \delta \cos^2 \phi\\
   % \propto  & -\cos i \cos (\delta-i) \sin \delta (1+\cos2\phi
\end{align}
 $\Delta f$ is the flux difference between the stellar surface and the spot and $f^\star$ is the mean stellar flux, $\omega$ is the local rotational velocity of the star and  $R^\star$ is the stellar radius, $\Delta V_{CB}$ is the difference between the velocity on the gas in the $u$ direction without and with convective blueshift inhibition. The + sign comes from the fact that we are projecting $u$ and $v$ onto $-x$, and the inhibition of convective blueshift has a positive velocity in the $-u$ direction. The photometric effect can be positive or negative depending on whether the area under consideration is brighter of darker than the continuum of the star.  
 The symbol $l$ represents the limb-darkening, function of the distance to the center of the star normalised by the stellar radius $\mu$. In our parametrisation, $\mu = (\phi,\delta,\phi)$. Indeed, the spherical trigonometry law of cosines yields $\mu = \cos \delta_0 \cos \phi_0$ where $\phi_0$ and $\delta_0$ are the spherical coordinate in the frame where the $x$ axis points to the observer, and applying the rotation about the $y$ axis of $i$ yields $\mu = \cos \delta_0 \cos \phi_0 = J$.
 For an equatorial spot and $i=0$, we simply have 
 \begin{align}
    y_{ph}=  \cos \phi \sin \phi  = \frac{1}{2} \sin 2 \phi \\
    y_{cb}= \cos^2 \phi = \frac{1}{2} (1+\cos 2 \phi)
\end{align}

 In the case of a non constant limb-darkening law $l(J)$, the expressions Eq.~\ref{eq:yph} and Eq.~\ref{eq:ycb} is particularly simple If $l$ is of the form
 \begin{align}
l(J) = \sum\limits_{k=0}^d a_k J^k
%\label{eq:ldlaw}
 \end{align}
 In the general case the photometric and convective blueshift inhibition effect are summed. For the sake of simplicity we write their contribution up to a multiplicative factor and denote the relative contribution of the photometric effect compared to the convective blueshift inhibition as $B$. The combined effect on the RV is then
\begin{align}
    g(t) = \sum\limits_{k=0}^d a_k J^{k+1}(J(i,\delta,\phi) + \beta \cos i  \sin \phi )
    \label{eq:g_limbdark}
    %\sum\limits_{k=0}^d a_k (J^{k+2} + J^{k+1} \beta \cos i  \sin \phi )
\end{align}
for $\phi$ satisfying \eqref{eq:seespot} and 0 otherwise.

When the limb darkening law is taken as a constant, the general formula for the radial velocity effect  of a spot is
%\begin{align}
%g(\phi) \propto & \sin^2 i \sin^2 \delta + \frac{\sin 2 i \sin 2\delta }{2} \cos \phi + \cos^2 i \cos^2 \delta \cos^2 \phi \\
%& +B \left( \frac{\sin 2 i \sin \delta }{2} \sin \phi + \frac{\cos^2 i  \cos \delta }{2} \sin 2\phi \right)
% \end{align} 
% which can  also be written 
 \begin{align}
g(\phi) \propto & \sin^2 i \sin^2 \delta + \frac{\cos^2 i \cos^2 \delta }{2}   \\& + \frac{\sin 2 i \sin 2\delta }{2} \cos \phi + \frac{\cos^2 i \cos^2 \delta}{2}  \cos 2 \phi \\
& -B \frac{\sin 2 i \sin 2\delta }{4} \sin \phi - B\frac{\cos^2 i  \cos^2 \delta }{2} \sin 2\phi  \label{eq:ggeneral}
 \end{align} 

\iffalse
  \begin{align}
g(\phi) \propto & \sin^2 i \sin^2 \delta + \frac{\cos^2 i \cos^2 \delta }{2}   \\& + \frac{\sin 2 i \sin 2\delta }{2} \cos \phi + \frac{\cos^2 i \cos^2 \delta}{2}  \cos 2 \phi \\
& -B \frac{\sin 2 i \sin \delta }{2} \sin \phi - B\frac{\cos^2 i  \cos \delta }{2} \sin 2\phi  \label{eq:ggeneral}
 \end{align} 
 \fi

  For a constant rotation rate, constant $\delta$, we have $\phi = \omega t$.  
  We can rewrite $g$ as 
 \begin{align}
g(t, i,\delta, B) = c_0 + c_1 \cos \omega t + c_2 \cos 2\omega t + s_1 \sin \omega t + s_2 \sin 2\omega t
\label{eq:gcos}
 \end{align}
 where $c_0,c_1,c_2,s_1,s_2$ are coefficients that depend on $i$, $\delta$ and $\beta$.   As a concluding remark, let us note that because the limb-Darkening is in power of $J$, itself an affine function of $\cos \omega t$. Regardless of the order of the Limb-Darkening law chosen, we can always  write $g$ in the form 
 \begin{align}
g(t, i,\delta, B) = c_0 + \sum_{k=1}^d c_k \cos(k\omega t) + s_k \sin(k\omega t) 
\label{eq:gcos}
 \end{align}

  \subsection{CCF-related channels}
  \label{app:ccf}
  
  In the previous section, we assume that the measured RV is the  sum of the local stellar RVs weighted by their relative flux. However, this is a simplistic assumption. In the present section we derive the expressions of RV, as well as ancillary indicators considering that the measured cross correlation function (CCF) is a weighted sum of the local stellar CCFs. 
  
  When extracting radial-velocity measurements through the CCF technique, one models
  the CCF with a Gaussian function
  \begin{equation}
   m_\mathrm{CCF}(v; \eta) = a \left(1 - c \exp\left(-\frac{1}{2}\left(\frac{v-v_0}{\sigma}\right)^2\right)\right)
  \end{equation}
  where $a$ is the continuum flux,
  $c$ is the contrast,
  $v_0$ is the radial velocity,
  $2\sqrt{2\ln(2)}\sigma$ is the FWHM.
  The parameters $\eta = (a,c,v_0,\sigma)$ are typically adjusted using a least-square estimator.
  We denote by $\eta_0$ the parameters obtained by fitting a CCF which is not affected by the activity contribution.
  We now consider the contribution $\delta\mathrm{CCF}$ of a small feature (spot/faculae) on the CCF.
  The impact of this feature on the parameters $\eta$ can be estimated by linearizing
  the Gaussian model in the vicinity of $\eta_0$
  \begin{equation}
      m_\mathrm{CCF}(v; \eta_0+\delta\eta) = m_\mathrm{CCF}(v; \eta_0) + \nabla_\eta m_\mathrm{CCF}(v; \eta_0) \delta\eta,
  \end{equation}
  with
  \begin{equation}
    \label{eq:gradCCF}
    \nabla_\eta m_\mathrm{CCF}(v; \eta_0) =
    {\scriptsize \begin{pmatrix}
      1 - c G(v)
      & - a G(v)
      & - \frac{a c(v-v_0)}{\sigma^2}  G(v)
      & - \frac{a c(v-v_0)^2}{\sigma^3}  G(v)
    \end{pmatrix}}
  \end{equation}
  and
  $G(v) = \exp\left(-\frac{1}{2}\left(\frac{v-v_0}{\sigma}\right)^2\right)$.
  The parameters are then obtained by finding the value of $\delta \eta$
  minimizing
  \begin{equation}
      \int_{v_\mathrm{min}}^{v_\mathrm{max}} \left(\delta\mathrm{CCF}(v)-\nabla m . \delta\eta\right)^2\mathrm{d}v,
  \end{equation}
  where the CCF is computed on a velocity interval $[v_\mathrm{min},v_\mathrm{max}]$.
  This yields the parameters estimate
  \begin{equation}
      \delta\eta = \alpha^{-1}\beta,
  \end{equation}
  with $\alpha$ the $4\times 4$ matrix, and $\beta$ the vector of size 4 given by
  \begin{align}
    \alpha &= \int_{v_\mathrm{min}}^{v_\mathrm{max}} \nabla m(v)^T \nabla m(v) \mathrm{d}v,\nonumber\\
    \beta &= \int_{v_\mathrm{min}}^{v_\mathrm{max}} \nabla m(v)^T \delta\mathrm{CCF}(v) \mathrm{d}v.
  \end{align}
  For on-ground spectroscopy, 
  the continuum flux $a$ is actually strongly affected by the Earth's atmosphere
  and other observational circumstances,
  which makes this parameter useless for stellar characterization.
  Moreover, the correlation between the continuum flux $a$,
  and the other parameters can be shown to be negligible,
  such that we can actually ignore this parameter and
  approximate the variations of the three other parameters
  $\eta_{1:} = (c, v_0, \sigma)$ with
  \begin{equation}
    \delta\eta_{1:} = \alpha_{1:,1:}^{-1}\beta_{1:}.
  \end{equation}
  Indeed, for a sufficiently large interval $[v_\mathrm{min}, v_\mathrm{max}]$,
  the continuum flux $a$ is dominated by the information located on the tails of the CCF,
  while all other parameters use the information contained in the center of the CCF
  (see Eq.~(\ref{eq:gradCCF})).
  We can verify this by increasing the length of the interval $[v_\mathrm{min}, v_\mathrm{max}]$
  toward $]-\infty,+\infty[$.
  As we do so, all the components of the matrix $\alpha$ converge, except for $\alpha_{0,0}$
  (which corresponds to the continuum flux $a$),
  which is asymptotically proportional to the interval length ($\Delta v=v_\mathrm{max}-v_\mathrm{min}$).
  We can then deduce the asymptotic behavior of the covariance matrix of the parameters $\alpha^{-1}$,
  as well as the correlation matrix.
  We find that the correlation between $a$ and the other parameters is asymptotically
  proportional to $1/\sqrt{\Delta v}$, and thus vanishes for a sufficiently large interval.
  
  In the limit of $[v_\mathrm{min}, v_\mathrm{max}] \longrightarrow ]-\infty, \infty[$,
  we obtain
  \begin{equation}
      \alpha_{1:,1:} = a^2 \sqrt{2\pi}\sigma \begin{pmatrix}
          1 & 0 & \frac{c}{2\sigma}\\
          0 & \frac{c^2}{2\sigma^2} & 0\\
          \frac{c}{2\sigma} & 0 & \frac{3c^2}{4\sigma^2}\\
      \end{pmatrix},
  \end{equation}
  therefore we have
  \begin{equation}
      \alpha_{1:,1:}^{-1} = \frac{2\sigma}{a^2 c^2 \sqrt{2\pi}} \begin{pmatrix}
          \frac{3c^2}{4\sigma^2} & 0 & -\frac{c}{2\sigma}\\
          0 & 1 & 0\\
          -\frac{c}{2\sigma} & 0 & 1\\
      \end{pmatrix},
  \end{equation}   
  and finally
  \begin{equation}
    \label{eq:deltacvsb}
      \begin{pmatrix}
          \frac{\delta c}{c}\\
          \frac{\delta v}{\sigma}\\
          \frac{\delta \sigma}{\sigma}
      \end{pmatrix}
      = - \frac{1}{2ac}\begin{pmatrix}
          3 & 0 & -2\\
          0 & 4 & 0\\
          -2 & 0 & 4
      \end{pmatrix} b,
  \end{equation}
  with $b$ given by ($k=0,1,2$)
  \begin{equation}
    \label{eq:CCFbk}
      b_k = \frac{1}{\sigma\sqrt{2\pi}}\int_{-\infty}^{+\infty} \left(\frac{v-v_0}{\sigma}\right)^k G(v) \delta\mathrm{CCF}(v) \mathrm{d}v.
  \end{equation}

  We now need to specify the effect $\delta\mathrm{CCF}$ of a spot on the CCF to be able to compute $b$.
  We assume that in the absence of the spot,
  the contribution to the CCF of the stellar surface at the position of the
  spot would have been
  \begin{equation}
      \mathrm{CCF}_l(v) = a_l \left(1 - c_l \exp\left(-\frac{1}{2}\left(\frac{v-v_l}{\sigma_l}\right)^2\right)\right).
  \end{equation}
  Because of the presence of the spot, the local CCF is altered.
  As a first approximation, we assume that it remains approximately Gaussian but with altered parameters $a_s, c_s, v_s, \sigma_s$:
  \begin{equation}
    \mathrm{CCF}_s(v) = a_s \left(1 - c_s \exp\left(-\frac{1}{2}\left(\frac{v-v_s}{\sigma_s}\right)^2\right)\right),
  \end{equation}
  where the flux $a_s < a_l$ for spots and $a_s > a_l$ for faculae.
  Due to the inhibition of the convective blue-shift, which results in a net red-shift,
  we expect $v_s > v_l$ for both spots and faculae.
  Moreover, the Zeeman broadening effect in both kinds of active regions tends to decrease the contrast $c_s < c_l$,
  and to increase the FWHM ($\sigma_s > \sigma_l$).
  Finally, the effect of the spot on the CCF can be modeled as
  \begin{equation}
      \delta\mathrm{CCF}(v) = \mathrm{CCF}_s(v) - \mathrm{CCF}_l(v).
  \end{equation}
  Replacing this expression in Eq.~(\ref{eq:CCFbk}), we find
  \begin{align}
      b &= (a_s-a_l)
      \begin{pmatrix}
          1\\
          0\\
          1
      \end{pmatrix}\\
      & + a_l \frac{c_l \sigma_l}{\sqrt{\sigma^2+\sigma_l^2}}
      \exp\left(-\frac{1}{2}\frac{\delta v_l^2}{\sigma^2+\sigma_l^2}\right)
      \begin{pmatrix}
        1\\
        \frac{\sigma}{\sigma^2+\sigma_l^2} \delta v_l\\
        \frac{\sigma_l^2}{\sigma^2+\sigma_l^2}
        + \frac{\sigma^2}{\left(\sigma^2+\sigma_l^2\right)^2} \delta v_l^2
      \end{pmatrix} \nonumber\\
      & - a_s \frac{c_s \sigma_s}{\sqrt{\sigma^2+\sigma_s^2}}
      \exp\left(-\frac{1}{2}\frac{\delta v_s^2}{\sigma^2+\sigma_s^2}\right)
      \begin{pmatrix}
        1\\
        \frac{\sigma}{\sigma^2+\sigma_s^2} \delta v_s\\
        \frac{\sigma_s^2}{\sigma^2+\sigma_s^2}
        + \frac{\sigma^2}{\left(\sigma^2+\sigma_s^2\right)^2} \delta v_s^2
      \end{pmatrix},\nonumber
  \end{align}
  with $\delta v_l = v_l - v_0$, $\delta v_s = v_s - v_0$.
  At leading order in $\delta v$, we find
  \begin{align}
    b_0 &\approx a_s \left(1 - \frac{c_s \sigma_s}{\sqrt{\sigma^2+\sigma_s^2}}\right) 
    - a_l \left(1 - \frac{c_l \sigma_l}{\sqrt{\sigma^2+\sigma_l^2}}\right),\nonumber\\ 
    b_1 &\approx \frac{a_l c_l \sigma_l \sigma}{\left(\sigma^2+\sigma_l^2\right)^{3/2}} \delta v_l
    - \frac{a_s c_s \sigma_s \sigma}{\left(\sigma^2+\sigma_s^2\right)^{3/2}} \delta v_s,\nonumber\\ 
    b_2 &\approx a_s \left(1 - \frac{c_s \sigma_s^3}{\left(\sigma^2+\sigma_s^2\right)^{3/2}}\right) 
    - a_l \left(1 - \frac{c_l \sigma_l^3}{\left(\sigma^2+\sigma_l^2\right)^{3/2}}\right),
  \end{align}
  and thus (see Eq.~(\ref{eq:deltacvsb}))
  \begin{align}
      2 a \times \delta c &\approx a_l
      \left(1 - c_l \frac{\sigma_l \left(3\sigma^2+\sigma_l^2\right)}{\left(\sigma^2+\sigma_l^2\right)^{3/2}}\right)
      - a_s
      \left(1 - c_s \frac{\sigma_s \left(3\sigma^2+\sigma_s^2\right)}{\left(\sigma^2+\sigma_s^2\right)^{3/2}}\right)
      ,\nonumber\\
      \frac{a c}{2 \sigma^2} \times \delta v &\approx 
      \frac{a_s c_s \sigma_s}{\left(\sigma^2+\sigma_s^2\right)^{3/2}} \delta v_s
      - \frac{a_l c_l \sigma_l}{\left(\sigma^2+\sigma_l^2\right)^{3/2}} \delta v_l,\\      
      \frac{ac}{\sigma} \times \delta \sigma &\approx 
      a_l \left(1 - c_l \frac{\sigma_l \left(\sigma_l^2-\sigma^2\right)}{\left(\sigma^2+\sigma_l^2\right)^{3/2}}\right)
      - a_s \left(1 - c_s \frac{\sigma_s \left(\sigma_s^2-\sigma^2\right)}{\left(\sigma^2+\sigma_s^2\right)^{3/2}}\right).\nonumber      
  \end{align}
  We thus find that, in addition to the velocity variations
  induced by the photometric effect (proportional to $a_s-a_l$)
  and the convective blue-shift inhibition effect ($v_s-v_l$),
  changes in the equivalent width $c_s\sigma_s - c_l\sigma_l$
  might also affect the velocity.
  Moreover, we observe that at leading order the contrast variations $\delta c$
  and the FWHM variations $\delta \sigma$
  are mainly proportional (with negative and positive signs respectively)
  to the flux variations due to the spot ($a_s-a_l$),
  with some corrections coming from the shape variations ($c_s-c_l$, $\sigma_s-\sigma_l$).
  These indicators might thus be modeled in first approximation with the photometric component of our model if we neglect the correction terms for the shape variations.
  
\section{A computationally efficient GP framework}

\label{app:gaussian_process}

\subsection{Outline}

In the present Appendix, we outline the methodology to pass from assumptions on the impulse responses in the different channels, statistical distribution and rate --- $g(t,\gamma),h(t, \gamma)$, $p(\gamma \mid t,\eta)$ and $\lambda(t)$ in Eq. \eqref{eq:crosscova_body}, respectively --- to an efficient evaluation of the likelihood (Eq. \eqref{eq:gauss_likelihood}). 

To evaluate the covariance matrix, we need to compute the integral \eqref{eq:crosscova_body}. Given hypothesis 1, we can write the impulse response as a product of the window function and a periodic part expanded in Fourier series. This done by, multiplying $g_0$ in Eq. \eqref{eq:feature_g0} by the limb darkening law expressed in Eq. \eqref{eq:ldlaw}, and approximating the visibility function $\mathbf{1}_{\mathrm{vis}}(t)$ with a Fourier expansion. Thanks to hypothesis 2 above, we can write the covariance as a product of the covariance of the window function $W(t)$ and the covariance of the periodic part.  Eq. \eqref{eq:crosscova_body} has integrals over the time $t$ and the impulse response parameters $\gamma$. The integral over time is easy to perform either on the window function or or the periodic part. After this integration on $t$, the autocorrelation of the periodic part is a truncated Fourier series such that only the coefficients depend on $\gamma$. We compute numerically the integral over these coefficients and  interpolate these coefficients and integrate over $\gamma$

    \subsection{Separation of the window and the periodic components}
    \label{sec:sepwinper}
    We consider here the stationary case, where the rate of spot appearance $\lambda$ is constant,
    and the spots properties $\gamma$ do not depend on time $p(\gamma|\eta, t) = p(\gamma|\eta)$.
    The covariance between two time series $y(t)=\sum_k I(t-t_k, \gamma_k)$ and $z(t)=\sum_k J(t-t_k, \gamma_k)$,
    is
    \begin{equation}
    	k^{y,z}(\tau; \eta) = \lambda \iint I(t, \gamma) J(t+\tau, \gamma)\d t p(\gamma | \eta)\d \gamma,
    \end{equation}
    where $\eta$ is the set of hyper-parameters describing the spots population properties,
    $I(t,\gamma) = W(t,\gamma) i(t, \gamma)$ and $J(t,\gamma) = W(t,\gamma) j(t, \gamma)$.
    Up to now, we did not specify how the time $t_k$ of appearance of the spots are defined.
    In the following we define it as the time of maximum "activity" of the spot.
    This means that $W$ reaches its maximum at $t-t_k = 0$.
    We assume that the longitude $\phi_0$ of the spots at $t=t_k$
    (which is one of the parameters $\gamma$)
    is uniformly distributed in $[0,2\pi]$ and independent of $t_k$ and of the other spots properties.
    We thus take this parameter out of $\gamma$, and rewrite $I, J$ as
    $I(t,\phi_0,\gamma) = W(t,\gamma) i(\phi(t), \gamma)$ and $J(t,\phi_0,\gamma) = W(t,\gamma) j(\phi(t), \gamma)$,
    with $\phi(t) = \phi_0 + \omega t$,
    such that
    \begin{equation}
    	k^{y,z}(\tau; \eta) = \lambda \iint I(t, \phi_0, \gamma) J(t+\tau, \phi_0, \gamma)\d t \frac{\d\phi_0}{2\pi} p(\gamma | \eta)\d \gamma.
    \end{equation}
    We first aim at computing
    \begin{equation}
    	k^{I,J}(\tau, \phi_0, \gamma) = \int_{-\infty}^{+\infty} I(t,\phi_0,\gamma) J(t+\tau,\phi_0,\gamma)\d t.
    \end{equation}
    In the case $I=J$, the Fourier transform of $k^{I,I}$ 
    is the power spectral density of $I$ (Wiener–Khinchin theorem)
    \begin{equation}
        \hat{k}^{I,I} = |\hat{I}|^2.
    \end{equation}
    In the more general case,
    \begin{equation}
    	\hat{k}^{I,J} = \bar{\hat{I}} \hat{J}
    \end{equation}
    Moreover, since $I$ is the product of $W$ and $i$,
    its Fourier transform is written as a product of convolution
    \begin{equation}
        \hat{I}(\nu,\phi_0,\gamma) = \sum_{k\in\mathbb{Z}} \hat{W}(\nu-k\omega,\gamma) i_k(\gamma) \exp(\mathrm{i} k\phi_0),
    \end{equation}
    where $i_k$ is the Fourier coefficient
    \begin{equation}
        i_k(\gamma) = \frac{1}{2\pi} \int_{0}^{2\pi} i(\phi, \gamma) \exp(-\mathrm{i} k \phi)  \d \phi.
    \end{equation}
    The same is true for $J$ and the Fourier transform of $k^{I,J}$ is thus written as
    \begin{align}
        \hat{k}^{I,J}(\nu,\phi_0,\gamma) = &\ \sum_{k,l\in\mathbb{Z}}
        \overline{\hat{W}}(\nu-k\omega,\gamma)\hat{W}(\nu-l\omega,\gamma)\nonumber\\
        & \ \times \bar{i}_k(\gamma) j_l(\gamma) \exp(\mathrm{i} (l-k) \phi_0).
    \end{align}
    We now integrate this over the longitude at maximum activity $\phi_0$.
    This longitude only appears in the term $\exp(\mathrm{i} (l-k) \phi_0)$, and we have
    \begin{equation}
        \int_0^{2\pi} \exp(\mathrm{i} (l-k) \phi_0) \frac{\d\phi_0}{2\pi} = \delta_{k,l},
    \end{equation}
    thus the integral of $\hat{k}^{I,J}$ simplifies to
    \begin{equation}
        \int_0^{2\pi} \hat{k}^{I,J}(\nu,\phi_0,\gamma) \frac{\d\phi_0}{2\pi}
        = \sum_{k\in\mathbb{Z}} \left|\hat{W}\right|^2(\nu-k\omega, \gamma) \bar{i}_k(\gamma) j_k(\gamma).
    \end{equation}
    Finally $k^{y,z}$ simplifies to
    \begin{equation}
        \label{eq:kyzsep}
        k^{y,z}(\tau; \eta) = \lambda \int k^W(\tau, \gamma) k^{i,j}(\tau, \gamma) p(\gamma|\eta) \d \gamma,
    \end{equation}
    with
    \begin{align}
        \label{eq:kijsep}
        k^W(\tau, \gamma) &= \int_{-\infty}^{+\infty} W(t, \gamma) W(t+\tau, \gamma) \d t\nonumber\\
        k^{i,j}(\tau, \gamma) &= \frac{1}{2\pi}\int_{0}^{2\pi} i(\phi, \gamma) j(\phi+\omega\tau, \gamma) \d \phi\nonumber\\
        &= \sum_{k\in\mathbb{Z}} \bar{i}_k(\gamma) j_k(\gamma) \exp\left(\mathrm{i} k\omega \tau\right).
    \end{align}

    \subsection{Fourier decomposition of the periodic part}
    We first consider the periodic part components $i(\phi,\gamma)$ and $j(\phi,\gamma)$,
    which can be written in the form:
    \begin{align}
        i(\phi,\gamma) &= \mathbf{1}_\mathrm{vis.}(\phi,\gamma) i_0(\phi,\gamma),\nonumber\\
        j(\phi,\gamma) &= \mathbf{1}_\mathrm{vis.}(\phi,\gamma) j_0(\phi,\gamma),
    \end{align}
    where $\mathbf{1}_\mathrm{vis.}$ is the indicator function of the spots visibility
    (equals to 1 when the spot is on the observer's side of the star, 0 otherwise),
    and $i_0$, $j_0$ describe the effect of the feature without visibility considerations.
    The Fourier expansions of the effect of the feature, including the limb-darkening effect,
    are provided in Appendix~\ref{app:spoteffect}.
    Overall, this can be very well approximated with a few harmonics of the rotation period (typically less than 5).
    We denote by $i_{0,k}$, $i_{0,k}$ these expansions up to a given harmonics $k_m$:
    \begin{align}
        i_0(\phi,\gamma) &= \sum_{k=-k_m}^{k_m} i_{0,k}(\gamma) \exp\left(\mathrm{i} k\phi\right),\nonumber\\
        j_0(\phi,\gamma) &= \sum_{k=-k_m}^{k_m} i_{0,k}(\gamma) \exp\left(\mathrm{i} k\phi\right).
    \end{align}
    As shown in Appendix~\ref{app:spoteffect}, the feature is visible when its longitude $\phi$
    is such that
    \begin{equation}
        \cos\phi \geq - \tan i \tan \delta.
    \end{equation}
    We denote by $\phi_\mathrm{m}$ the maximum allowed value of $\phi$ in the range $[0,\pi]$, i.e.:
    \begin{equation}
        \phi_\mathrm{m} = \begin{cases}
            0,\quad \text{if the spot is never visible, i.e.,} -\tan i \tan \delta \geq 1,\\
            \pi,\quad \text{if the spot is always visible, i.e.,} -\tan i \tan \delta \leq -1,\\
            \arccos(-\tan i \tan\delta),\quad \text{otherwise.}\\
            \end{cases}
    \end{equation}
    The Fourier coefficients of $i$ are then given by
    \begin{align}
        i_l(\gamma) &= \frac{1}{2\pi} \int_{-\pi}^{\pi} \mathbf{1}_\mathrm{vis.}(\phi,\gamma) i_0(\phi,\gamma) \exp(-\mathrm{i} l\phi) \d\phi\nonumber\\
        &= \frac{1}{2\pi} \int_{-\phi_\mathrm{m}}^{\phi_\mathrm{m}} i_0(\phi,\gamma) \exp(-\mathrm{i} l\phi) \d\phi\nonumber\\
        &= \frac{1}{2\pi} \sum_{k=-k_m}^{k_m} i_{0,k}(\gamma) \int_{-\phi_\mathrm{m}}^{\phi_\mathrm{m}}  \exp(\mathrm{i} (k-l)\phi) \d\phi\nonumber\\
        &= \frac{\phi_\mathrm{m}}{\pi} \sum_{k=-k_m}^{k_m} \sinc((k-l)\phi_\mathrm{m}) i_{0,k}(\gamma).
    \end{align}
    Similarly, we have
    \begin{equation}
        j_l(\gamma) = \frac{\phi_\mathrm{m}}{\pi} \sum_{k=-k_m}^{k_m} \sinc((k-l)\phi_\mathrm{m}) j_{0,k}(\gamma). 
    \end{equation}
    Since the sinus-cardinal function rapidly decreases, the Fourier expansions of $i$ and $j$
    can still be limited to a few harmonics.
    We note that configurations where the spots are visible for a very short fraction of the period
    ($\phi_\mathrm{m} \ll \pi$) would require to include more harmonics since the sinus-cardinal will decay more slowly with $k-l$.
    However, such configurations also correspond to spots that always remain close to the limb of the star,
    which typically lead to a lower amplitude in the RV and indicators,
    due to projection and limb-darkening effects.
    The kernel $k^{i,j}$ can thus be approximated with few harmonics:
    \begin{equation}
        \label{eq:kijtrunc}
        k^{i,j}(\tau,\gamma) = \sum_{k=-k_m}^{k_m} \bar{i}_k(\gamma) j_k(\gamma) \exp(\mathrm{i} k \omega \tau).
    \end{equation}
    
    \subsection{Efficient modeling with \spleaf{}}
    
    The kernel of Eq.~(\ref{eq:kijtrunc}) can be very efficiently modeled using \spleaf{} 
    \citep{delisle2019b,delisle2022} or \celerite{} \citep{foremanmackey2017}.
    Indeed, in the general case, the cost of likelihood evaluations
    of a GP is proportional to the number of measurements cubed.
    For a restricted class of kernel functions,
    the \spleaf{} and \celerite{} GP frameworks allow to significantly reduce this cost,
    to a linear scaling with the number of points.
    In order to be able to use these frameworks, the covariance between the measurements $Y=(Y_1, \dots, Y_n)$
    must be semiseparable:
    \begin{equation}
    \cov(Y_a, Y_b) = \sum_{s=1}^r U_{a,s} V_{b,s},\quad \text{for $a>b$,}
    \end{equation}
    where $r$ is the rank of the semiseparable representation.
    In particular, sums and products of exponential kernels, sinusoidal kernels, and Matérn kernels
    are semiseparable.
    As shown in \citet{delisle2022}, this reasoning remains valid when considering several
    time series affected by the same GP but with different coefficients.
    In this case, the different time series
    should be merged in a single heterogeneous time series $Y$, ordered by increasing time,
    and whose covariance should be semiseparable.
    
    Let us consider $m$ time series $y^{(1)}(t),\dots,y^{(m)}(t)$, which we merge and sort
    by increasing time in a single heterogeneous time series $Y$.
    Considering only the periodic part $k^{i,j}$ for a fixed set of spot parameters $\gamma$,
    and ignoring the window part $k^W$ for now,
    the covariance matrix of the merged time series is
    \begin{equation}
        \label{eq:covijsep}
        \cov(Y_a, Y_b) = \sum_{k=-k_m}^{k_m} \left(i_{\mathcal{I}_a,k}\exp(\mathrm{i}k\omega t_a)\right)
        \left(\bar{i}_{\mathcal{I}_b, k}\exp(-\mathrm{i}k\omega t_b)\right),
    \end{equation}
    for $a>b$ and where $\mathcal{I}_k$ is the index associating to each point in the merged time series $Y$
    the identifier of the original time series it comes from
    and $i_{s,k}$ is the $k$-th Fourier coefficient of the time series $s$.
    The covariance of Eq.~(\ref{eq:covijsep}) is thus semiseparable with rank $r = 2k_m+1$.
    However, to obtain Eq.~(\ref{eq:covijsep}), we neglected the window part $k^W$, and more importantly,
    we assumed fixed parameters $\gamma$ for the spots.
    
    \subsection{Window function}
    \label{sec:winfunc}

    Up to now we did not specify the shape $W$ of the spots' appearing and disappearing process, which models the evolution of the combined effect of area and temperature of the feature as a function of time. 
    In Section \ref{sec:windowfunction}, we mentioned that on the Sun, spots and faculae tend to appear faster than they decay. To have a S+LEAF representation of the noise, in the present work the window function is such that both its increase and decrease in intensity are exponential, potentially with different time-scales. 
    
    %????????????? Physical justification ?????????????
    
    We provide in Table~\ref{tab:windowkernel} the kernel $k^W$ obtained when assuming
    different window shapes $W$: sudden spot appearing and exponential decay, symmetric or asymmetric exponential appearing and disappearing.
    In these three cases, the obtained kernel is semiseparable with low rank
    \citep[see][]{foremanmackey2017,delisle2022}.
    More generally, the Matérn kernel family can be used for efficient modeling of $k^W$ with \spleaf{} or \celerite{}.
    
    \begin{table*}
        \centering
        \begin{tabular}{ccccc}
             \hline
             \hline
             & $W(t)$ & $k^W(\tau)$ & kernel type & rank \\
             \hline
             Sudden appearing and exp. decay & $\mathbf{1}_{t>0} \exp(-t/\rho)$ & $\exp(-|\tau|/\rho)$ & Matérn 1/2 & 1\\
             Asymmetric exp. (dis)appearing & $\mathbf{1}_{t<0} \exp(t/\rho_-)+\mathbf{1}_{t>0} \exp(-t/\rho_+)$ 
             & $\frac{\rho_+\exp(-|\tau|/\rho_+) - \rho_-\exp(-|\tau|/\rho_-)}{\rho_+ - \rho_-}$
             & Sum of Matérn 1/2 & 2\\
             Symmetric exp. (dis)appearing & $\exp(-|t|/\rho)$ & $ \left(1+\frac{|\tau|}{\rho}\right) \exp(-|\tau|/\rho)$ & Matérn 3/2 & 3\\             
             \hline
        \end{tabular}
        \caption{Kernels ($k^W$) obtained for different window shapes ($W$),
        normalized such that $k^W(0)=1$.}
        \label{tab:windowkernel}
    \end{table*}
    
    Finally, for fixed spot properties $\gamma$, the covariance matrix of the merged time series $Y$, 
    including the window contribution,
    is semiseparable with rank $r = r_W r_\mathrm{per.} = r_w (2k_m+1)$ 
    since the kernel is the product of two semiseparable kernels (see Eq.~(\ref{eq:kyzsep})).
    
    \subsection{Distribution of spot properties}
    \label{sec:spotpropdist}
    We now aim at integrating our kernel over the distribution of spot parameters $\gamma$.
    We assume that the spot parameters $\gamma$ can be split in two sets of independent parameters,
    $\gamma_W$ (only affecting $W$) and $\gamma_\mathrm{per.}$ only affecting the periodic part.
    By independent we mean
    \begin{equation}
        p(\gamma_W,\gamma_\mathrm{per.} | \eta) = p(\gamma_W | \eta)p(\gamma_\mathrm{per.} | \eta),
    \end{equation}
    such that the integral of Eq.~(\ref{eq:kyzsep}) can be written as
    \begin{equation}
        \label{eq:kyzsepprior}
        k^{y,z}(\tau; \eta) = \lambda k^W(\tau; \eta) k^{i,j}(\tau; \eta),
    \end{equation}
    with
    \begin{align}
        k^W(\tau; \eta) &= \int k^W(\tau, \gamma_W) p(\gamma_W|\eta)\d\gamma_W,\nonumber\\
        k^{i,j}(\tau; \eta) &= \int k^{i,j}(\tau, \gamma_\mathrm{per.})p(\gamma_\mathrm{per.}|\eta)\d\gamma_\mathrm{per.}.
    \end{align}
    In the following, we assume that the integral of the window kernel $k^W(\tau; \eta)$ can still be approximated
    with a Matérn kernel, such that it remains semiseparable.
    For the periodic part, the Fourier coefficients of $k^{i,j}(\tau; \eta)$ are given by
    \begin{equation}
        k^{i,j}_k(\eta) = \int \bar{i}_k(\gamma) j_k(\gamma) p(\gamma|\eta)\d\gamma.
    \end{equation}
    These integral might be performed analytically for some specific choices of
    distributions $p(\gamma|\eta)$.
    However, this is beyond the scope of this article and we rather turn here to numerical integrals.
    We thus need to perform an integral for each harmonics $k$ and each couple $i,j$ of original time series
    (RV and indicators).
    For each harmonics $k$ we obtain a Hermitian positive definite matrix $C_k$ of size $m\times m$ of these integrated coefficients,
    where $m$ is the number of original time series.
    We then compute a square root of the matrix $C_k$ (e.g., by Cholesky decomposition or eigendecomposition)
    \begin{equation}
       C_k = R_k R_k^*,
    \end{equation}
    such that the coefficients $k^{i,j}_k$ are given by
    \begin{equation}
        k^{i,j}_k(\eta) = \sum_{s=1}^{m} R_{k, i,s} \bar{R}_{k, j,s}.
    \end{equation}
    With such a decomposition, we obtain a semiseparable representation for $k^{i,j}$
    with rank $m (2k_m+1)$, where $m$ is the number of considered time series (RV and indicators)
    and $k_m$ is the harmonics at which the Fourier series is truncated.
    Finally, the full covariance matrix -- including the window part $k^W$ and integrated over
    the distribution of spots parameters -- is semiseparable with rank $r = m r_W (2k_m +1)$.
    
    We note that the integrals over the distribution of spot parameters required to compute the matrices $C_k$ and $R_k$
    might present a significant computational cost.
    For better performances, we precompute the matrices $R_k$ on a grid of values for $\eta$,
    and then interpolate on this grid to get an estimate of $R_k$ for a given $\eta$.

    \subsection{Variable rate $\lambda$}
\label{app:variable_rate}
    
    We now consider the case of a variable spot appearance rate $\lambda$.
    We assume that $\lambda$ is a stationary random process with expectation $\mu_\lambda(\eta)$,
    and covariance function $k^\lambda(\tau; \eta)$.
    Following previous sections, we obtain the following conditional expectations (given $\lambda$):
    \begin{align}
        \E\{y(t_i) | \lambda, \eta\} &= \int_{-\infty}^{+\infty} \lambda(t) \E_\gamma\{I(t_i-t,\gamma)\}\d t,\nonumber\\
        \E\{y(t_i)z(t_j) | \lambda, \eta\} &= \int_{-\infty}^{+\infty} \lambda(t)\E_\gamma\{I(t_i-t,\gamma) J(t_j-t,\gamma)\}\d t\nonumber\\
        & + \E\{y(t_i) | \lambda, \eta\}\, \E\{z(t_j) | \lambda, \eta\}.
    \end{align}
    Moreover, the reasoning of Sect.~\ref{sec:sepwinper} holds and we have
    \begin{align}
        \E_\gamma\{I(t_i-t,\gamma)\} &= \E_\gamma\{W(t_i-t,\gamma)\} \E_\gamma\{i_0(\gamma)\},\\
        \E_\gamma\{I(t_i-t,\gamma) J(t_j-t,\gamma)\} &= \E_\gamma\{W(t_i-t,\gamma)W(t_j-t,\gamma)\} k^{i,j}(t_i-t_j; \eta).\nonumber
    \end{align}
    By the law of total expectation, we obtain
    \begin{align}
        & \E\{y(t_i)z(t_j) | \lambda, \eta\} = \mu_\lambda(\eta) k^W(t_i-t_j; \eta) k^{i,j}(t_i-t_j; \eta)\\
        & + \E\{y(t_i) | \eta\}\, \E\{z(t_j) | \eta\}\nonumber\\
        & + \E_\gamma\{i_0(\gamma)\} \E_\gamma\{j_0(\gamma)\} \nonumber\\
        & \times \int_{-\infty}^{+\infty}\int_{-\infty}^{+\infty} k^\lambda(t-s) \E_\gamma\{W(t_i-t,\gamma)\} \E_\gamma\{W(t_j-s,\gamma)\} \d t \d s.\nonumber
    \end{align}
    The last line is the convolution product of $k^\lambda$
    and the autocorrelation of the mean window function $k^{\mu_W}$.
    This autocorrelation $k^{\mu_W}$ differs from the mean autocorrelation $k^W$,
    since the mean over the spot properties is taken before computing the autocorrelation.
    
    Finally, the kernel function in the case of a variable spot appearance rate is given by
    \begin{equation}
        \label{eq:kyzvarla}
        k^{y,z}(\tau; \eta) = \mu_\lambda k^W(\tau;\eta) k^{i,j}(\tau;\eta) + \mu_i(\eta)\mu_j(\eta) \left(k^\lambda * k^{\mu_W}\right)(\tau; \eta),
    \end{equation}
    where $\mu_i(\eta) = \E_\gamma(i_0(\gamma))$, $\mu_j(\eta) = \E_\gamma(j_0(\gamma))$.
    The first term in Eq.~(\ref{eq:kyzvarla}) is exactly the same as
    in the case of a constant rate $\lambda = \mu_\lambda$,
    but we have an additional term capturing the variations of $\lambda$.
    
%    If we assume that both $k^\lambda$ and $k^{\mu_W}$ can be approximated by Matérn 1/2

    %in Sect.~\ref{sec:spotpropdist}.

    \subsection{Granulation covariance}
    \label{app:granulation}

    In this section, we establish the expression of the covariance for the model of granulation presented in Section \ref{sec:granulation}. Let us recall that in our model, we suppose that granule packets appear at times following a Poisson process $\lambda_0$. If a granule packet appears at time $t_0$, then several granules appear at $N$ times $t_i, i=1..N$ distributed as a Poisson process of rate $\lambda$ on the time interval $[t_0, t_0+T]$. As a result, $N$ follows a Poisson distribution of parameter $\lambda T$. The impulse response is then
\begin{align}
     g_{gran}(t,\gamma) &= \sum\limits_{i=0}^N  P(t, \gamma) W(t-t_i) \label{eq:ggran}
\end{align}
where $P(t, \gamma)$ is the effect of the granules moving as the star rotates and $W(t-t_i)$ the window functions of granules growing and vanishing, assumed to be identical for all granules for simplicity. The reasoning is very similar if this assumption is dropped.

We have established the expression of the covariance in Eq. \eqref{eq:covar}, and we replace the impulse response by the expression of Eq \eqref{eq:ggran}. Since the time of appearance of granules $t_i$ follow a Poisson process, we can write them $t_i = t_0 + u_i$ where $u_i$ is uniformly distributed on $[0,T]$.
To simplify notations, we write $P(t,\gamma) = P(t)$ and $W(t,\gamma) = W(t)$. 
The covariance between the value of the granulation process at $t$ and $t+\tau$ is 
\begin{align}
k_{gran}(\tau)= \frac{1}{T^2} \sum\limits_{i,j=0}^N \int_{-\infty}^{\infty}  \dd t P(t) P(t+\tau) \int_{0}^T\int_{0}^T W(t  -u_i) W(t + \tau -u_j) \dd u_i \dd u_j  \label{eq:kgran1}
\end{align}
$k_{gran}(\tau)$ can be decomposed in two terms
\begin{align}
D &=  \frac{1}{T} \sum\limits_{i=0}^N \int_{-\infty}^{\infty} \dd t P(t) P(t+\tau)  \int_{0}^T W(t  -u_i) W(t + \tau -u_i) \dd u_i  \\
 & = \frac{N}{T} \int_{-\infty}^{\infty} \dd t P(t) P(t+\tau)  \int_{0}^T W(t  -u) W(t + \tau -u) \dd u \\
 & =  N \int_{-\infty}^{\infty} \dd t P(t) P(t+\tau)  w_2(t,\tau) \dd t
\end{align}
where
\begin{align}
    w_2(t, \tau) = \frac{1}{T} \int_{0}^T W(t  -u) W(t + \tau -u) \dd u
\end{align}
and 
\begin{align}
C &=  \frac{1}{T^2} \sum\limits_{i,j=0, i \neq j}^N \int_{-\infty}^{\infty}  \dd t P(t) P(t+\tau)  \int_{0}^T\int_{0}^T W(t  -u_i) W(t + \tau -u_j) \dd u_i \dd u_j \\
 & = N(N-1) \int_{-\infty}^{\infty} P(t)    P(t + \tau)  w(t) w(t+\tau)  \dd \tau
\end{align}
where
\begin{align}
    w(t) = \frac{1}{T} \int_{0}^T W(t-u) \dd u
\end{align}
Now marginalizing on $N$ we obtain  
\begin{align}
    k_{gran}(\tau) = \lambda T \int_{-\infty}^{\infty} P(t)    P(t + \tau)  w_2(t,\tau) \dd t  + (\lambda T)^2 \int_{-\infty}^{\infty} P(t)    P(t + \tau)  w(t) w(t+\tau) \dd t . \label{eq:kgranul}
\end{align}
Since granule packets lifetimes are typically much lower than the rotation period (one hour vs 25 days on the Sun), it can be assumed that $P(t)$ is approximately constant. Let us recall that we dropped the dependency of $P$ 
on the parameters $\gamma$. We here re-introduce
equal to $P(t_0, \gamma)$ and \ref{eq:kgranul} simplifies to
\begin{align}
    k_{gran}(\tau,\gamma) = &P(t_0,\gamma)  \lambda T \int_{-\infty}^{\infty} W(t)   W(t + \tau)   \dd t  + P(t_0,\gamma) \\ & +(\lambda T)^2 \int_{-\infty}^{\infty}  w(t) w(t+\tau) \dd t . \label{eq:kgranul}
\end{align}
Now marginalizing on $\gamma$, we have 
\begin{align}
    k_{gran}(\tau,\gamma) \propto \lambda T \int_{-\infty}^{\infty} W(t)   W(t + \tau)   \dd t  + (\lambda T)^2 \int_{-\infty}^{\infty}  w(t) w(t+\tau) \dd t . \label{eq:kgranul}
\end{align}

\end{document}